\documentclass[11pt]{article}
\usepackage{jheppubmodnew}
\usepackage{float}
\pdfoutput=1
\usepackage{amssymb}
\usepackage{amsmath}
\usepackage{soul}
\usepackage{amsthm}
\usepackage{subfig}
\usepackage{mathtools}
\usepackage{graphicx}
\usepackage{xcolor}
\usepackage{cite}
\usepackage{epsfig}
\usepackage{braket}
\usepackage[colorlinks=true]{hyperref}
\usepackage{simpler-wick}

\newcommand{\alset}{{\cal A}}
\newcommand{\altrunc}{\alset_{\text{trunc}}}
\newcommand{\alcoarse}{\alset_{\text{coarse}}}
\newcommand{\albartrunc}{\overline{\alset}_{\text{trunc}}}
\newtheorem*{question}{Question}

\newtheorem*{conclusion}{Conclusion}
\newtheorem*{definition}{Definition}

\def\al{A}

\def\tphi{\widetilde{\phi}}

\def\Or[#1]{{\text{O}}\left({#1}\right)}

\newcommand{\be}{\begin{equation}}
\newcommand{\ee}{\end{equation}}
\newcommand{\ba}{\begin{align}}
\newcommand{\ea}{\end{align}}
\newcommand{\bs}{\begin{split}}
\def\sess\end{split}
\def\tr{\mathrm{tr}}

\usepackage{tikz}
\usetikzlibrary{matrix}
\usetikzlibrary{decorations.markings,calc,shapes,decorations.pathmorphing,patterns,decorations.pathreplacing,arrows.meta}
\tikzset{snake it/.style={decorate, decoration=snake}}
\usetikzlibrary{patterns}
\usetikzlibrary{positioning}
\usetikzlibrary{decorations.pathreplacing,calligraphy}
\title{\mbox{Seeing Page Curves and Islands with Blinders On}}
\author{Hao Geng$^{a}$, Andreas Karch$^{b}$, Carlos Perez-Pardavila$^{d}$, Suvrat Raju$^{c}$, Lisa Randall$^{a}$ and Marcos Riojas$^{a}$}
\affiliation{$^a$Gravity, Spacetime, and Particle Physics (GRASP) Initiative, Harvard University, 17 Oxford St., Cambridge, MA, 02138, USA.}
\affiliation{$^b$Theory Group, Department of Physics, University of Texas, Austin, TX 78712, USA.}
\affiliation{$^c$International Centre for Theoretical Sciences, Tata Institute of Fundamental Research, Shivakote, Bengaluru 560089, India.}
\affiliation{$^d$Department of Physics and Beyond: Center for Fundamental Concepts in Science, Arizona State University, Tempe, AZ 85281, USA.}
\emailAdd{haogeng@fas.harvard.edu, karcha@utexas.edu, cperezpa@asu.edu}
\emailAdd{suvrat@icts.res.in, randall@g.harvard.edu}
\emailAdd{marcos\_riojas@fas.harvard.edu}
\date{}
\abstract{This paper summarizes recent discussions of the Page curve and the information paradox, and responds to the reasoning and examples from \texttt{arXiv:2506.04311}. We review arguments demonstrating that in quantum gravity the algebra of observables at infinity is complete, both in AdS and in asymptotically flat space. This completeness implies that the bulk Hilbert space in quantum gravity does not factorize along the radial direction, undermining a key common assumption in Hawking's argument for information loss and in initial derivations of the Page curve. As a consequence, in a standard theory of gravity, information does not ``emerge'' from a black hole in the manner suggested by the Page curve; rather, it is already encoded in asymptotic observables. Relatedly, the full black hole interior, and not just an ``island'', can be reconstructed from exterior data. Page curves and islands can be obtained by removing the Hamiltonian from the exterior algebra. This may be implemented operationally by restricting access to part of the asymptotic region (a detector with a ``blind spot'') or, in the special case of null infinity in asymptotically flat spacetimes, by formally discarding the Hamiltonian from the set of observables despite its physical accessibility. Such Page curves describe only the redistribution of information between measured and unmeasured degrees of freedom, rather than fundamental information recovery. Finally, Page curves and islands also arise when a black hole is coupled to a nongravitational bath, a setup that yields  a nonstandard theory of gravity.  We show how, even in this setting, the unusual localization of information in gravity provides a concrete physical mechanism for information transfer from the gravitational system into the bath.
}

\usepackage{tikz}
\usetikzlibrary{arrows.meta, positioning, shapes.geometric}

\tikzset{
  box/.style={
    draw, rounded corners=2pt, very thick,
    minimum height=9mm, minimum width=32mm,
    align=center, font=\sffamily,
    text width=36mm, inner sep=5pt
  },
  >={Stealth[length=3mm,width=2mm]},
  arrow/.style={-Stealth, very thick},
}
\begin{document}
\maketitle
\flushbottom
\section{Introduction}
In the past few years, considerable attention has been devoted to toy models of AdS black holes coupled to nongravitational baths. The Page curve \cite{Page:1993df,Page:1993wv} describing the entanglement between two parts of the bath has been derived using holographic computations  \cite{Almheiri:2019psf,Penington:2019npb, Almheiri:2019qdq, Almheiri:2019hni,Almheiri:2020cfm,Geng:2024xpj} in which a compact entanglement wedge called an ``island'' plays a key role. It has been suggested that such computations provide a resolution of the black hole information paradox \cite{Hawking:1976ra}.  

In previous work \cite{Geng:2020qvw,Geng:2020fxl,Geng:2021hlu} we showed that this Page curve trivializes in the absence of a bath or when the bath is made gravitational. We also argued that compact entanglement wedges do not appear in standard theories of gravity. This is consistent with a simpler resolution of the information paradox: in a theory of quantum gravity, information about the black hole interior is always accessible in the exterior \cite{Laddha:2020kvp,Raju:2020smc,Raju:2021lwh} --- a feature called the ``principle of holography of information.''   Framed in terms of observables, this implies that observables in the exterior of the black hole are always sensitive to the entire interior and not just to an island. Equivalently, no change of the state can be made to the black hole interior without affecting the exterior.

Recently, \cite{Antonini:2025sur} argued that the Page curve and islands remain relevant to the information paradox even in standard theories of gravity. We respond to their arguments in this paper. For more discussions and  detailed analysis, we refer the reader to \cite{Geng:2025byh}.

We note that \cite{Antonini:2025sur} does not challenge the technical arguments underlying our previous results. Our argument for the triviality of the Page curve and the inconsistency of islands in standard gravity crucially relied on the fact that the Hamiltonian is a boundary term in gravity. The Page curves in  \cite{Antonini:2025sur}  are obtained by artificially removing the Hamiltonian from the set of observables in the exterior. 

In asymptotically AdS spacetimes, the Hamiltonian can be made inaccessible by dividing the boundary into two parts and restricting observations to one part. This can be thought of as a detector with a ``blind spot'' and \cite{Antonini:2025sur} obtained a Page curve by further studying an atypical initial state with a black hole that is localized near the blind spot.  In asymptotically flat spacetimes, since the algebra in the strict asymptotic limit at future null infinity (${\cal I}^{+}$)  becomes free, one can formally drop the Hamiltonian from the algebra, even though it is at least as accessible as other observables.

Said another way, to see a Page curve in a standard theory of gravity, we must effectively put on our blinders by choosing to artificially neglect some accessible degrees of freedom.  The Page curves obtained in such cases correspond to the flow of information between the degrees of freedom that we choose to measure and those that are not measured. Therefore, they do not measure information emerging from the black hole and are not directly relevant to the information paradox. 

We also note that such possibilities for obtaining a Page curve by dropping the ADM Hamiltonian from the set of allowed observables were earlier discussed in \cite{Laddha:2020kvp,Raju:2020smc,Raju:2021lwh}.

Our objective in this paper is not simply  to point out deficiencies in extant computations of the Page curve but also to explain why the principle of holography of information is physically important and provides an elegant resolution to the information paradox. We review how this principle reveals a key error in Hawking's argument for information loss.  Moreover, we will show that even in models with a nongravitational bath the principle of holography of information  provides a physical explanation for how information about the black hole enters the bath.

Turning to islands, we provide additional arguments to show that islands, which we defined in \cite{Geng:2021hlu} to be entanglement wedges that do not reach the asymptotic part of the  gravitational region,  cannot appear in standard theories of gravity provided the global state is pure. These arguments refine the observation made in  \cite{Geng:2021hlu} that a nonzero commutator between observables inside the putative compact wedge and observables outside (including the Hamiltonian) makes the wedge inconsistent. 

The ``islands'' described in \cite{Antonini:2025sur} are not compact entanglement wedges. They all include a portion of the asymptotic boundary. This is related to our observation above that the radiation region is defined to be only part of the asymptotic region rather than the entire asymptotic region. This type of entanglement wedge was explicitly constructed in \cite{Geng:2020fxl} and was shown to be consistent with the observations of \cite{Geng:2021hlu}. 

It might appear that a compact entanglement wedge can be made consistent by defining appropriate ``relational observables''.  Relational observables lead to approximately-local algebras of observables that can be identified with a compact region.  These constructions are important because they ensure that the holographic nature of gravity is not in contradiction with our mundane experience of locality. They also ensure that there is no subtlety in measuring a Page curve for qubits in a lab.

However, these approximate algebras are insufficient to define a fine-grained entropy for the black hole exterior that follows the Page curve, or to make islands consistent in standard gravitational theories. We discuss this limitation in detail in a specific construction of relational observables discussed in \cite{Antonini:2025sur} --- which is based on a construction due to Papadodimas and Raju \cite{Papadodimas:2015xma,Papadodimas:2015jra} as generalized recently in \cite{Bahiru:2023zlc,Bahiru:2022oas,Jensen:2024dnl}.  We also show that the same limitation is expected to apply to  any  relational construction.

We show that compact entanglement wedges can appear behind double horizons when the gravitational system is in a mixed state. However, such wedges are inaccessible to an infalling observer and are not germane to the information paradox, which is conventionally framed as a question about black holes formed from the collapse of pure states.

We now discuss aspects of the Page curve and islands in more detail.

\subsection{Page curve} 
\paragraph{Hawking's paradox and Page's argument.}
Before we turn to the Page curve, it is helpful to review Hawking's argument for information loss \cite{Hawking:1976ra}. Hawking started with the assumption that the set of all observables can be factorized into the set of observables on ${\cal I}^{+}$ (corresponding to information available outside the black hole) and a commuting set of observables inside the black hole. Hawking argued that the gravitational constraints were unimportant since the classical no hair theorem suggests that observables outside the black hole retain information only about the mass, charge, and angular momentum of the interior.  Hawking concluded that black hole evaporation through Hawking radiation\cite{Hawking:1974sw} would lead to the loss of all other information about the interior, leaving the state on ${\cal I}^{+}$ in a mixed state obtained by tracing over the interior.

Page \cite{Page:1993df,Page:1993wv} accepted Hawking's assumption of factorization, but posited that information would be transferred from the interior of the black hole to ${\cal I}^{+}$  gradually, as one might expect in an evaporating nongravitational system. This led Page to argue that the entanglement entropy on ${\cal I}^{+}$ would follow a Page curve  that first rises and then comes down to zero if the black hole was formed from the collapse of a pure state.

Historically,  the bias within the high-energy community has been that Hawking's argument is incorrect but Page's argument is valid. However, {\em both} these arguments are based on the same key assumption that observables outside the black hole commute with observables inside. This assumption can be shown to be incorrect in a theory of quantum gravity \cite{Marolf:2008mf,Laddha:2020kvp,Raju:2021lwh,Raju:2020smc} as we explain below. Therefore, the error in Hawking's argument is also the error in Page's argument. Removing the incorrect assumption resolves  Hawking's information paradox, but simultaneously trivializes the Page curve. Conversely, retaining this assumption leads directly to the information paradox.

\paragraph{Holography of information.}
In section \ref{secoutsidecomplete}, we review how the gravitational constraints lead to the principle of ``holography of information''  \cite{Laddha:2020kvp,Raju:2020smc,Raju:2021lwh}. 

A short argument provides intuition for  this phenomenon \cite{Marolf:2013iba}.  In a theory of gravity, the constraints imply that the Hamiltonian is available to the  observer outside the black hole as the integral of a specific metric component over a Gaussian sphere at infinity.   In the quantum theory, the observer can use the Hamiltonian to evolve observables at infinity into the past. Thus, observables in the exterior (including the Hamiltonian)  can probe the black hole interior since the matter that constitutes the black hole was outside it in the past. We provide a more precise argument in section \ref{secoutsidecomplete} that leads to the same conclusion.  

The principle of holography of information implies that observables that appear to be localized in the black hole interior can be expressed in terms of observables at infinity. This immediately implies that the Hilbert space cannot be factorized into a subspace corresponding to the interior and another corresponding to the exterior.

\paragraph{Page curves.}
It is for this reason that the Page curves in the literature do not correspond to the straightforward setting where one surrounds the black hole with a sensitive detector and then quantifies the information available outside the black hole.  Instead, Page curves are usually obtained by adding a nongravitational bath and measuring the flow of information across an imaginary interface in the bath. However, in our world, gravity does not ``switch off'' abruptly at some point. 

Moreover, in the presence of a bath, the gravitational system is an open quantum system, in which nonstandard bulk dynamics applies, and we will refer to this as a massive theory of gravity consistent with the terminology used in \cite{Karch:2000ct,Porrati:2001gx,Aharony:2006hz,Geng:2020qvw,Geng:2023ynk,Geng:2023zhq,Geng:2025rov,Geng:2025byh}.

In \cite{Antonini:2025sur}, it was noted that a Page curve can be obtained in a standard theory of gravity  by dividing the asymptotic boundary into two parts and restricting observations to one part.  Physically, this corresponds to a detector with a ``blind spot'' since it is insensitive to degrees of freedom on a part of the asymptotic boundary. Mathematically, such a detector cannot measure the Hamiltonian, which requires an integral of a metric fluctuation on the entire boundary.  If one further positions a black hole so that it is close to the blind spot initially, then its evaporation leads to the spread of information over the rest of the detector, leading to a Page curve.

\begin{figure}[H]
\begin{center}
\includegraphics[width=\linewidth]{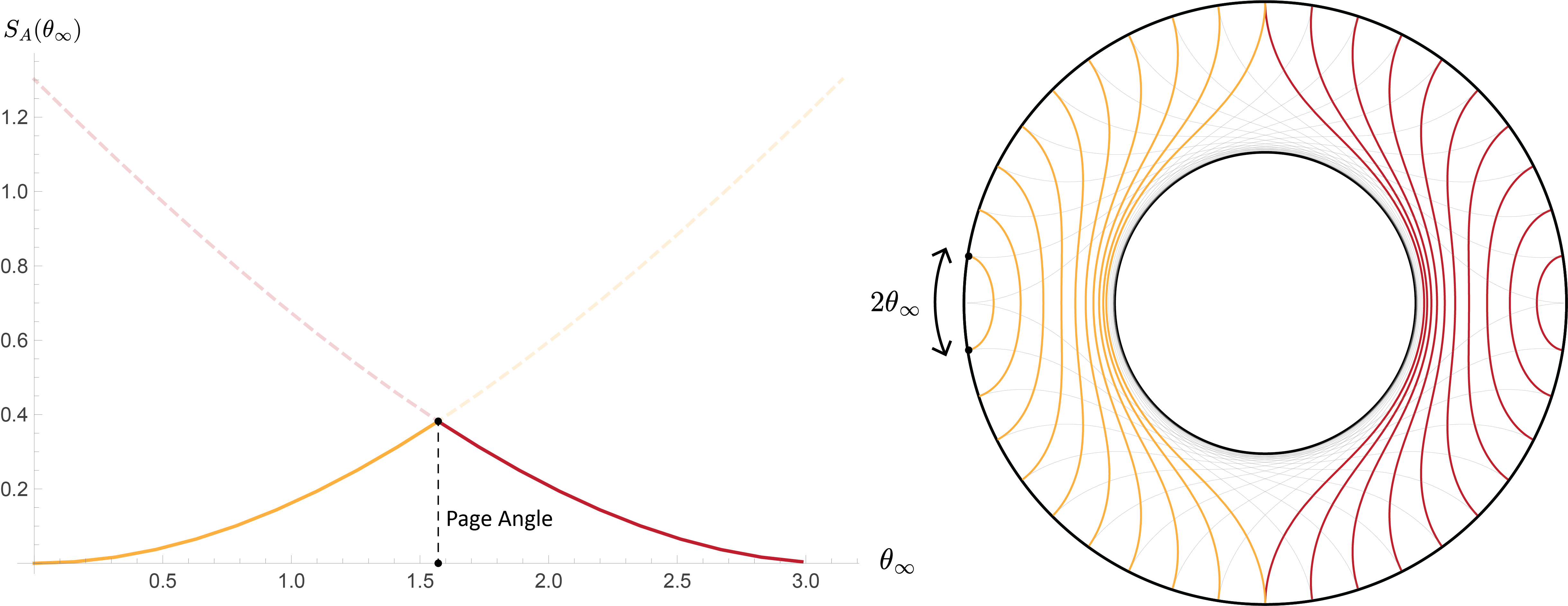}
\caption{Entanglement entropy of a boundary region plotted against its angular size for a  single-sided black hole in AdS dual to a pure state. See section~\ref{subsectrivialpage} for details of the simple computation.}  
\label{figtrivialpage} 
 \end{center}
\end{figure}

Even more simply, if one is allowed to divide the asymptotic boundary into two parts and restrict observations to one part, then the following Page curve can be obtained. Consider a large static black hole in AdS corresponding to a pure state. The entanglement entropy of a part of the boundary obeys a Page curve as a function of the solid angle it occupies.  See Figure \ref{figtrivialpage}. 

Although this static example is technically trivial,  it is conceptually equivalent to the dynamic example described in \cite{Antonini:2025sur}. Both these examples simply measure how information is distributed between different parts of the asymptotic region and they do not correspond to a Page curve that measures information emerging from a black hole.

Finally, as first discussed in detail in  \cite{Laddha:2020kvp,Raju:2020smc,Raju:2021lwh} and then reviewed in  \cite{Antonini:2025sur}, a Page curve may be obtained by discarding the components of the metric that correspond to the Hamiltonian from the set of observables  at ${\cal I}^{+}$ in asymptotically flat spacetimes. This truncation of the algebra is mathematically consistent strictly at ${\cal I}^{+}$, where the algebra becomes free. But as we elaborate below, this division is physically unnatural since natural observables such as the Riemann tensor simultaneously encode information about both the Hamiltonian and other observables. As in the case of AdS in Figure \ref{figtrivialpage}, the Page curve obtained here is simply a measure of the information that we lose on part of ${\cal I}^{+}$ because we chose not to look at it.

One might hope that a Page curve could be found for a detector that surrounds the black hole through some kind of ``natural'' division of the degrees of freedom. We cannot rule out this possibility in general but no such proposal has been advanced in the literature, and we discuss some possible obstacles below.

\subsection{Islands} 
A novel entanglement wedge called an ``island"  makes an  appearance in the calculations of the Page curve with a nongravitational bath. Since this term is not used uniformly, in \cite{Geng:2021hlu} we clarified that we define islands as entanglement wedges that do not extend to the asymptotic region of the gravitating spacetime. These are the kinds of islands that appear in theories with a nongravitational bath. 

Islands are inconsistent in standard gravity for the same reason that there is no Page curve. Observables in the asymptotic region have access to the entire bulk. Therefore, it is not possible to define commuting algebras for a compact bulk region and its complement.

Islands also provide a  physical picture that is misleading in two respects.  First, they suggest that it is only in special configurations (where islands arise) that there is a redundancy between asymptotic gravitational degrees of freedom and bulk degrees of freedom. However, the principle of holography of information tells us that this redundancy is always present, regardless of whether islands appear. Second, islands  suggest that  an observer making measurements on the asymptotic part of a Cauchy slice can reconstruct only part of the bulk. But the principle of holography of information tells us that, in a standard theory of gravity,  asymptotic measurements are sufficient to reconstruct the entire bulk.

Our arguments about the inconsistency of islands do not apply when the entanglement wedge and its complement both contain an asymptotic gravitational region and a disconnected compact part. This is obvious, since excluding part of the asymptotic region  excludes the Hamiltonian, which requires an integral over a complete sphere in the asymptotic region. 

In this paper, we will continue to use the term ``island'' in accordance with our previous definition \cite{Geng:2021hlu}, which excludes entanglement wedges that include an asymptotic piece.\footnote{We note that the component of the entanglement wedge purely inside the bulk was dubbed the ``pseudo island" in previous work \cite{Geng:2025byh,Geng:2025gqu}.} While such wedges have not been found by explicit calculations, \cite{Antonini:2025sur} plausibly argued that they should be relevant in the time-dependent states alluded to above when one studies information transfer between two parts of the asymptotic region or, said more physically, when one studies detectors with ``blind spots''.

\paragraph{Summary and organization of paper.}
To summarize: our claim is not (and has never been) that existing computations of the Page curve or of islands are technically incorrect. Our point is that these computations obscure the essential gravitational physics of holography that is crucial to resolving the information paradox. The new  examples of Page curves described in \cite{Antonini:2025sur} obtained by discarding the Hamiltonian from the algebra have an unclear relationship to the information paradox. While the Page curve is of historical interest in the context of the information paradox, and important in nongravitational systems, it is not evident what physical insight we gain by constructing artificial systems that obey a Page curve.

Our paper is organized as follows.
\begin{enumerate}
\item
In  section \ref{sechawkingpage}, we review Hawking's argument for information loss and also Page's argument leading to a tent-shaped  Page curve for the von Neumann entropy on ${\cal I}^{+}$.  We emphasize how both these arguments rely on the assumption that observables in the black hole spacetime can be factorized into a subalgebra that can be associated with the exterior and a commuting subalgebra associated with the interior. 
\item \label{complalgebra}
In section \ref{secoutsidecomplete}, we review the arguments that show that the set of observables at infinity, including the Hamiltonian, are complete in a standard theory of gravity both in asymptotically flat spacetimes and in spacetimes that are asymptotically AdS.  Therefore, in a standard theory of gravity, information about the black hole interior is always available outside the black hole. 
\item
In section \ref{secpagecurves}, we review how a Page curve is obtained in standard theories of gravity. The essential step is to make the Hamiltonian inaccessible ---  in AdS, this is done by dividing the asymptotic region into two parts, and in flat space by discarding some gravitational degrees of freedom by hand. We describe an important open question about Page curves for small AdS black holes.
\item
In section \ref{secinconsistency} we explain why islands --- entanglement wedges with a compact overlap with the gravitational region ---  are inconsistent in standard theories of gravity. In section \ref{secrelational}, we explain why relational observables cannot be used to make islands consistent or define a fine-grained entropy for the black hole exterior that follows the Page curve.
\item
In section \ref{secgoodislands}, we show how adding a nongravitational bath allows islands to be consistent. Islands can also be consistent when they are behind double horizons. We also review consistent entanglement wedges that are a union of a disconnected compact piece and an asymptotic piece, although these are not islands according to our definition. 
\item
In section \ref{secinformtransfer} we show that the principle of holography of information is important even in the presence of a nongravitational bath: it helps to resolve potential puzzles and explains how information is transferred from the black hole interior to the bath.
\item
Section \ref{secdiscussion} contains a summary and describes some open questions.
\end{enumerate}

We use physical terms, such as ``detectors'', only to provide a physical interpretation for mathematical results about well-defined algebras of observables. Since black hole information is a subtle subject, we eschew the use of words that cannot be made precise in this manner.

\section{The information paradox and the Page curve \label{sechawkingpage}}
In this section we review the arguments advanced by Hawking leading to the information paradox, and those by Page that suggested that information should emerge from the black hole according to a Page curve. Our objective in reviewing this standard material is to emphasize that {\em both} these arguments share the assumption that the set of observables factorize into a part that can be associated with the interior and another part that is associated with the exterior. It is now understood that this assumption is incorrect.

Before we describe Hawking's argument in section \ref{subsechawkingarg} we will review a naive but common argument for information loss in section \ref{secbhrad} that only relies on the spectrum of radiation emitted by a black hole. The flaw in this naive argument is evident from simple considerations of quantum mechanics and does not even require us to examine the issue of factorization in the bulk.

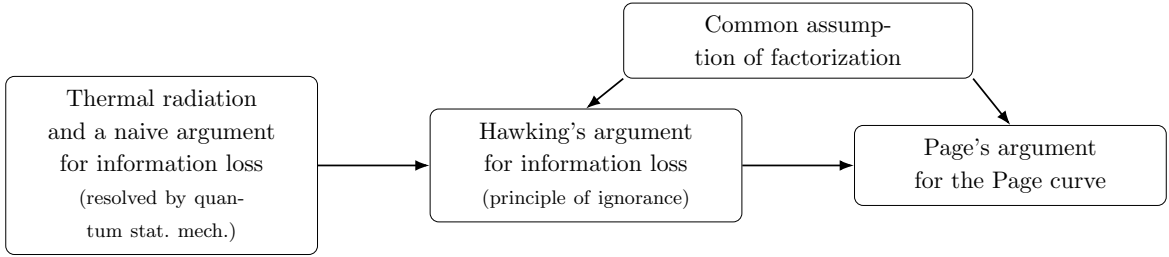
\begin{figure}
\begin{center}
\resizebox{\textwidth}{!}{
\begin{tikzpicture}[
  box/.style={
    draw,
    rounded corners,
    align=center,
    inner sep=6pt,
    text width=4.6cm
  },
  arrow/.style={-Latex, thick}
]

\node[box] (thermal)
{Thermal radiation and a naive argument for information loss\\
\footnotesize (resolved by quantum stat.\ mech.)};

\node[box, right=18mm of thermal] (hawking)
{Hawking's argument for information loss\\
\footnotesize (principle of ignorance)};

\node[box, right=18mm of hawking] (page)
{Page's argument for the Page curve};

\draw[arrow] (thermal) -- (hawking);
\draw[arrow] (hawking) -- (page);

\coordinate (midHP) at ($(hawking)!0.5!(page)$);

\node[box, above=14mm of midHP, text width=5.2cm] (factor)
{Common assumption of factorization};

\draw[arrow] (factor.south west) -- (hawking.north);
\draw[arrow] (factor.south east) -- (page.north);

\end{tikzpicture}}

\end{center}
\caption{Flowchart for section \ref{sechawkingpage}. \label{flowchartsechawkpage}}
\end{figure}

Figure \ref{flowchartsechawkpage} depicts the logical flow of this section.
\subsection{Radiation from a black hole \label{secbhrad}}

Consider a field propagating in the spacetime of an evaporating black hole in an asymptotically flat spacetime. At late times, we have an emergent time translational symmetry.  The field can be expanded in terms of modes that are ingoing at the horizon, and those that also have an outgoing component.

Let $N_{\omega}$ be the number operator that measures the occupancy of outgoing modes with frequencies that are close to $\omega$. Hawking \cite{Hawking:1974sw} showed that  the process of black hole formation and evaporation would lead to a state where
\be
\label{thermalrad}
\langle N_{\omega} \rangle = {e^{-\beta \omega}  \over 1 - e^{-\beta \omega}},
\ee
where $\beta$ is the inverse-temperature of the black hole.

The precise radiation seen at ${\cal I}^{+}$ also depends on the greybody factors that arise from the scattering of radiation in the black hole geometry. Setting this simple effect aside, \eqref{thermalrad} implies that the black hole emits a flux of thermal radiation.

It is sometimes believed that this computation, by itself, leads to a paradox. The formula \eqref{thermalrad} suggests that the radiation at ${\cal I}^{+}$ depends only on the black hole temperature and is therefore independent of the initial state of the black hole. So, one might naively conclude that \eqref{thermalrad} implies that the final state observed at ${\cal I}^{+}$ is independent of the initial state.

Our first point is that this naive argument is flawed. Moreover, it is not the argument that Hawking used to argue for the loss of information. 

The flaw in the above naive argument is that in any quantum-mechanical system with a large number of degrees of freedom, a typical pure state is virtually indistinguishable from a mixed state up to exponentially small corrections. Therefore, many pure states can satisfy \eqref{thermalrad} to great accuracy and \eqref{thermalrad} cannot be used, by itself, to conclude that the state at ${\cal I}^{+}$ is thermal.

This can be made more precise as follows. Consider a system with a large number of states --- denoted by $e^{S}$ --- in a band of energies spanned by eigenstates $|E_i \rangle$.  A pure state from this band takes the form
\be
|\Psi \rangle = \sum_{i} a_{i} |E_i \rangle.
\ee
Any choice of $a_i$ subject to the constraint $\sum |a_i|^2 = 1$ yields a valid state. But we can define a notion of a typical state using the ``Haar measure'', defined by
\be
d \mu = {1 \over {\cal V}} \delta(\sum_{i} |a_i|^2 - 1) \prod_{i} d^2 a_{i},
\ee
where the normalization ${\cal V} = {\pi^{e^{S}} \over \Gamma(e^{S})} $ is set by demanding that 
\be
\int d \mu = 1.
\ee
 
On the other hand, compare the microcanonical density matrix drawn from the same band of energies 
\be
\label{rhomixed}
\rho = {1 \over e^{S}} \sum_{i} |E_i \rangle \langle E_i |.
\ee

As quantum-mechanical states, $\rho$ and $|\Psi \rangle$ are very different. However, from the point of view of a typical physical observation, they look almost identical. More precisely, for any projector $P$, we have
\be
\label{meanissame}
\int (\langle \Psi | P |\Psi \rangle  - \tr(\rho P)) d \mu = 0,
\ee
and
\be
\label{stddevissmall}
\int (\langle \Psi | P | \Psi \rangle - \tr(\rho P))^2 d \mu \leq {1 \over e^{S} + 1}.
\ee
Since the probability of any outcome of a measurement is always given by the expectation value of a projector, this tells us that typical pure and mixed states differ only by $\Or[e^{-{S \over 2}}]$.

The results \eqref{meanissame} and \eqref{stddevissmall} were obtained by Lloyd \cite{lloyd1988black}. A review can be found in \cite{Raju:2018xue}.

Note that for any given state, $|\Psi \rangle$, it is always possible to find a specific observable, attuned to the state, with the property that it takes different values for $|\Psi \rangle$ and $\rho$. But if we fix our observable ahead of time then, with overwhelming probability, the observable will give  exponentially close values in the pure state and the mixed state. 

This is a simple result from quantum statistical mechanics. However, it is relevant for the naive argument for the information paradox described above because it tells us the formula for thermal radiation \eqref{thermalrad} cannot, by itself, be used to distinguish between a pure and a mixed state unless we measure the expectation value in \eqref{thermalrad} to exponential precision. If we believe that, in a UV complete theory of quantum gravity, a black hole is drawn from an ensemble of $e^{S_{\text{bh}}}$ states, where $S_{\text{bh}}$ is the Bekenstein-Hawking entropy \cite{Bekenstein:1973ur,Gibbons:1976ue} of the black hole, then even in a typical pure state one would expect \eqref{thermalrad} to hold to excellent accuracy. 

The parameter ${1 \over S_{\text{bh}}}$ also controls the size of gravitational effects since $G {1 \over \beta^{d-1}} \propto {1 \over S_{\text{bh}}}$ for a black hole in $d+1$ dimensions. In order to extend \eqref{thermalrad} to obtain a paradox, one would have to refine the calculation to all orders in ${1 \over S_{\text{bh}}}$ and further to nonperturbative order to rule out the possibility that corrections of $\Or[e^{-{S_{\text{bh}} \over 2}}]$  restore purity to the state.

\paragraph{Small corrections theorem.} We would also like to briefly address the ``small corrections theorem'' of \cite{Mathur:2009hf}. Equations \eqref{meanissame} and \eqref{stddevissmall} show that a pure state and a mixed state resemble each other closely in almost all physical experiments. The small corrections theorem is the observation that
\be
\label{tracedistance}
\tr\left(|\Psi \rangle \langle \Psi| - \rho\right)^2 = 1 - e^{-S},
\ee
for any state $|\Psi \rangle$.  Therefore, when measured in terms of the trace-distance between density matrices, the mixed-state density matrix \eqref{rhomixed} cannot be corrected by small corrections to obtain a pure-state density matrix $|\Psi \rangle \langle \Psi | $.

However, Hawking's computation only yields the expectation values of low-point correlators such as \eqref{thermalrad}. These expectation values are expected to receive exponentially small corrections in a pure state.  The trace distance \eqref{tracedistance}, like the entanglement entropy, is a fine-grained observable and Hawking's computation \eqref{thermalrad} is not precise enough to predict such observables at ${\cal I}^{+}$.

\subsection{Information paradox \label{subsechawkingarg}}
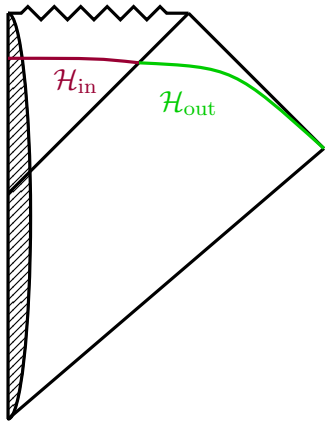
\begin{figure}[!ht]
\begin{center}
\begin{tikzpicture}[scale=0.6,decoration=snake]
    \draw[-,very thick] 
       decorate[decoration={zigzag,pre=lineto,pre
       length=5pt,post=lineto,post length=5pt}] {(-2,0) to (2,0)};
       \draw[-,very thick] (-2,0) to (-2,-9);
       \draw[-,very thick] (2,0) to (-2,-4);
       \draw[-,very thick] (2,0) to (5,-3);
       \draw[-,very thick] (5,-3) to (-2,-9);
       \draw[-,very thick,black] (-2,0) arc (90:-90:0.5 and 4.5);
       \draw[thick,black,pattern=north east lines,pattern color=black] (-2,0) arc (90:-90:0.5 and 4.5)--(-2,0);
       \draw[-,very thick,black!20!green] (5,-3) ..controls (3,-1.2).. (0.9,-1.1);
        \draw[-,very thick,black!20!purple] (-2,-1) ..controls (0,-1).. (0.9,-1.1);
        \node at (2,-2) {\textcolor{black!20!green}{$\mathcal{H}_{\text{out}}$}};
        \node at (-0.5,-1.5) {\textcolor{black!20!purple}{$\mathcal{H}_{\text{in}}$}};
\end{tikzpicture}
\caption{A nice slice in a black hole geometry. Hawking assumed that the Hilbert space would factorize into a part associated with the interior ${\cal H}_{\text{in}}$ and another part associated with the exterior ${\cal H}_{\text{out}}$. \label{figniceslice}}
\end{center}
\end{figure}

Hawking's original argument leading to the information paradox was more refined \cite{Hawking:1976ra} and did not rely purely on \eqref{thermalrad}.
In modern terminology, Hawking's assumption was that the Hilbert space could be written as
\be
\label{hilbfact}
{\cal H}_{\text{full}} = {\cal H}_{\text{in}} \otimes {\cal H}_{\text{out}}, 
\ee
where ${\cal H}_{\text{in}}$ and ${\cal H}_{\text{out}}$ correspond  to degrees of freedom that live on the part of a nice slice that are, respectively, inside and outside the black hole. See Figure \ref{figniceslice}.

It is convenient to rephrase this assumption in terms of the algebra of observables. In this language,  one might equivalently phrase Hawking's assumption as
\be
\label{alfact}
\al_{\text{full}} = \al_{\text{in}} \otimes \al_{\text{out}},
\ee
i.e. that the full algebra of observables factorizes into two commuting subalgebras, one of which can be associated with the interior and the other with the exterior.

Hawking termed the part of the nice slice in the black hole interior a ``hidden surface'' and asserted that \eqref{hilbfact} is a ``basic assumption of quantum theory'' --- although we will see below that it is really a much stronger assumption about how degrees of freedom are localized.  

Furthermore, Hawking posited that the observer outside should adopt a ``principle of ignorance'' with respect to the quantum state  on the nice slice inside the black hole. In \cite{Hawking:1976ra}, this principle was stated as: ``All data on a hidden surface compatible with the observer's limited information are equally probable.'' 

We interpret this to mean, in modern terminology, that an observer outside should democratically trace over ${\cal H}_{\text{in}}$, while imposing a constraint only  on the total ``energy, angular momentum, or charge'' in ${\cal H}_{\text{in}}$. Since there are many possible states in ${\cal H}_{\text{in}}$ that are expected to be compatible with this small number of constraints, one is led to the conclusion that the observer outside would measure a mixed state on ${\cal H}_{\text{out}}$  even if the initial state was pure.

This argument has the virtue that it does not rely on formula \eqref{thermalrad} holding precisely. The formula for black hole radiation is  used only to conclude that the black hole eventually evaporates. Therefore, even if \eqref{thermalrad} receives corrections at higher orders in perturbation theory,  the ``principle of ignorance'' would ensure that the state on ${\cal H}_{\text{out}}$ would remain mixed.

\subsection{Page curve}
We now turn to the Page curve. 
This curve was first introduced by Page in \cite{Page:1993df}. This paper deals with a very general setting in which the Hilbert space of a quantum mechanical system was assumed to factorize into subspaces of dimension $m$ and $n$ as
\be
\label{pagedivision}
{\cal H} = {\cal H}_{m} \otimes {\cal H}_{n}.
\ee
This setting, a priori, is not restricted  to black holes  and, indeed, the Page curve is important in nongravitational systems where the degrees of freedom can be divided as in \eqref{pagedivision}. 

Page studied a typical pure state in the joint system --- where typicality is defined using the Haar measure, as we discussed previously --- and argued that the von Neumann entropy of both subsystems is expected to be
\be
S_{m,n} = \log \left(\text{min}(m,n) \right) - {\text{min}(m,n) \over 2 \text{max}(m,n)},
\ee
up to small corrections.

In a separate paper \cite{Page:1993wv}, Page applied this result to black holes. In order to do so, Page also explicitly assumed that the Hilbert space factorizes into the interior and the exterior. Page stated  that ``the radiation and the black hole are subsystems of a total system of Hilbert space dimension $m n$'' where the ``radiation subsystem has dimension $m \sim e^{s_r}$ where $s_r$ is the thermodynamic radiation entropy, and the black hole subsystem has a Hilbert space dimension $n \sim e^{s_h}$ where $s_h$ \ldots is the semiclassical Hawking entropy.''

\begin{figure}
    \centering
\includegraphics[width=0.7\linewidth]{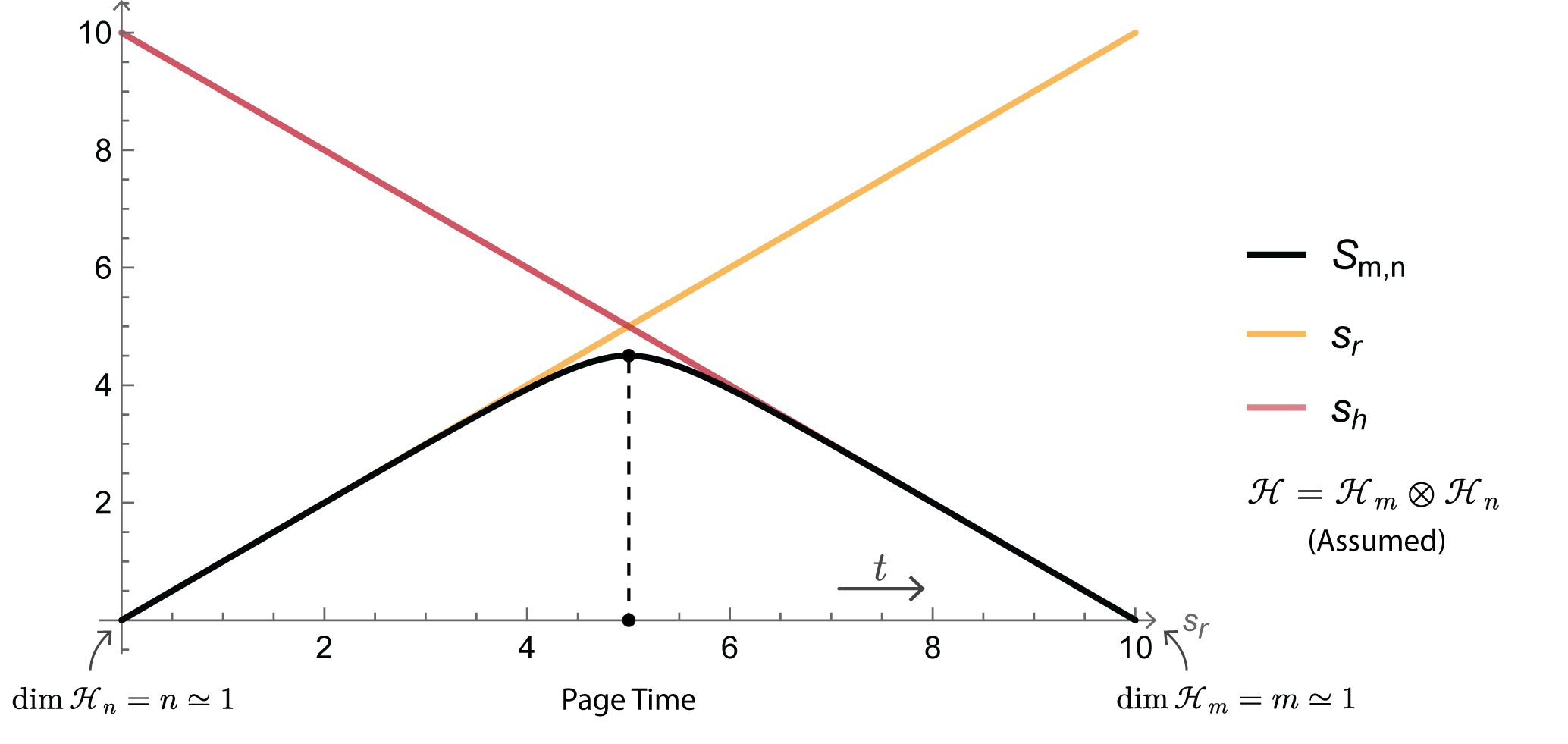}
    \caption{The Page curve \cite{Page:1993df,Page:1993wv} was obtained by modeling the black hole and its radiation as a bipartite system $\mathcal{H} \simeq \mathcal{H}_m \otimes \mathcal{H}_{n}$, with $\operatorname{dim} \mathcal{H}_m=m$ and $\operatorname{dim} \mathcal{H}_{n}=n$ for the black hole and radiation subsystems, respectively. Page expected the subsystems to be nearly maximally mixed, so that the fine-grained entropy nearly saturates the bound set by the smaller Hilbert space $\log(\text{min}(m,n))$. By assuming the joint system is in a Haar random pure state, Page modeled the fine-grained entropy: for $n \ge m$, $S_{m,n} \approx \log m-\frac{m}{2 n}$; for $m \ge n$, $S_{m,n} \approx \log n-\frac{n}{2 m}$. 
    The coarse-grained entropy of the radiation $s_r \approx \log m$ increases with time, while the coarse-grained entropy of the black hole  $s_{h} \approx \log n$ decreases; the fine-grained entropy $S_{m,n}$ begins to decrease around when the coarse-grained entropies become comparable at the Page time.}
    \label{fig:prototypepage}
\end{figure}

Following this reasoning, Page argued that the entanglement entropy of the radiation should obey a Page curve that, qualitatively, follows Figure \ref{fig:prototypepage}.

Page's analysis is different from Hawking's in that Page posited,  based on unitarity, that an unknown mechanism would transfer information from the black hole to the radiation. Nevertheless, Page accepted Hawking's central premise of locality and the factorization of the Hilbert space and used {\em precisely} the same assumption to argue for the Page curve.

In the next section, we will argue that, in quantum gravity, the assumption of factorization fails. This resolves the information paradox, as formulated by Hawking, but also invalidates the argument leading to a tent-shaped curve. Instead, a more-natural answer is that the entanglement entropy of the exterior region is always zero.

On the other hand, by means of various artifices, one can invent systems that involve evaporating black holes in which a Page curve can be obtained. However, such Page curves do not describe the physical process suggested by Page's original calculation --- where information emerges from the black hole at a particular rate.

Here to be clear we want to distinguish between what is in principle measurable and what an observer has chosen to measure. In a standard theory of gravity, the exterior always has zero entropy. However, an observer can measure a Page curve by changing the subsystem size as we discuss later. This description is observer dependent instead of being intrinsic to the black hole. The ``artifices" are systems in which one can precisely measure a curve that follows the Page shape by artificially dropping some operators from the measurable set. These observer-dependent Page curves similarly don't correspond to the original Page curve intent.

\section{Holography of information \label{secoutsidecomplete}}
In this section, we show that the assumption of the independence of observables inside and outside the black hole --- made by both Hawking and Page --- is incorrect. It is possible to argue, with only weak assumptions about the UV-complete theory of gravity, that observables outside the black hole can also probe the interior \cite{Laddha:2020kvp,Raju:2020smc}. We refer the reader to  \cite{Jacobson:2012gh,Jacobson:2019gnm,deMelloKoch:2022sul,deMelloKoch:2024juz} for closely related discussion.

This should not be surprising from the point of view of AdS/CFT \cite{Maldacena:1997re,Witten:1998qj,Gubser:1998bc}. The ``extrapolate dictionary'' tells us that boundary observables are limits of bulk observables. Since boundary observables on a time slice furnish a complete set of observables, asymptotic bulk observables on a single Cauchy slice must also be complete. Below, we show how a similar conclusion can be reached from the bulk without assuming AdS/CFT. 

This result was termed the principle of ``holography of information'' \cite{Raju:2020smc} to distinguish it from a full-fledged holographic dual like AdS/CFT. The arguments below suffice to show that operators at infinity can be used to represent operators in the bulk both in asymptotically AdS and in asymptotically flat spacetimes. But they do not tell us how to recast the dynamics of those operators into an autonomous theory.

This section is structured as follows. In subsection \ref{secasympcomplete}, we  present a full-fledged nonperturbative argument that proves the principle of holography of information, subject to weak and clearly specified assumptions about the UV-complete theory of quantum gravity. This argument might appear to be somewhat formal, but we present it to dispel the common misconception that the principle of holography of information applies only to the vacuum or within perturbation theory. We then present additional discussion to elucidate aspects of the result. In section \ref{secperturbative}, we explain the implications of this principle in the perturbative regime. In section \ref{secmundane}, we delineate the regime in which this principle is unimportant, which we term the ``QFT Limit''.  Black holes {\em cannot} be studied in the ``QFT Limit'', and holography of information is crucial to ensure that black hole evaporation is unitary as we explain in section \ref{secfinegrained}.

The results in this section are meant to provide a constructive resolution of the information paradox. Therefore, this section should be read independently, and not as a response to any specific point made in \cite{Antonini:2025sur}.

\subsection{Asymptotic observables are complete \label{secasympcomplete}}

We will argue that, in a theory of quantum gravity in an asymptotically AdS spacetime, any operator in the theory can be represented on an infinitesimal time band on the AdS boundary. A similar argument holds in asymptotically flat spacetimes, and we quote the result at the end of this subsection. This automatically means that operators sensitive to the black hole interior can be represented in the black hole exterior.

This result requires certain assumptions about the UV-complete theory of gravity that we explain in turn. The first assumption is that near the asymptotic boundary, the language of quantum field theory makes sense even in a theory of quantum gravity. This conforms to the broader principle that, even in quantum gravity, we keep the asymptotic region of spacetime fixed while allowing the bulk to fluctuate. 

We demand that the asymptotic metric takes the form
\be
\label{adsmetric}
d s^2 \underset{r \rightarrow \infty}{\longrightarrow} -(r^2 + 1) dt^2 + {d r^2 \over r^2 + 1} + r^2 d \Omega^2,
\ee
where $d \Omega^2$ is the metric on $S^{d-1}$. 
Let $\phi(r,t,\Omega)$ be a bulk field. The assumption above implies that it makes sense to study the operator, $O(t,\Omega)$, obtained as the  limit of this field as it approaches the boundary 
\be
O(t,\Omega) = \lim_{r \rightarrow \infty} r^{\Delta} \phi(r, t, \Omega),
\ee
where $\Delta$ is related to the mass of the field $\phi$ by the standard dictionary: $\Delta (\Delta - d) = m^2$  \cite{Gubser:1998bc,Witten:1998qj}.

Following the usual rules of quantum mechanics, we can multiply these operators together and take linear combinations. This leads to the asymptotic {\em algebra} of the bulk theory
\be
\begin{split}
\alset  = ~&\text{span}\{O(t_1,\Omega_1), O(t_1,\Omega_1) O(t_2, \Omega_2), \ldots O(t_1, \Omega_1) O(t_2, \Omega_2) \ldots O(t_n, \Omega_n), \ldots\},\\
&\text{with}~~t_i \in (-\infty, \infty).
\end{split}
\ee

\begin{figure}[!ht]
\begin{center}
\includegraphics[width=\linewidth]{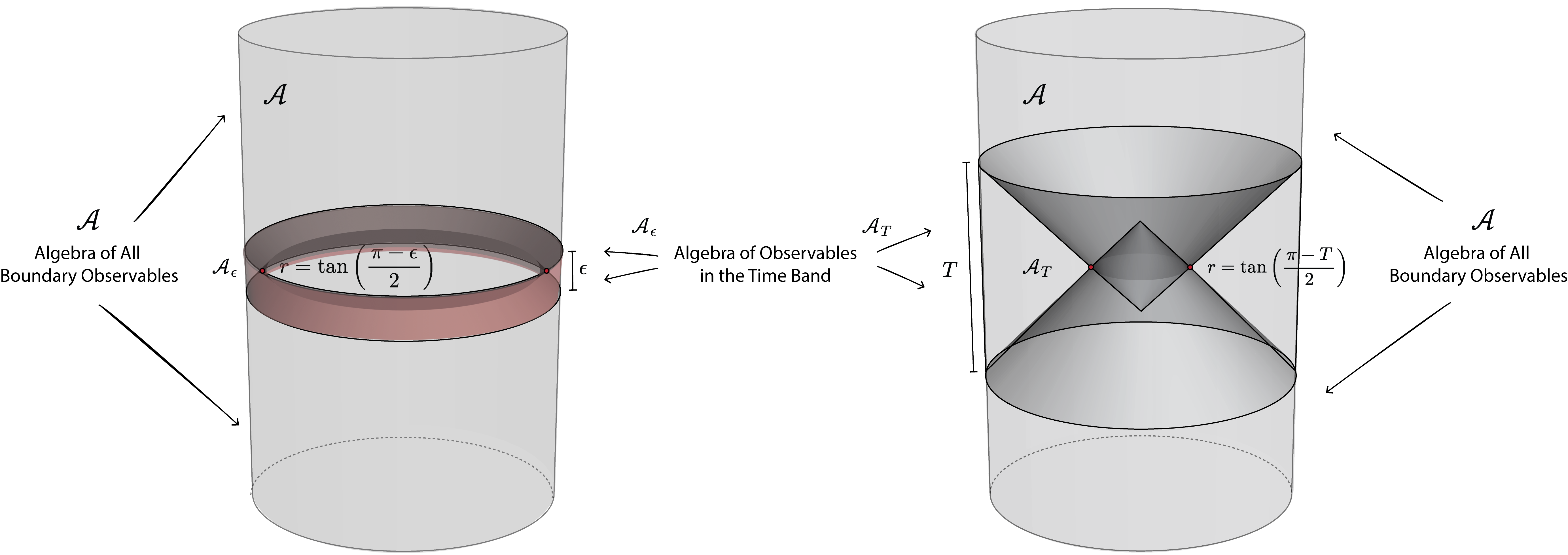}
\caption{The left subfigure shows an asymptotic time band on the AdS boundary. In a gravitational theory, the algebra of a time band has no commutant. The right subfigure is the familiar figure in a {\em nongravitational} system in which the algebra of a time band corresponds to the bulk algebra on the $t=0$ slice for $r \in (\tan{\pi - T \over 2}, \infty)$ but has a nonzero commutant corresponding to the algebra of a diamond near the middle of AdS.\label{figalgs}}
\end{center}
\end{figure}

Similarly, we can define the algebra of operators confined to an infinitesimal time band on the boundary
\be
\label{adsalgtimeband}
\begin{split}
\alset_{\epsilon}  = ~&\text{span}\{O(t_1,\Omega_1), O(t_1,\Omega_1) O(t_2, \Omega_2), \ldots O(t_1, \Omega_1) O(t_2, \Omega_2) \ldots O(t_n, \Omega_n), \ldots\}, \\
&\text{with}~~t_i \in (-{\epsilon \over 2}, {\epsilon \over 2}).  
\end{split}
\ee
The algebras $\alset$ and $\alset_{\epsilon}$ are depicted in Figure \ref{figalgs}.

When the bulk theory is nongravitational, $\alset$ is a larger algebra with observables that are distinct from those of $\alset_{\epsilon}$.  We will show that, however, in a theory of gravity, any element of $\alset$ can be approximated arbitrarily well by an element of $\alset_{\epsilon}$.

The second assumption we will need is that the Hamiltonian of the system is bounded below.  While we cannot prove this, it is reasonable to believe that it will hold in any UV-complete theory of quantum gravity. If so, given that the AdS radius provides an IR-cutoff, there is a gapped lowest energy state that we denote by $|0 \rangle$.  We consider the Hilbert space obtained by starting with the vacuum and then acting with all possible boundary operators.  
\be
\label{superselectedh}
{\cal H} = \alset |0 \rangle.
\ee
This corresponds to states that can be obtained by exciting the vacuum using boundary operators.  This Hilbert space includes all black holes formed from collapse.

It may be shown that, subject to the positivity of the Hamiltonian, any state in ${\cal H}$ can be approximated arbitrarily well by the action of an element of ${\alset}_{\epsilon}$. This result \cite{Haag:1992hx} is not a result that is specific to gravitational theories.  It holds even in  nongravitational quantum field theories and is a consequence of entanglement within the vacuum.  We refer the reader to \cite{Laddha:2020kvp} (see the Appendix) for details of the proof. This means that $\forall |n \rangle \in {\cal H}$ and $\forall \delta > 0,  ~~\exists X_n \in \alset_{\epsilon}$ such that
\be
\label{rsresult}
\left\lVert X_n | 0 \rangle  - |n \rangle \right\rVert^2  < \delta.
\ee
We introduce some notation to write this result as
\be
|n \rangle \doteq X_n | 0 \rangle,
\ee
where $\doteq$ means ``can be approximated arbitrarily well.''

In a theory of gravity, the Hamiltonian can be measured at infinity, as the asymptotic limit of a bulk observable,  and so is an element of $\alset_{\epsilon}$.   In the Fefferman-Graham gauge, we have
\be
\label{admham}
H = {d \over 16 \pi G} \lim_{r \rightarrow \infty} r^{d-2} \int d^{d-1} \Omega h_{t t},
\ee
where $h_{tt }$ is the time-component of the metric fluctuation \cite{Skenderis:2000in}. Note that this expression involves an integral over the entire $S^{d-1}$ at the boundary and so the Hamiltonian is not accessible if one restricts the algebra to only part of the boundary. 

The rules of quantum mechanics tell us that, upon measuring an observable, the probabilities of different outcomes are given by the corresponding projectors onto eigenstates. In particular, an observer at infinity with access to the Hamiltonian could ask about the probability that the measurement of the energy yields the lowest possible value.   This probability is given by the expectation value of the projector on the vacuum
\be
\label{projvac}
P_0 = |0 \rangle \langle 0 |.
\ee

In a gravitational theory, $P_0 \in \alset_{\epsilon}$ since the Hamiltonian is an element of $\alset_{\epsilon}$. This is a unique property of gravity that is not shared by nongravitational theories.

The conclusions above immediately imply that any operator in the Hilbert space ${\cal H}$  can be approximated arbitrarily well by an operator from $\alset_{\epsilon}$.

Consider the operator 
\be
Q = | n \rangle \langle m|.
\ee
Any operator ${\cal H} \rightarrow {\cal H}$ is a linear combination of operators of this form.

Using the result above \ref{rsresult} we find that
\be
Q \doteq  X_n |0 \rangle \langle 0| X_m^{\dagger}.
\ee
Here $\doteq$ is used in the same sense as above and means that the right-hand side approximates the left hand side to any desired precision in the operator norm.  But now we can write
\be
Q \doteq  X_n P_0 X_m^{\dagger}.
\ee
In a gravitational theory, all three operators on the right-hand side are elements of $\alset_{\epsilon}$. This establishes the desired result. We emphasize that the result above is a statement about operators, and not just their action on one particular state.

The result above can also be derived in another way \cite{Marolf:2008mf,Marolf:2013iba,Laddha:2020kvp}. Any operator in the time band can be evolved forward or backward in time using the boundary Hamiltonian.
\be
\label{evolvewithH}
O(t + \tau, \Omega) = e^{i H \tau} O(t, \Omega) e^{-i H \tau}.
\ee
Since in the presence of gravity the Hamiltonian is an element of the algebra of the time band, it is clear that the entire asymptotic algebra is contained in the time band. A common concern with \eqref{evolvewithH} is that the gravitational Hamiltonian might differ slightly from the UV Hamiltonian, and those small effects might accumulate in the evolution \eqref{evolvewithH}. The advantage of the proof given previously is that it relies on robust physical properties: bulk entanglement, which gives rise to \eqref{rsresult},  and the physical ability to measure the energy, which gives rise to \eqref{projvac}. 

The result above is formulated in terms of asymptotic observables, since our first assumption was that such observables make sense even in the UV-complete theory of quantum gravity. Physically, it might be more natural to think of an observer ``near'' infinity with a detector that occupies some range of $r$ near infinity at a constant time slice. See Figure \ref{figalgs}.

\subsubsection{Asymptotically flat space}
A similar result can be proven in asymptotically flat spacetimes. We start by reviewing the algebra of observables on ${\cal I}^{+}$ --- the future null boundary of an asymptotically flat spacetime. We work in four dimensions, where this can be parameterized by retarded time, $u = t - r \in (-\infty, \infty)$ and a celestial 2-sphere, $\Omega$. 

As above, it is important to fix the asymptotic boundary conditions. It is standard to demand \cite{Strominger:2017zoo} that, in Bondi gauge, the asymptotic metric takes the form
\be
\label{bondicoords}
ds^2 \underset{r \rightarrow \infty}{\longrightarrow}  - du^2 - 2 du dr + r^2 \gamma_{AB} d \Omega^{A} d \Omega^{B} + r C_{AB}(u,\Omega) d \Omega^{A} d \Omega^{B} + {2 m(u,\Omega) \over r} d u^2 + \ldots,
\ee
where $\gamma_{AB}$ is the standard metric on the 2-sphere, i.e.
\be
 ds^{2}_{S^{2}}=\gamma_{AB}d\Omega^{A}\Omega^{B}=\frac{4}{(1+z\bar{z})^4}dzd\bar{z}\,.
\ee

Here we see the metric of flat space followed by ${1 \over r}$ terms that correspond to metric fluctuations. These are parameterized by the {\em shear} $C_{AB}(u, \Omega)$ and the Bondi mass aspect $m(u,\Omega)$. The Bondi news is defined by
\be
N_{AB}(u, \Omega) = \partial_{u} C_{A B}(u,\Omega).
\ee

The ADM Hamiltonian \cite{Arnowitt:1962hi,Regge:1974zd} is the limit of the integrated Bondi mass aspect at the past boundary of ${\cal I}^{+}$\footnote{It has been argued \cite{Bousso:2017xyo} that the Bondi mass at finite $u$ --- $m(u) = \int \sqrt{\gamma} m(u, \Omega) d^2 \Omega$ ---  is not a good observable because it has infinite fluctuations, although bounded functions of $m(u)$ might still be good observables. However, the ADM Hamiltonian is well defined, and a procedure to damp out the fluctuations in the ADM Hamiltonian was discussed in \cite{Laddha:2020kvp}.}
\be
H = \lim_{u \rightarrow -\infty} {1 \over 4 \pi G} \int m(u, \Omega) \sqrt{\gamma} d^2 \Omega.
\ee

If there are other massless fields, it is possible to similarly extrapolate them to ${\cal I}^{+}$. For instance, we can demand that a massless scalar field has a falloff
\be
\label{matterlimit}
\phi(r, u, \Omega)  \underset{r \rightarrow \infty}{\longrightarrow} {O(u, \Omega) \over r},
\ee
where $O$ is the boundary value of the field. 

Let $\alset$ be the algebra of all operators on ${\cal I}^{+}$. This means that we simply take all possible combinations of the boundary operators above
\be
\label{alsetdef}
\begin{split}
\alset =~&\text{span}\{C_{AB}(u_1, \Omega), m(u_1, \Omega), O(u_1, \Omega), C_{A B}(u_1,\Omega) O(u_2,\Omega'), C_{A B}(u_1,\Omega) m(u_2,\Omega'), \ldots \}, \\
&\text{with}~~u_i \in (-\infty, \infty). 
\end{split}
\ee
This set of operators furnishes a complete basis of observables for massless particles. 

This algebra leaves out operators at future timelike infinity, denoted by $i^+$. Those operators appear in the S-matrix of massive particles. But in the context of black hole evaporation, these operators are not expected to be relevant since the evaporation of a large black hole is not expected to produce a significant number of massive particles.

Similarly, we can define the algebra of the region near the past boundary of ${\cal I}^{+}$, which we denote by $\alset_{\epsilon}$. This is defined as the same set of polynomials as in Equation \eqref{alsetdef} but with $u_i \in (-\infty, -{1 \over \epsilon})$.

\begin{figure}[!ht]
\begin{center}
\begin{tikzpicture}
    \draw[-,very thick,black] (0,3) to (3,0);
    \draw[-,very thick,black] (3,0) to (0,-3);
    \draw[-,very thick,black] (0,-3) to (-3,0);
     \draw[-,very thick,black] (-3,0) to (0,3);
     \node at (0.2,3.3) {\textcolor{black}{$i^{+}$}};
     \node at (0.2,-3.2) {\textcolor{black}{$i^{-}$}};
     \node at (3.2,0) {\textcolor{black}{$i^{0}$}};
       \node at (2,1.7) {\textcolor{black}{$\mathcal{I}^{+}$}};
        \node at (2,-1.7) {\textcolor{black}{$\mathcal{I}^{-}$}};
        \node at (-2.6,2.2) {\textcolor{black}{$\mathcal{A}$}};
        \draw [-,very thick,black, decorate,
    decoration = {brace,mirror,raise=5pt,
        amplitude=7pt}] (-0.2,3.2)--(-3.2,0.2);
     \draw[-,very thick,black] (-3,0) arc (180:360:3 and 0.3);
     \draw[-,very thick,dashed,black] (-3,0) arc (180:0:3 and 0.3);
     \draw[-,very thick,red] (-2.6,0.4) arc (180:360:2.6 and 0.3);
     \draw[-,very thick,dashed,red] (-2.6,0.4) arc (180:0:2.6 and 0.3);
    \draw[red,pattern=north east lines,pattern color=red] (-2.6,0.4)-- (-3,0) arc (180:360:3 and 0.3)--(2.6,0.4) arc (0:-180:2.6 and 0.3);
    \draw[red,pattern=north east lines,pattern color=red] (-2.6,0.4)-- (-3,0) arc (180:0:3 and 0.3)--(2.6,0.4) arc (0:180:2.6 and 0.3);
    \node at (3.1,0.5) {\textcolor{red}{$\mathcal{A}_{\epsilon}$}};
\end{tikzpicture}
\caption{All operators on ${\cal I}^{+}$  can be approximated arbitrarily well by operators from the retarded time interval $(-\infty, -{1 \over \epsilon})$.}
\end{center}
\end{figure}
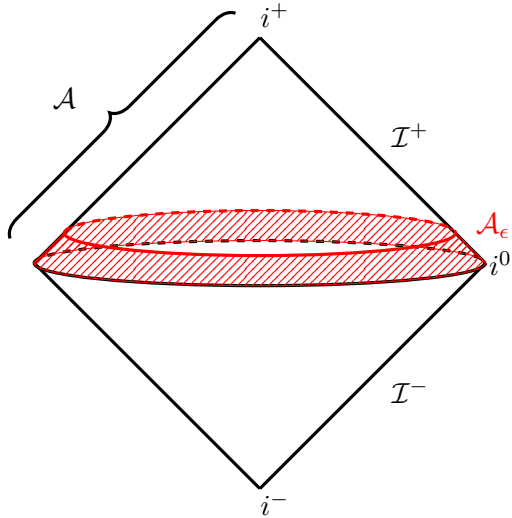

Then, using a chain of logic very similar to the one above, it may be shown that all operators in $\alset$ can be approximated arbitrarily well by operators in $\alset_{\epsilon}$.  We refer the reader to \cite{Laddha:2020kvp} for details.

The vacuum structure of flat space is more subtle than AdS due to the presence of soft modes that cause the vacuum to be infinitely degenerate. This subtlety is discussed in greater detail in \cite{Laddha:2020kvp}. On the other hand, the algebra $\alset$ in \eqref{alsetdef} is simpler than the algebra on the conformal boundary of AdS since the commutation relations between the operators at ${\cal I}^{+}$ are independent of interactions \cite{Ashtekar:1981sf,Ashtekar:1981hw,Ashtekar:1987tt,Ashtekar:2018lor}.  This feature will be important below.

\subsection{Implications and limitations of the result}
The results above have been studied in three regimes. 
\begin{enumerate}
\item
{\bf Perturbative regime:} Here, we consider low-energy excitations about the vacuum, where gravitational perturbation theory applies. In this regime, low-point correlators of the Hamiltonian and other asymptotic observables can be used to access information from the bulk. This provides a perturbative check of the principle of holography of information.
\item
{\bf QFT Limit:} A trivial QFT limit is obtained by taking $G \to 0$. But even at nonzero $G$, it is possible to study an intermediate regime where (a) we have a time-dependent background in the bulk, and (b) we restrict ourselves to coarse-grained asymptotic observables (see \cite{Geng:2024dbl,Geng:2025gqu} for explicit examples). In this limit, mundane notions of locality are restored and holography is obscured. 
\item
{\bf Fine-grained regime:} In the fine-grained regime, the asymptotic observer can access all information in the bulk even in the presence of time-dependent backgrounds or black holes. 
\end{enumerate}

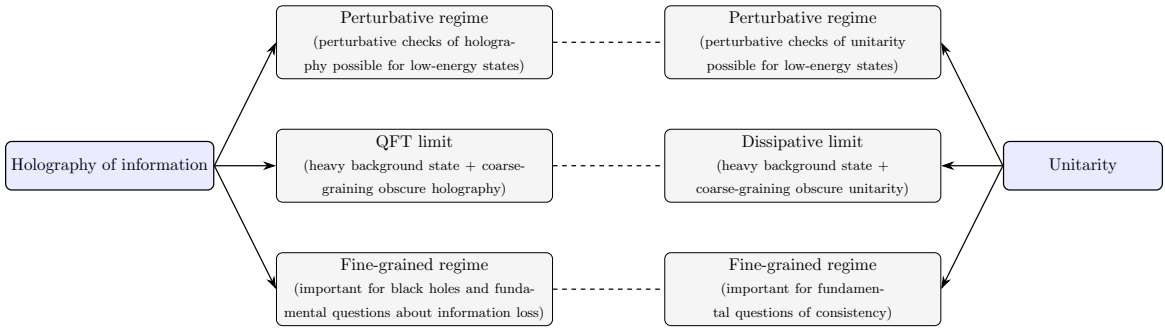
\begin{figure}[ht]
\centering
\resizebox{\textwidth}{!}{
\begin{tikzpicture}[
  font=\normalsize,
  every node/.style={align=center},
  title/.style={draw, thick, rounded corners, fill=blue!8, minimum width=3.6cm, minimum height=1.1cm},
  reg/.style={draw, rounded corners, fill=gray!8, text width=6.0cm, minimum height=1.25cm},
  link/.style={dashed, thick},
  arr/.style={->, thick}
]

\node[reg] (h1) at (0,  2.8) {Perturbative regime\\
{\footnotesize (perturbative checks of holography possible for low-energy states)}};

\node[reg] (h2) at (0,  0.0) {QFT limit\\
{\footnotesize (heavy background state + coarse-graining obscure holography)}};

\node[reg] (h3) at (0, -2.8) {Fine-grained regime\\
{\footnotesize (important for black holes and fundamental questions about information loss)}};

\node[reg] (u1) at (8.8,  2.8) {Perturbative regime\\
{\footnotesize (perturbative checks of unitarity possible for low-energy states)}};

\node[reg] (u2) at (8.8,  0.0) {Dissipative limit\\
{\footnotesize (heavy background state + coarse-graining obscure unitarity)}};

\node[reg] (u3) at (8.8, -2.8) {Fine-grained regime\\
{\footnotesize (important for fundamental questions of consistency)}};

\node[title, left=1.4cm of h2] (holo) {Holography of information};
\node[title, right=1.4cm of u2] (unit) {Unitarity};

\draw[arr] (holo.east) -- (h1.west);
\draw[arr] (holo.east) -- (h2.west);
\draw[arr] (holo.east) -- (h3.west);

\draw[arr] (unit.west) -- (u1.east);
\draw[arr] (unit.west) -- (u2.east);
\draw[arr] (unit.west) -- (u3.east);

\draw[link] (h1.east) -- (u1.west);
\draw[link] (h2.east) -- (u2.west);
\draw[link] (h3.east) -- (u3.west);

\end{tikzpicture}
}
\caption{Holography of information and unitarity can be studied in parallel regimes; dashed lines indicate the correspondence between regimes. \label{holinforegimes}}
\end{figure}

It is very helpful to think of the analogy with unitarity in nongravitational QFT. Unitarity can be explicitly checked in perturbative QFT --- an exercise that appears in QFT textbooks. However, in our everyday experience, the world around us appears to be dissipative rather than unitary. Dissipation arises when (a) we have a background with many degrees of freedom that can absorb energy from excitations and (b) we coarse-grain the set of observables. Unitarity is obscured in the dissipative regime. Nevertheless, in a fine-grained regime, where we keep track of all observables, unitarity still holds as a fundamental result.  This parallel is shown in Figure \ref{holinforegimes}.

We elaborate on all three regimes in the subsections below. We again emphasize that the discussion of these three regimes is an independent discussion of the principle of holography of information and not directly related to the Page curves or islands described in \cite{Antonini:2025sur}. Those are discussed in the next section.

\subsection{The perturbative regime. \label{secperturbative}}
If we consider simple perturbative excitations about the vacuum or other backgrounds that have high symmetry, then it is possible to show, within perturbation theory, that information about the bulk can be obtained from boundary observables.

A simple but illuminating example is as follows \cite{Laddha:2020kvp,Raju:2024gvc,Geng:2025gqu}. (See \cite{Chowdhury:2020hse,Chowdhury:2021nxw,Chowdhury:2022wcv,Gaddam:2024mqm} for related discussion.)  Consider a single massless scalar field $\phi(u,r,\Omega)$ in the Bondi coordinates of \eqref{bondicoords} weakly coupled to gravity and consider the state
\be
\label{statedef}
|f \rangle = e^{-i \lambda \phi(f)} | 0 \rangle,
\ee
with
\be
\label{phifdefn}
\phi(f) \equiv \int f(r) \phi(u=0,r, \Omega) d^2 \Omega d r,
\ee
where  $|0 \rangle$ is a normalized flat-space vacuum state and $f(r)$ has compact support for $r = (0,1)$.  This is a spherically symmetric excitation of the vacuum near the ``middle'' of flat space.

In a nongravitational quantum field theory, this state corresponds to a local unitary excitation on top of the vacuum that is spacelike separated to the past boundary of ${\cal I}^{+}$.   Therefore, all operators in $\alset_{\epsilon}$ are left invariant by the action of the unitary
\be
\langle f | a | f \rangle = \langle 0 | a | 0 \rangle, \qquad \forall a \in \alset_{\epsilon} ~\text{without~gravity}.
\ee

However, in a gravitational theory, since the Hamiltonian is an element of $\alset_{\epsilon}$,  we can consider the correlator
\be
G(u') = \lim_{r' \rightarrow \infty} r' {\partial \over \partial \lambda} \langle f | H \phi(u', r',  \Omega') | f \rangle \big|_{\lambda=0}.
\ee
This correlator is the expectation value of an element of $\alset_{\epsilon}$ when $u' \in (-\infty, -{1 \over \epsilon})$. 

To evaluate this correlator, we expand both and bra and the ket to first order in $\lambda$ and use the relation $\langle 0 | H = 0$. This leads to
\be
G(u') = \lim_{r' \rightarrow \infty} i r'   \langle 0 | \phi(f) H \phi(u',r',\Omega') | 0  \rangle.
\ee
Now after using the standard expansion of a  massless field, and the expression for the energy of one-particle states, some algebra yields
\be
\label{twoptfn}
G(u') =\int f(r) d r {-1 \over 2\pi (2 r u' - (u')^2)}.
\ee

It may be shown that a knowledge of the right-hand side of \eqref{twoptfn} is sufficient to completely reconstruct $f$. This follows from expanding $G(u')$ as
\be
\label{gupexpansion}
G(u') = \sum_{n=0}^{\infty} {2^n \over 2 \pi (u')^{n+2}} \int r^n f(r) d r.
\ee
The observer at large negative $u'$ needs to measure the correlator at different values of $u'$ and develop an expansion in ${1 \over u'}$, which approaches zero from the negative axis as $|u'|$ becomes large. The coefficients of different orders in this expansion give us different moments of $f(r)$. The observer can determine arbitrarily many moments of $f(r)$ by determining higher and higher powers of  ${1 \over u'}$ in $G(u')$. Therefore, they can determine $f(r)$ to arbitrary accuracy.

\paragraph{Distinguishing symmetric states.}
Note that this result does {\em not} rely on simply measuring the energy from infinity which is, in general, insufficient to determine the function $f$. This is clearly illustrated by considering the case where there is a {\em second} scalar field in the theory, $\tphi(t,r, \Omega)$ that is related by a global symmetry to $\phi$.\footnote{It is believed that unitary theories of quantum gravity do not have exact global symmetries \cite{Harlow:2018tng,Harlow:2018jwu,Geng:2025gns} but this consideration is irrelevant for the simple perturbative example here.} Now consider the state
\be
|\widetilde{f} \rangle = e^{-i \lambda \tphi(f)} | 0 \rangle,
\ee
where $\tphi$ is defined precisely as in \eqref{phifdefn}.

The state $|f \rangle$ and $|\widetilde{f} \rangle$ are related by a global symmetry. So no measurement of the energy can distinguish them. However,  we find that 
\be
{\partial \over \partial \lambda} \langle \widetilde{f} | H \phi(u', r',  \Omega') | \widetilde{f} \rangle \big|_{\lambda=0} = {\partial \over \partial \lambda} \langle f | H \tphi(u', r',  \Omega') | f \rangle \big|_{\lambda=0} = 0,
\ee
but
\be
{\partial \over \partial \lambda} \langle \widetilde{f} | H \tphi(u', r',  \Omega') | \widetilde{f} \rangle \big|_{\lambda=0} = {\partial \over \partial \lambda} \langle f | H \phi(u', r',  \Omega') | f \rangle \big|_{\lambda=0},
\ee
where the value of the right-hand side was computed above. 
Both operators  $H \tphi(u', r',  \Omega')$ and $H \phi(u', r',  \Omega')$ are elements of $\alset_{\epsilon}$. Therefore, by measuring the expectation values of both operators it is possible to distinguish these two states.

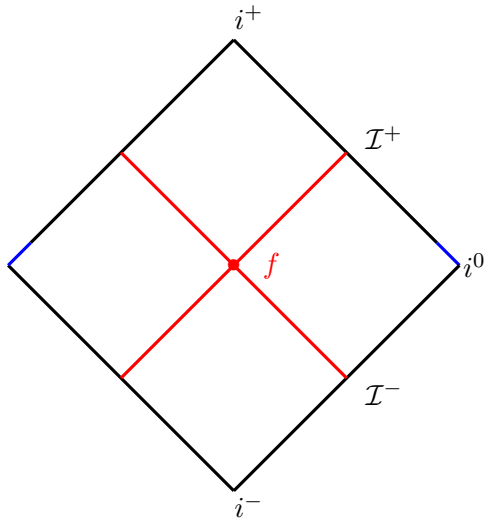
\begin{figure}[h]
\begin{center}
\begin{tikzpicture}
    \draw[-,very thick,black] (0,3) to (2.7,0.3);
    \draw[-,very thick,blue] (2.7,0.3) to (3,0);
    \draw[-,very thick,black] (3,0) to (0,-3);
    \draw[-,very thick,black] (0,-3) to (-3,0);
     \draw[-,very thick,black] (-2.7,0.3) to (0,3);
      \draw[-,very thick,blue] (-2.7,0.3) to (-3,0);
     \node at (0.2,3.3) {\textcolor{black}{$i^{+}$}};
     \node at (0.2,-3.2) {\textcolor{black}{$i^{-}$}};
     \node at (3.2,0) {\textcolor{black}{$i^{0}$}};
       \node at (2,1.7) {\textcolor{black}{$\mathcal{I}^{+}$}};
        \node at (2,-1.7) {\textcolor{black}{$\mathcal{I}^{-}$}};
        \draw[-,very thick,red] (1.5,-1.5) to (-1.5,1.5);
        \draw[-,very thick,red] (-1.5,-1.5) to (1.5,1.5);
        \node at (0,0)
        {\textcolor{red}{$\bullet$}};
        \node at (0.5,0) {\textcolor{red}{$f$}};
\end{tikzpicture}
\caption{The state $|f \rangle$ is seemingly created by a local unitary that acts near the middle of space. However, in gravity, the constraints imply that the excitation must originate on ${\cal I}^{-}$. So it is not a violation for causality for information about this excitation to be present near ${\cal I}^{+}_{-}$ \label{figcausalshockwave}}
\end{center}
\end{figure}

\paragraph{Causality. }
This is a surprising result since it indicates that, even in perturbation theory, an asymptotic observer can use correlators of the Hamiltonian and other fields to reconstruct the bulk state. However, this is not inconsistent with causality as Figure \ref{figcausalshockwave} shows. In a theory of quantum gravity, it is impossible to locally create or destroy energy. Therefore, the excitation $|f \rangle$ can be thought of as a wave that enters spacetime from ${\cal I}^{-}$. However, as it enters the bulk, it modifies some observables at asymptotic infinity and it is these observables that are picked up by the correlators above. 

Although classical gravity is not holographic, it is nevertheless a useful setting to think about questions of causality. Even in classical gravity, the Gauss law tells us that the energy of the bulk can be measured from infinity by integrating the flux of the gravitational field. This is because the energy in the bulk entered from infinity and while doing so, it permanently modified the component of the metric that corresponds to the integrated flux. This component cannot be modified unless the energy exits the bulk. Therefore, the classical Gauss law can also be thought of as a manifestation of the ``memory'' of spacetime.

The difference between nongravitational and gravitational theories is that, in a gravitational theory, the asymptotic detectors can switch off beyond $u = -{1 \over \epsilon}$ even if it takes additional time to process those measurements. In a nongravitational theory, the detectors must continue to make measurements at later times to reconstruct the bulk excitation.

\subsection{QFT limit \label{secmundane}}
In the subsection above we have explained that, in quantum gravity, observables on the boundary of a Cauchy slice can be used to obtain information in the bulk of the slice and, moreover, that this can be verified in gravitational perturbation theory. But clearly, in nongravitational QFT, this is impossible. Here, we briefly explain how one should take the ``QFT limit'' of a gravitational theory to obtain approximately local observables.

We can always switch off gravitational effects by taking $G \to 0$. In this limit, the Hamiltonian effectively ceases to be a boundary term since the fluctuation of the metric is related to the Hamiltonian via a factor of ${1 \over G}$ as shown in \eqref{admham}. However, even at nonzero $G$, in the presence of a carefully chosen background, it is possible to define approximately local state-dependent observables that commute with asymptotic observables to  good accuracy \cite{Papadodimas:2015xma,Papadodimas:2015jra,Bahiru:2023zlc,Bahiru:2022oas,Jensen:2024dnl,Geng:2024dbl,Geng:2025bcb}.

The physical idea of this construction is quite simple. In the presence of a background state that can function as a clock, it is possible to define observables that are ``dressed'' to the clock. Such observables approximately commute with boundary observables in the limit where the clock has a large number of degrees of freedom. However, if we probe the system with observables that are sufficiently fine-grained to be sensitive to the finite entropy of the clock, this construction, like any other construction, breaks down.  

Another way to state the relevant physics is that a clock that keeps track of time accurately must correspond to a state with large fluctuations in the energy. One can then design operators that resemble local bulk operators but take advantage of the fluctuations to ``hide'' their effect from the asymptotic observer. If the asymptotic observer is restricted from making sufficiently precise measurements of the energy, they cannot detect the action of these operators. However, if the asymptotic observer is allowed to make sufficiently fine-grained measurements of the energy and other observables, the construction breaks down. 

 We present a  detailed technical discussion of relational observables in section \ref{secrelational}, and so we eschew further technical discussion here. 

We would like to make two comments.
\begin{enumerate}
\item
The QFT limit discussed here is also the limit in which ``mundane'' notions of locality are recovered. It is an important limit because even though we live in a world where gravity is quantized, we do not find it convenient to use a holographic description for mundane phenomenon. This is analogous to the fact that although we live in a world where unitarity is an exact principle, mundane phenomena are dissipative. 
\item
The QFT limit should be distinguished from constructions that we will describe in section \ref{secpagecurves} where the Hamiltonian is simply dropped from the algebra. As we explain in section \ref{secrelational}, this limit cannot be used to define Page curves or islands.
\end{enumerate}

\subsection{The fine-grained regime. \label{secfinegrained}}
 As we saw in section \ref{sechawkingpage}, pure states and mixed states can be exponentially close. So, it is necessary to examine fine-grained observables to distinguish between these two cases. Therefore, the QFT limit --- where we are required to coarse-grain our observables --- is not the appropriate regime in which to ask questions about the unitarity of black hole evaporation.  Instead, one must consider the full asymptotic algebra of observables discussed in section \ref{secasympcomplete}. As we have already stated,  the Hilbert space ${\cal H}$ includes black holes formed from collapse and so the result of section \ref{secasympcomplete} can be applied to study their evaporation.

We start by considering black holes in asymptotically flat space. 
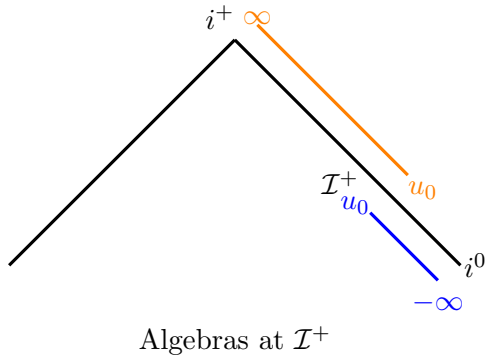
\begin{figure}
\begin{center}
\begin{tikzpicture}
    \draw[-,very thick,black] (0,3) to (3,0);
    \draw[-,very thick,blue] (1.8,0.7) to (2.7,-0.2);
    \draw[-,very thick,orange] (2.3,1.2) to (0.3,3.2);
     \draw[-,very thick,black] (-3,0) to (0,3);
     \node at (-0.2,3.3) {\textcolor{black}{$i^{+}$}};
     \node at (3.2,0) {\textcolor{black}{$i^{0}$}};
       \node at (1.4,1.1) {\textcolor{black}{$\mathcal{I}^{+}$}};
       \node at (0,-1) {\textcolor{black}{Algebras at $\mathcal{I}^{+}$}};
       \node at (2.7,-0.5)
       {\textcolor{blue}{$-\infty$}};
       \node at (1.6,0.8)
       {\textcolor{blue}{$u_{0}$}};
       \node at (2.5,1)
       {\textcolor{orange}{$u_{0}$}};
       \node at (0.3,3.3)
       {\textcolor{orange}{$\infty$}};
\end{tikzpicture}
\caption{The algebra for $u \in (-\infty, u_0)$. \label{figu0algebra}}
\end{center}
\end{figure}
The von Neumann entropy on ${\cal I}^{+}$ is a fine-grained observable that can distinguish between pure and mixed states. Consider the algebra of observables in the retarded time interval $(-\infty, u_0)$ as shown in Figure \ref{figu0algebra}. This algebra is defined precisely as in \eqref{alsetdef} but with the restriction that the $u_i$ that appear there obey $u_i < u_0$. We denote this algebra by $\alset(u_0)$. In any state, we can define the ``density matrix'' for this algebra, $\rho$ to satisfy the property that
\be
\label{rhodef}
\tr(\rho(u_0) a) = \langle a \rangle, \qquad \forall a \in \alset(u_0),
\ee
with $\rho(u_0) \in \alset(u_0)$.  After suitable regulation, the entropy of this algebra is defined as
\be
\label{vndef}
S(u_0) = -\tr(\rho(u_0) \log \rho(u_0)).
\ee

The arguments of section \ref{sechawkingpage} tell us that every element of $\alset(u_0)$ can be approximated arbitrarily well by an element of the algebra near the past boundary of ${\cal I}^{+}$ that was denoted by $\alset_{\epsilon}$. Therefore, we can always choose $\rho(u_0) \in \alset_{\epsilon}$ which is independent of $u_0$ and,  as a consequence, we find
\be
{\partial S(u_0) \over \partial u_0} = 0.
\ee
In words, because every element of $\alset(u_0)$ can be approximated arbitrarily well by an element of the fine-grained time-band algebra $\alset_\epsilon$, no new degrees of freedom are added when we move the endpoint $u_0$ on $\cal{I}^+$.

This implies that rather than following a tent-shaped Page curve, this fine-grained entropy is flat as a function of $u_0$. 

A similar result can be proven for black holes in AdS. Consider a large black hole formed from collapse in AdS. Define the algebra in the time band $(t-{\epsilon \over 2},t+{\epsilon \over 2})$ as $\alset_{\epsilon}(t)$. The definition is precisely as in \eqref{adsalgtimeband}.   Probe the state with $\alset_{\epsilon}(t)$  and consider the fine-grained entropy, $S(t)$. This is defined by means of the analogues of \eqref{rhodef} and \eqref{vndef}. We first fix $\rho(t) \in \alset_{\epsilon}(t)$ via
\be
\tr(\rho(t) a ) = \langle a \rangle, \qquad \forall a \in \alset_{\epsilon}(t),
\ee
and then define
\be
S(t) = -\tr(\rho(t) \log(\rho(t))).
\ee
This black hole does not evaporate but applying Hawking's argument one might have reached the conclusion that this entropy rises from zero and then saturates. However, the same argument as above tells us that 
\be
\label{flatinads}
{d S(t) \over d t} = 0.
\ee
Of course this result follows from the AdS/CFT duality. But \eqref{flatinads} does not rely on AdS/CFT and follows from the result of section \ref{secasympcomplete} that we proved independently.

\section{Page curves in standard theories of gravity \label{secpagecurves}}
In the section above, we have shown that the entanglement entropy on the asymptotic boundary is constant for an evaporating black hole. This means that, in a standard theory of gravity, if one measures all possible observables outside the black hole in a sufficiently fine-grained sense, there is no gain or loss of information as the black hole forms and evaporates.

This should not be misinterpreted as the incorrect statement that ``there is `no way' to find a Page curve in a standard theory of gravity.'' In this section, we review several possibilities for Page curves in a standard theory of gravity, including those discussed in \cite{Antonini:2025sur}. These Page curves are obtained by changing the question: rather than asking the natural  question that Page asked, we ask a different question whose answer is a tent-shaped Page curve.

These proposals are largely not new. Several possibilities  described in  \cite{Antonini:2025sur} had been discussed earlier in detail in \cite{Raju:2020smc,Raju:2021lwh}.

A technical point that is common to each example below is that, in order to observe a Page curve, one must make the Hamiltonian inaccessible to the observer outside the black hole. In AdS this is done by restricting observations to one part of the asymptotic boundary and in flat space this is done by manually discarding the Hamiltonian. We show that these restrictions are not natural, which is why we suggest that it is necessary for the observer to put on their blinders in order to see the Page curve. 

In section \ref{subsecopenquestion}, we describe an open question: can one find a ``natural'' algebra of observables --- where one does not artificially discard some observables while keeping others that are accessible at the same level of accuracy --- for small black holes in AdS, whose entropy follows a Page curve as a function of time? We describe some potential obstacles but do not settle the question. 

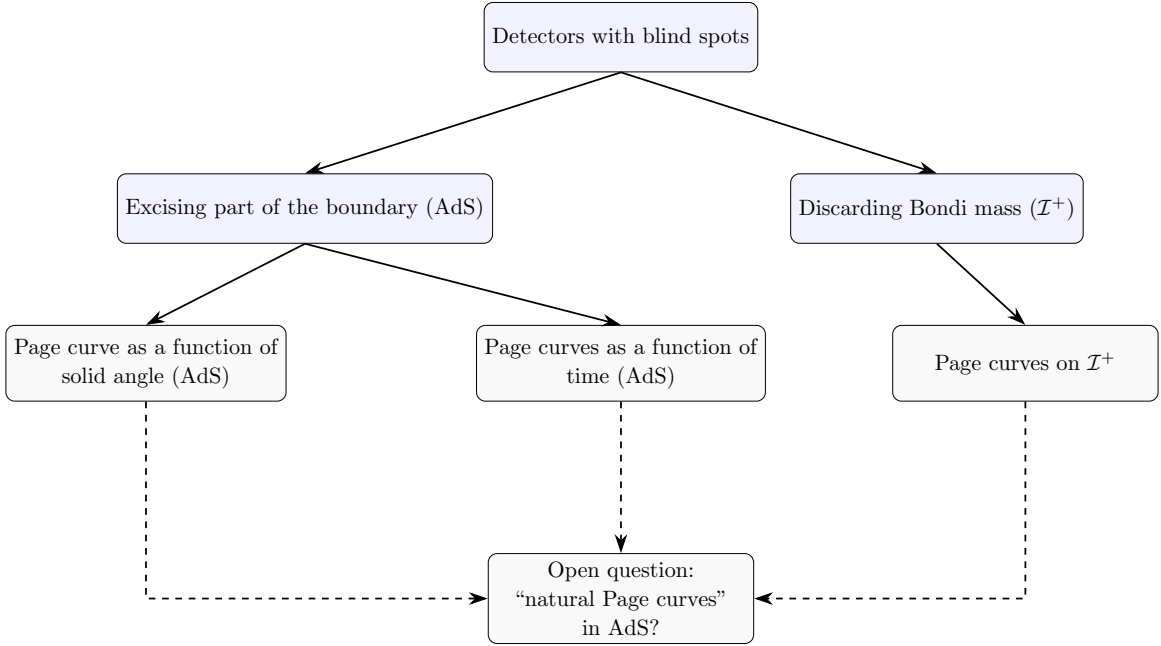
\begin{figure}[t]
\centering
\resizebox{\textwidth}{!}{
\begin{tikzpicture}[
    box/.style={
        draw, rounded corners, align=center,
        minimum width=4.2cm,
        minimum height=1.2cm,
        font=\small,
        fill=gray!5
    },
    sidebox/.style={
        draw, rounded corners, align=center,
        minimum width=4.2cm,
        minimum height=1.1cm,
        font=\small,
        fill=blue!5
    },
    arrow/.style={->, thick},
    dashedarrow/.style={->, thick, dashed}
]

\node[box] (angle)
{Page curve as a function of \\ solid angle (AdS)};

\node[box, right=3.0cm of angle] (time)
{Page curves as a function of \\ time (AdS)};

\node[box, right=2.0cm of time] (flat)
{Page curves on ${\cal I}^{+}$};

\node[box, below=2.4cm of time] (small)
{Open question:\\
``natural Page curves''\\
in AdS?};

\node[sidebox, above=4cm of time] (detectors)
{Detectors with blind spots};

\node[sidebox, below=1.6cm of detectors, xshift=-5.0cm] (excising)
{Excising part of the boundary (AdS)};

\node[sidebox, below=1.6cm of detectors, xshift=+5.0cm] (hamiltonian)
{Discarding Bondi mass (${\cal I}^{+}$)};

\draw[arrow] (detectors.south) -- (excising.north);
\draw[arrow] (detectors.south) -- (hamiltonian.north);

\draw[arrow] (excising.south) -- (angle.north);
\draw[arrow] (excising.south) -- (time.north);
\draw[arrow] (hamiltonian.south) -- (flat.north);

\draw[dashedarrow] (angle.south) |- (small.west);
\draw[dashedarrow] (time.south) -- (small.north);
\draw[dashedarrow] (flat.south) |- (small.east);

\end{tikzpicture}
}
 \caption{Flowchart for section \ref{secpagecurves} \label{flowsecpagestandard}.}
\end{figure}

The organization of this section is described in the Figure \ref{flowsecpagestandard}.

\subsection{Page curve for a non-evaporating AdS black hole \label{subsectrivialpage}}
Consider a single-sided large black hole in global AdS formed from the collapse of a pure state.  This black hole never evaporates. Nevertheless, it is possible to obtain a Page curve in the late-time geometry as follows.

Divide the boundary into two parts so that part $A$ contains a fraction $\eta$ of the full solid angle of the boundary $S^{d-1}$ and part $B$ contains the fraction $1 - \eta$.   We can now compute the entanglement entropy of $A$ (which is  the same as that of $B$ when the state is pure) and regulate it to obtain $S_{A}^{\text{reg}}$.   This computation can be performed,  using the Ryu-Takayanagi formula \cite{Ryu:2006ef,Ryu:2006bv,Hubeny:2007xt}, although one must ensure that the homology constraint appropriate for collapsing black holes is used. Regardless of how the computation is done, it is clear that $S_{A}^{\text{reg}}$ follows a Page curve as a function of $\eta$.

The computation is particularly easy for a BTZ black hole with horizon radius $r_{+}$. We find that as a function of the angle $\theta_{\infty}$ spanned by the region $A$,
\be
S_A\left(\theta_{\infty}\right)=\frac{c}{3} \log \left(\frac{2 r_{\infty}}{r_{+}} \sinh \left(r_{+} \theta_{\infty}\right)\right), \qquad \theta_{\infty} < {\pi \over 2},
\ee
where $r_{\infty}$ is a cutoff at large $r$ that corresponds to the standard UV cutoff that is necessary to obtain a finite entanglement entropy and $c={3 \over 2 G}$ is the central charge in units where the AdS radius is set to unity. It is useful to study the regulated quantity
\be
S_{A}^{\text{reg}}(\theta_{\infty}) = S_A\left(\theta_{\infty}\right) - S_{A}(0) = \frac{c}{3} \log \left(\sinh \left(r_{+} \theta_{\infty}\right)\right), \qquad \theta_{\infty} < {\pi \over 2}. 
\ee

However, for $\theta_{\infty} > {\pi \over 2}$, a different RT surface dominates as shown in Figure \ref{figtrivialpage}. Note that the disconnected surface, where one component wraps the horizon, does not appear here as it does for a mixed state \cite{Hubeny:2013gta}. This is clear from the fact that the entropy of $A$ must coincide with the entropy of its complement, $B$ for a pure state. The smaller-area RT surface is consistent with the homology constraint for a black hole formed from collapse because, in the far past, the black hole did not exist and so the RT surface can be ``slipped off'' the horizon and contracted to the anchoring asymptotic region. Therefore, we find that
\be
S_{A}^{\text{reg}}(\theta_{\infty})  = \frac{c}{3} \text{min}\left[ \log \left(\sinh \left(r_{+} \theta_{\infty}\right) \right), \log \left(\sinh \left(r_{+} (\pi - \theta_{\infty})\right) \right) \right].
\ee
This is the quantity plotted in Figure \ref{figtrivialpage}.

The entropy of $A$ starts decreasing once we cross the halfway point  even though the black hole does not evaporate. This is precisely because information is already present in the asymptotic region, and an observer with access to more than half of the asymptotic region can see signals of the purity of the state. When $A$ expands to the entire boundary, we see that $S_{A}^{\text{reg}} = 0$.  This is consistent with the argument provided in section \ref{secoutsidecomplete} that the algebra on the entire boundary has zero entropy when the state is pure. 

As long as $A$ and $B$ are both smaller than the entire boundary, the entropy is not zero. The argument of section \ref{secoutsidecomplete} cannot be applied to the case where $A$ is not the entire boundary because the Hamiltonian is an integral over the entire boundary, and so it is not an element either of the algebra of $A$ or of $B$. To use words that will appear again below, the nonzero entropy of $A$ arises because we are probing the state with a detector that has a ``blind spot'':  we are prevented from making observations in $B$. 

Evidently, this Page curve tells us nothing about how information ``emerges'' from the black hole. It is simply a measure of how the degrees of freedom on the boundary subregion, $A$, are entangled with degrees of freedom in the boundary subregion, $B$.  However, the fact that the entropy is zero when $A$ is the entire boundary tells us that unitarity is preserved on the boundary.

\subsection{Page curve for an evaporating black hole in AdS \label{pageevaporatingads}}
In the example above, a  Page curve emerged even for a static black hole through the variation of the entanglement entropy as a function of the size of the boundary region.  We now consider a conceptually similar but technically more challenging Page curve.

We study a black hole that is ``localized'' in part of AdS. The asymptotic boundary, as a whole, always has complete information about this black hole. However, since the black hole is in an atypical state, this information is asymmetrically distributed on the boundary. 

As the black hole evaporates, and the state becomes more typical, this information spreads out. Therefore the entanglement entropy of the boundary region where the information is originally concentrated, which equals the entanglement entropy of its complement,  follows a Page curve as a function of time.  It is clear that this Page curve reflects redistribution of information in asymptotic region, and not information recovery from the black hole.

\begin{figure}[h]
    \centering
    \includegraphics[height=0.3\textheight]{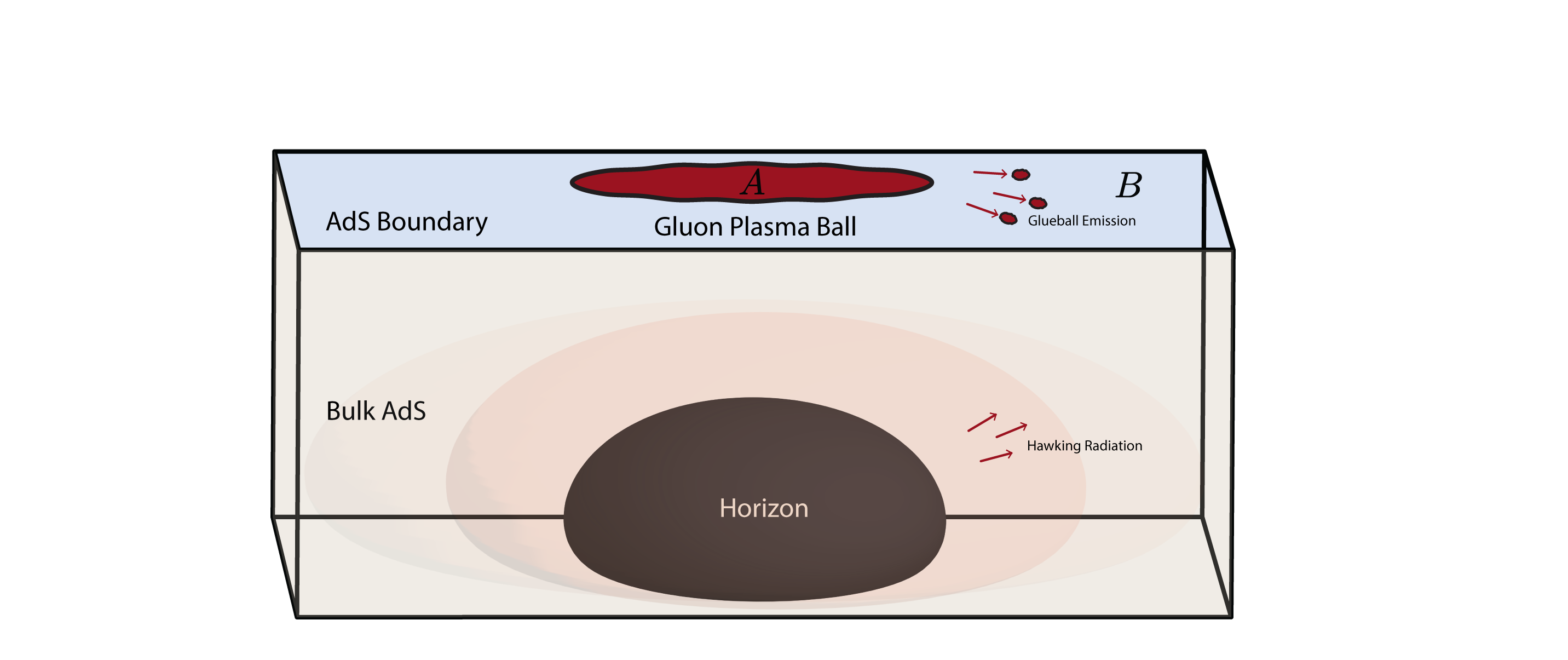}
\caption{A cartoon showing a black hole in AdS that is localized in the transverse coordinates. We use the embedding  computed in \cite{Emparan:2009dj, Plebanski:1976gy}, which corresponds to a ``plasma ball''\cite{Aharony:2005bm} on the boundary. In the bulk, the protuberance indicates the region where a black hole is present. In the boundary, there is a plasma present in the region ``A'' that gradually thermalizes by losing information and energy into the complementary region ``B''. \label{figplasmaball}}
\end{figure}

To our knowledge, such a process was first described using the  so-called ``plasma ball''  \cite{Aharony:2005bm,Emparan:2009dj}. Here one considers a Poincare AdS geometry  and a black hole that is localized in the transverse coordinates. On the boundary, one can consider a dual process where a ``quark gluon plasma'' is formed in a localized region, $A$, This plasma evaporates via the emission of glueballs into the complementary region, $B$,  and the bulk black hole evaporates via the emission of Hawking radiation in the transverse directions. The setup is shown in  Figure \ref{figplasmaball}.

The putative Page curve for plasma ball evaporation was outlined in the literature, at least as far back as \cite{Papadodimas:2013jku} and was again discussed in \cite{Laddha:2020kvp}.  The argument of section \ref{secoutsidecomplete} does not apply to the region $A$ or the region $B$ because the Hamiltonian is not an observable in either region individually. 

An asymptotic  detector localized purely in the region $B$  does not have complete information about the state because it has a ``blind spot'' in the region $A$ and moreover, the black hole is initially localized near that blind spot. So the Page curve followed by the entropy of the region $B$ is a measure of information flow between observables we choose to measure and those that we don't.

\begin{figure}[h]
\begin{center}
\includegraphics[height=0.3\textheight]{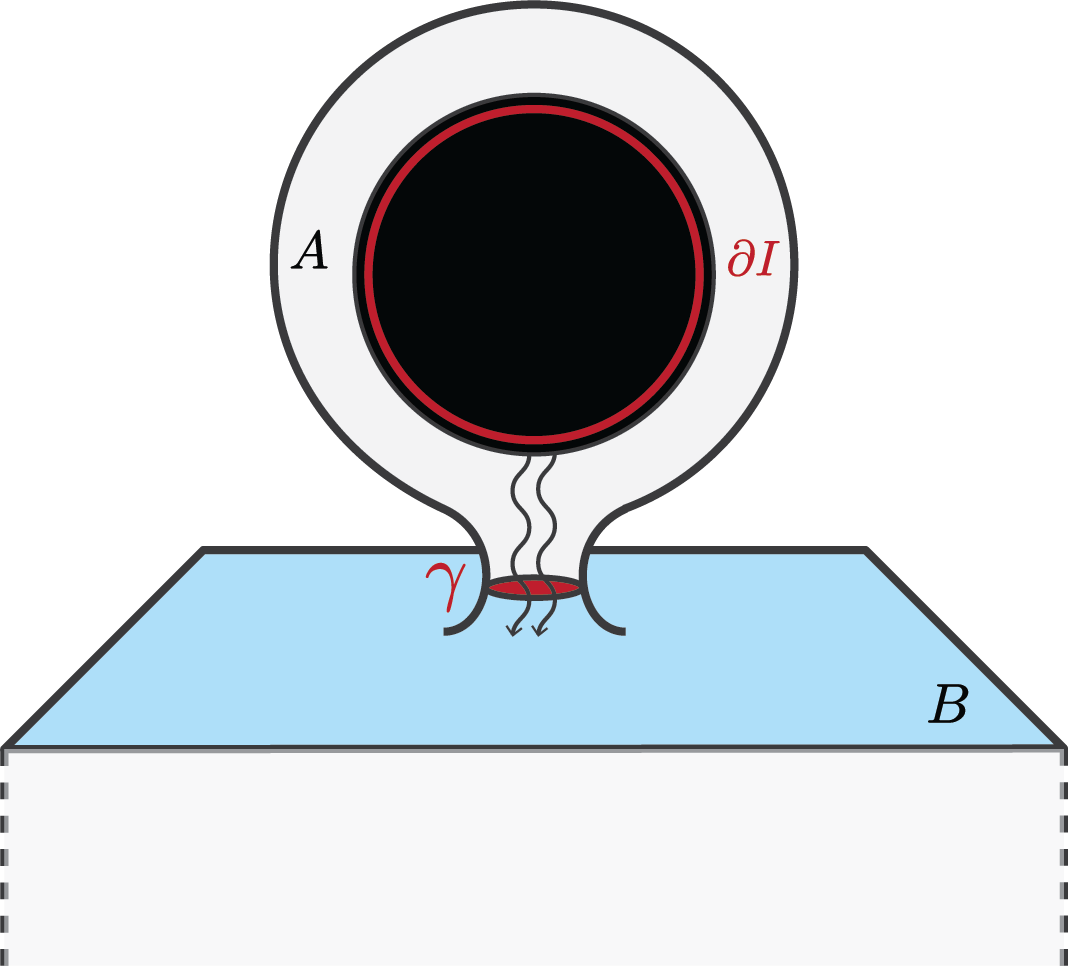}
\caption{An island produced by dividing the boundary into two parts. The boundary has a ``bubble'' and the bulk black hole is in the bubble. \label{figbubblebdry}}
\end{center}
\end{figure}

Another example of a Page curve that, conceptually, is almost identical to the plasma ball was studied in \cite{Antonini:2025sur}. This setup also involves a black hole that is localized on one part of the boundary. However, rather than studying a boundary $R^{d}$, the authors of \cite{Antonini:2025sur} study a boundary that has a ``neck'' and a ``bubble'' as shown in Figure \ref{figbubblebdry}. Moreover, rather than studying a generic state one considers a very atypical state with a black hole that is localized in the bubble. 

The exterior observer chooses to measure observables in the region $B$ (the complement of the bubble) but not in the region $A$. The entanglement entropy measured by such an observer  follows a Page curve.  This is because information about the black hole that is present on the asymptotic boundary, but localized in the region $A$, gradually spreads over the entire boundary.  As above, this is a purely nongravitational Page curve that arises due to redistribution of information on different parts of the boundary. 

It was argued in \cite{Antonini:2025sur} that the computation of the Page curve in this setting could give rise to an entanglement wedge that is the union of a disconnected piece and an asymptotic piece \cite{Antonini:2025sur}. (See \cite{Geng:2025byh} for additional discussion of candidate entanglement wedges.)  We return to this in section \ref{secinconsistency}. Here, we simply note that the arguments of \cite{Geng:2021hlu} do not rule out such a wedge because it contains an asymptotic piece. Therefore, from the bulk perspective, the Hamiltonian is not observable purely in the entanglement wedge or purely in its complement.

\subsection{Page curve in flat space}
We now explain how a Page curve can be obtained in flat space. The principle is the same as above: we choose to discard some observations, and the Page curve is a measure of information flow between the degrees of freedom we keep and those we discard.

The proposal for obtaining a Page curve discussed below was first discussed in \cite{Laddha:2020kvp,Raju:2021lwh,Raju:2020smc} and then reviewed in \cite{Antonini:2025sur}. 
 
Consider the truncated algebra that contains only news operators, and does not contain the shear or the Bondi mass.
\be
\label{alsettruncdef}
\begin{split}
\altrunc(u_0) = &~\text{span}\{N_{AB}(u_1, \Omega),  O(u_1, \Omega), N_{AB}(u_1, \Omega)  O(u_2, \Omega'),  N_{A B}(u_1,\Omega) N_{A'B'}(u_2,\Omega') \ldots \},\\
&\text{with}~~u_i \in (-\infty, u_0),
\end{split}
\ee
where the boundary values of the metric and of matter fields are defined as in \eqref{bondicoords} and \eqref{matterlimit}. Apart from the truncation in the range of $u$, this differs structurally from the definition \eqref{alsetdef} because we have discarded the constrained operator $m(u, \Omega)$ but retained the news. Therefore, this algebra should be distinguished from the algebra defined in section \ref{secfinegrained}. 

This is an unconstrained free-field algebra. We can define a commuting algebra of news operators
\be
\label{complementalsettruncdef}
\begin{split}
\albartrunc(u_0) =&~\text{span}\{N_{AB}(u_1, \Omega), O(u_1, \Omega), N_{AB}(u_1, \Omega) O(u_2, \Omega'), N_{A B}(u_1,\Omega) N_{A'B'}(u_2,\Omega') \ldots \},\\
&\text{with}~~u_i \in (u_0, \infty).
\end{split}
\ee
The elements of \eqref{complementalsettruncdef} commute with those of \eqref{alsettruncdef} because of the commutation relations
\be
 [N_{AB} (u, \Omega), N_{CD}(u',\Omega') ] = i 16 \pi G \partial_{u'} \delta(u-u') {\delta(\Omega, \Omega') \over \sqrt{\gamma}} \left[\gamma_{A(C}\gamma_{D)B} - {1 \over 2} \gamma_{A B} \gamma_{CD} \right].
\ee
These relations follow from canonical commutation relations between $N_{AB}$ and $C_{AB}$ 
\be
[ N_{AB} (u, \Omega), C_{CD}(u',\Omega') ] = i 16 \pi G \delta(u-u') {\delta(\Omega, \Omega') \over \sqrt{\gamma}} \left[\gamma_{A(C}\gamma_{D)B} - {1 \over 2} \gamma_{A B} \gamma_{CD} \right].
\ee

The algebras $\altrunc(u_0)$ and $\albartrunc(u_0)$ can be thought of as containing the radiative degrees of freedom without the constrained degrees of freedom. 

Although the elements of  $\altrunc(u_0)$ and $\albartrunc(u_0)$ commute, these subalgebras are not commutants of each other.  The commutant of the algebra $\altrunc(u_0)$ is an enlarged version of the algebra $\albartrunc(u_0)$ with $N_{AB}$ replaced by $C_{AB}$. This is a subtle point that we return to below. 

Both $\altrunc(u_0)$ and $\albartrunc(u_0)$ contain the news and not the shear. The news can be integrated to obtain the shear via
\be
\label{remainderc}
C_{A B}(u, \Omega) = C_{A B}(\infty, \Omega) - \int_{u}^{\infty} N_{AB}(u, \Omega).
\ee
The operator $m(u,\Omega)$ is not part of either algebra. It is related to the operators that appear via
\be
\partial_{u} m(u, \Omega) = {1 \over 4} \partial_{u} D^{A} D^{B} C_{AB}(u, \Omega) - {1 \over 8} N_{AB} (u, \Omega) N^{AB}(u, \Omega) - 4 \pi G T_{u u}^{M}(u, \Omega)\,,
\ee
where $T_{u u}^M(u, \Omega)$ is the coefficient of the asymptotic ${1 \over r^2}$ term of the matter stress tensor. 

It is possible to integrate this equation to relate $m(u,\Omega)$ to operators in $\albartrunc(u_0)$ and $\altrunc(u_0)$,
\be
m(u,\Omega) = {\cal R}(\Omega) +   {1 \over 4} D^{A} D^{B} C_{AB}(u, \Omega) + \int_{u}^{\infty} d u \left({1 \over 8} N_{AB}(u, \Omega) N^{AB}(u,\Omega) +  4 \pi G T_{u u}^{M}(u,\Omega) \right),
\ee
with
\be
\label{remainderm}
{\cal R}(\Omega) = m(\infty,\Omega) - {1 \over 4} D^{A} D^{B} C_{AB}(\infty,\Omega).
\ee

For now, let us neglect the operators  $m(\infty, \Omega)$ and $C_{AB}(\infty, \Omega)$ that appear in \eqref{remainderm} and \eqref{remainderc}; we will return to them below. These operators carry information about massive particles and soft charges. For configurations where the information in these degrees of freedom is negligible,  we can effectively reconstruct the Bondi mass aspect and the local shear at each point on ${\cal I}^{+}$ given the observables in $\albartrunc(u_0)$ and $\altrunc(u_0)$. Therefore, every observable in the theory can effectively be written as a linear combination of products of observables in these two subalgebras for such configurations.  

This  returns us to the situation studied by Page where the space of observables factorizes into observables from an algebra and observables from its complement. Therefore, Page's argument applies and, if a black hole can be thought of as a typical state, then it is reasonable to expect that the appropriately regularized entropy of $\altrunc(u_0)$ and of $\albartrunc(u_0)$ follows a Page curve.  

We would like to point out two subtleties.

\paragraph{What does a physical observer measure?}
The first subtlety is that the algebra above is somewhat artificial because we drop some components of the metric while retaining others. This can be seen by considering a natural observable like the Riemann tensor. The Riemann tensor sees {\em all} components of the metric and not simply the news operator. For instance, a computation of the Riemann tensor shows that in the Bondi frame (i.e. $\gamma^{AB}C_{AB}=0$)
\be
\label{riemanncomp}
R_{r u r u} = -\frac{2m(u, \Omega)}{r^3}+O(\frac{1}{r^4})\,.
\ee
A more elaborate computation of the Weyl scalars, as performed in \cite{Geiller:2024ryw} (see Equation 2.13) shows that the second Weyl scalar is given by
\be
\begin{split}
\Psi_2 &= {1 \over r^3} \left(m(u, \Omega) + {1 \over 16} \partial_{u} \big(C_{AB}(u,\Omega) C^{AB}(u,\Omega)\big)  \right) \\&+ {i \over 4 r^3} \left(D_{A} D_{B} - {1 \over 2} N_{A B}(u,\Omega) \right) \gamma^{A C} \gamma^{B D} C_{C D}(u,\Omega).
\end{split}
\ee
Therefore, natural gauge-invariant quantities at ${\cal I}^{+}$ depend on the Bondi mass.\footnote{The Weyl scalar computed in \cite{Geiller:2024ryw} and the Riemann tensor component quoted here depend on the mass aspect. The mass aspect can be changed via a supertranslation, which corresponds to a change of asymptotic Bondi frame. But the Bondi mass is invariant under supertranslations. In our arguments, we only need the ADM Hamiltonian, which is the limit of the Bondi mass as $u \rightarrow -\infty$.}

The paper \cite{Antonini:2025sur} advocated that one should examine $\altrunc(u_0)$ rather than $\alset(u_0)$ (which was defined in section \ref{secfinegrained} ) and suggested that this was ``the perspective of most of the field.''  

However, we believe that a common misunderstanding is that an observer making measurements {\em outside} the black hole will observe a Page curve as information emerges from the black hole. Such an observer would naturally measure observables such as \eqref{riemanncomp}, which contain information about the mass. 
This is true even in the classical theory. For instance, the mass of a  black hole is an important observable in astronomy.

By shutting our eyes to information about the mass, one may obtain a Page curve --- as was already noted in \cite{Laddha:2020kvp,Raju:2020smc,Raju:2021lwh}. But it is possible that the community might classify this as a mathematical curiosity, were this point to be clearly emphasized. 

Second, at any finite value of $r$, the OPE of components of the metric operator with themselves will generate other components of the metric operator. It is {\em only} strictly at ${\cal I}^{+}$ that it is possible to cleanly retain some components of the metric while dropping others. In reality, any physical observer can only assemble a Dyson sphere with finite radius.  At finite $r$, the algebra of operators is not free and we do not know of any way to discard the constrained components of the metric, while retaining the others. This provides an additional reason that the restriction of the algebra $\alset(u_0)$ to $\altrunc(u_0)$ is physically unnatural. 

\paragraph{Soft and hard entanglement.}

Setting aside the issue of naturalness, dropping the Hamiltonian leads to a Page curve subject to an additional potential subtlety that we now describe.  This issue was briefly discussed in \cite{Laddha:2020kvp} and pertains to the observables $m(\infty, \Omega)$ and $C_{AB}(\infty, \Omega)$ that were dropped above. 

These observables receive contributions from massive particles whose world lines end at $i^+$ rather than ${\cal I}^{+}$. For a large enough black hole it is reasonable to neglect this term. However, the operator $C_{AB}(\infty, \Omega)$  also receives a contribution from the zero mode of the shear that is not present in $N_{AB}(u,\Omega)$. This term contributes to the so-called {\em soft} supertranslation charge \cite{Strominger:2017zoo}. 

Therefore, the algebra of news operators does not have information about the soft charge. In \cite{Hawking:2016sgy}, it was suggested that the soft charge carries significant information about the black hole. If so, then the entanglement between the hard and soft degrees of freedom would prevent the algebra of news operators from following a Page curve.

We are sympathetic to the idea that this subtlety is not important. However, it is worth examining this issue in greater detail although we do not do so in this paper.

\subsection{Open question: Page curve for small AdS black holes? \label{subsecopenquestion}}
In the discussion above, we explained that the algebra at ${\cal I}^{+}$ is somewhat peculiar because of simplifications that happen as $r \rightarrow \infty$. This allows us to discard the ADM Hamiltonian from the algebra in a manner that it does not reappear in OPEs. 

It is not convenient to directly study the algebra at finite radius since finite-size effects are hard to control in a theory of quantum gravity. On the other hand, anti-de Sitter space provides a natural model of gravity in a box. So it is instructive to study ``small black holes'' in AdS, which are black holes that are small enough that they behave like flat-space black holes. The algebra on the AdS boundary is precisely defined but it is not free. In this sense, we can think of observations at the AdS boundary as providing a model of observations at finite radius in flat space. 

We first clarify the notion of a small AdS black hole. For a black hole to be amenable to semiclassical analysis, its horizon radius, $r_s$, must satisfy 
\be
r_s \gg l_{\text{pl}}\,,
\ee
where $l_{\text{pl}}$ is the Planck length.
But if the evaporation time of the black hole is much smaller than the AdS$_{d+1}$ radius, $l_{\text{ads}}$,  
\be
t = {r_s^d \over l_{pl}^{d-1}} \ll l_{\text{ads}}\,,
\ee
the black hole evaporates as if it is in flat space. When the AdS scale and the Planck scale are separated by a large parameter,
\be
{\cal N} = {l_{\text{ads}} \over l_{\text{pl}}},
\ee
this leaves a large window of masses where black holes can be semiclassical but yet evaporate as if they are in flat space. This is the window where the black hole radius satisfies 
\be
{l_{\text{ads}} \over {\cal N}} \ll r_s \ll {l_{\text{ads}} \over {\cal N}^{{d-1 \over d}}}.
\ee
The corresponding mass-parameter of the black hole, $\mu = {r_s^{d-2} \over l_{\text{pl}}^{d-1}}$ must satisfy
\be
{{\cal N} \over l_{\text{ads}}}  \ll \mu \ll \frac{\mathcal{N}^{\frac{2(d-1)}{d}}}{l_{\text{ads}}}.
\ee

We will now consider the following question.
\begin{question}
Is there a natural measure of the entropy of the entire AdS boundary that follows the Page curve as the black hole evaporates?
\end{question}
We will consider various possibilities for such an entropy and show that they run into difficulty. We provide some arguments that suggest that such a measure should not exist.  On the other hand, it is quite easy to find algebras whose entropy always vanishes and also coarse-grained algebras whose entropy follows the Hawking curve (i.e. rises monotonically). We also explain why these difficulties arise specifically for black holes and not when we are studying a lump of coal. 

\subsubsection{Coarse-grained entropy}

A first possibility is to define an entropy using a coarse-grained set of boundary observables. Let us define a set of coarse-grained observables that involve products of a ``few'' light operators
\be
\label{alcoarsedef}
\begin{split}
\alcoarse(t,n_c) =&~\text{span}\{O(t_1), O(t_1) O(t_2), \ldots, O(t_1) O(t_2) \ldots O(t_n) \},\\
&\text{with}~~n < n_c~~\text{and}~~t_i \in [t - {\epsilon \over 2}, t + {\epsilon \over 2}],
\end{split}
\ee
where $n_c$ is a cutoff.  We take $\epsilon$ to be a time-interval that is much smaller than the evaporation time of the black hole. This algebra should be distinguished from the algebra defined in section \ref{secfinegrained} because of the cutoff $n_c$. For the definition of $\alcoarse(t,n_c)$, we will assume that the cutoff does not scale with ${\cal N}$.  In contrast, the algebra in section \ref{secfinegrained} is defined in the limit where $n_c \rightarrow \infty$.

The algebra $\alcoarse(t,n_c)$  allows us to define an entropy for the state, $|\Psi \rangle$,  using the S-maximization procedure \cite{jaynes1957information,jaynes1957informationII}. We define the entropy to be the maximum of 
\be
\label{smax}
S_{\text{coarse}}(t,n_c) = \text{max}\left[-\tr(\rho(t,n_c) \log \rho(t,n_c)) \right],
\ee
where the maximum is taken by allowing $\rho(t,n_c)$ to range over operators (including operators that do not belong to $\alcoarse(t,n_c)$) that satisfy
\be
\label{rhosmax}
\tr(\rho(t,n_c) a) = \langle \Psi | a | \Psi \rangle, \qquad \forall a \in \alcoarse(t,n_c).
\ee
Note that \eqref{rhosmax} constrains $\rho(t,n_c)$ but does not fix it since the space of allowed $\rho(t,n_c)$ is larger than the size of $\alcoarse(t,n_c)$ and the maximization prescription \eqref{smax} instructs us to pick the $\rho(t,n_c)$ that yields the largest entropy.

As a function of $t$, this entropy is expected to have the same properties as the thermodynamic entropy. Therefore, before the black hole forms, we expect that such an entropy will be small because even low-point correlators are sufficient to identify the state precisely. After the black hole forms, we expect that such an entropy will be close to the Bekenstein-Hawking entropy of the black hole. After the black hole evaporates, this entropy increases even further to approach the final entropy of the Hawking radiation. Therefore, it follows the Hawking curve.

The entropy above does not measure the purity of the underlying state. So this is not a paradox. It is simply a manifestation of the second law of thermodynamics, which tells us that the coarse-grained entropy of the system always rises.\footnote{Similar conclusions apply to \cite{deBoer:2026cng}.}

\paragraph{Varying the size of the algebra.}
The coarse-grained entropy is defined with respect to the set of observables \eqref{alcoarsedef}. The size of this set increases with $n_c$.

Since the coarse-grained entropy is defined a maximization procedure, it is clear that, keeping $t$ fixed, we have
\be
\label{scoarsenc}
{\partial S_{\text{coarse}}(t,n_c) \over \partial n_c} \leq 0
\ee

\subsubsection{Fine-grained entropy}

The fine-grained entropy for boundary observables is obtained in the limit where we send $n_c \rightarrow \infty$ in \eqref{alcoarsedef}. This is the appropriate algebra for an observer who can make  fine-grained measurements of the state at the boundary. It was shown in section \ref{secfinegrained} that the entropy defined using this algebra vanishes: $\dot{S}(t) = 0$ as shown in \eqref{flatinads}.

From the point of view of AdS/CFT, this is simply the obvious statement that the boundary CFT is always in a pure state whose von Neumann entropy vanishes.

\subsubsection{Dropping the Hamiltonian?}
One might hope to move away from the restrictive result that the fine-grained entropy vanishes by dropping the Hamiltonian as can be done on ${\cal I}^{+}$. However, this is not straightforward at the AdS boundary. 
Even if we do not include the Hamiltonian in the algebra to start with, the Hamiltonian is {\em generated} via the OPE.
\be
\label{opebdry}
O(x) O(0) = {1 \over |x|^{2 \Delta}} + \ldots + {\Delta \over d C_{T}} {x^{\mu} x^{\nu} \over |x|^{2 \Delta - d + 2}} T_{\mu \nu}(0) + \ldots,
\ee
where $C_{T}$ is related to the two-point function of the stress tensor.
Here, the first term gives the contribution of the identity operator. The $\ldots$ at the end indicate terms that are less singular in the limit $|x| \rightarrow 0$.  If there are additional operators in the theory with a dimension between the identity operator and the stress-tensor, they appear in the first $\ldots$ and are more singular than the displayed term in the limit $|x| \rightarrow 0$. 

Since the number of light operators is limited in any theory with a gravitational dual, we can subtract off their contributions. This leads us to
\be
\left[O(x) O(0)\right]_{\text{subtracted}} = {\Delta \over d C_{T}} {x^{\mu} x^{\nu} \over |x|^{2 \Delta - d + 2}} T_{\mu \nu}(0) + \ldots,
\ee
where the $\ldots$ now indicates only less-singular terms. 

If we separate the two operators slightly in time by $\delta t$ and then take $\delta t \rightarrow 0$, we find that
\be
\lim_{\delta t \rightarrow 0} (\delta t)^{(2 \Delta - d)} \left[O(\delta t) O(0)\right]_{\text{subtracted}} = {\Delta \over d C_{T}} T_{0 0}.
\ee
By integrating the right-hand side we get the Hamiltonian. 
Therefore, in a local CFT, closure of the algebra of a time band requires us to include the Hamiltonian. 

The OPE \eqref{opebdry} reflects the bulk physics that the interaction of two nongravitational particles can produce a graviton. The difference between flat space and AdS is that due to the strict $r \rightarrow \infty$ limit that we take to reach ${\cal I}^{+}$, the algebra becomes free. This does not happen in AdS. A small black hole in AdS serves as a model of a black hole in flat space surrounded by a detector at large but finite radius, and this discussion shows that dropping the Hamiltonian from the set of observables is not straightforward in this setting.

\subsubsection{Intermediate entropy?}
To summarize, we have found above that there are two natural regimes in which to study the entropy of a black hole. One of them is the coarse-grained regime, where the entropy satisfies the Hawking curve. This is consistent with locality since Hawking's argument for information loss is based on locality.  The other is the fine-grained regime. In this regime, the holographic properties of gravity are important and bulk locality is not exact. For this reason, Hawking's argument fails. 

As far as we know, no third intermediate regime where the Page curve emerges naturally has been described in the extant literature.\footnote{We can exploit \eqref{scoarsenc} to get a Page curve by simply varying $n_c$ as a function of time. Such a Page curve, once again, measures the rate at which we {\em choose} to measure information.  Correspondingly, the ``Page time''  depends on how we vary $n_c$ rather than being an intrinsic property of the black hole. The question in this subsection is whether it is possible to fix $n_c$ once and for all so that the entropy obeys a Page curve with time.}

If the black hole had been an ordinary lump of coal in AdS, one might have hoped to find a Page curve by setting $n_{c}$ in \eqref{algcoarsedef} to be large enough to detect the purity of the state but small enough to be insensitive to the gravitational effects that lead to holography. Let us make this more precise.

If the lump of coal in the bulk emits particles with typical energy $E$ and has entropy $S$ then we keep operators in the algebra that can keep track of effects of size $e^{-S}$. But we drop nonperturbative effects in the dimensionless gravitational coupling 
\be
\lambda = G E^{d-1}.
\ee

On the other hand, this is not straightforward for a black hole. For a small black hole in AdS, the entropy, $S_{\text{bh}}$ and the dimensionless coupling are linked through 
\be
G E^{d-1} = {\zeta \over S_{\text{bh}}}; \qquad \zeta = {\Omega_{d-1} \over 4} \left({d - 2 \over 4 \pi} \right)^{d-1},
\ee
where $\Omega_{d-1}$ is the volume of a unit $S^{d-1}$ and $E=T_{H}$ with $T_{H}$ the temperature of the black hole, which is the same relation followed by flat-space black holes.
Thus, for the black hole, sending $G \rightarrow 0$ also sends $S_{\text{bh}} \rightarrow \infty$.

\section{Inconsistency of islands in standard gravity \label{secinconsistency}}
In this section, we explain why islands are not expected to be relevant when we study evaporating black holes in a pure state in a standard theory of gravity. 

We remind the reader that we strictly use the term ``island'' to refer to entanglement wedges whose intersection with the gravitational region is compact. This is also the definition of ``islands'' supplied in \cite{Geng:2021hlu}. In models with a nongravitational bath, this definition is broadly accepted. But, we caution the reader that in models without a nongravitational bath, the term ``islands'' might be used more broadly to refer to any entanglement wedge with a disconnected piece.

The argument is that the existence of islands implies the existence of an underlying algebra of operators that commutes with the algebra of the asymptotic boundary. We argue below that these commutators must vanish to nonperturbative accuracy for islands to be consistent. But this is impossible when the assumptions of section \ref{secoutsidecomplete} hold.

In the second part of this section, we examine ``relational observables''. Relational observables are important because they tell us how to define {\em approximately} local operators in the presence of a background. Relational observables  {\em cannot} be used to remove the potential contradiction between islands and the absence of a commutant for the asymptotic algebra because they do not give rise to a true algebra that can be used to define a fine-grained von Neumann entropy. We show this in one suggested realization of relational observables and explain why this is expected to be true for other realizations.

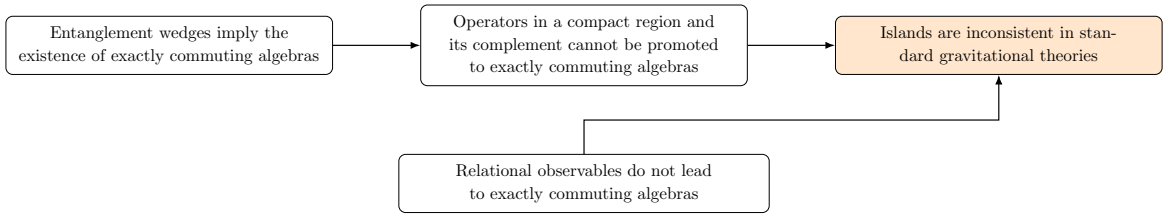
\begin{figure}[H]
\centering
\resizebox{\textwidth}{!}{
\begin{tikzpicture}[
  font=\small,
  box/.style={draw, rounded corners, align=center, inner sep=6pt, text width=6.3cm},
  arrow/.style={-{Latex[length=2mm]}, line width=0.7pt}
]
\node[box] (ew) {Entanglement wedges imply the existence of exactly commuting algebras};

\node[box, right=18mm of ew] (compact)
{Operators in a compact region and its complement cannot be promoted to exactly commuting algebras};

\node[box, right=18mm of compact,fill=orange!20] (islands)
{Islands are inconsistent in standard gravitational theories};

\node[box, below=14mm of compact, text width=7.2cm] (rel)
{Relational observables do not lead to exactly commuting algebras};

\draw[arrow] (ew) -- (compact);
\draw[arrow] (compact) -- (islands);

\draw[arrow] (rel.north) -- ++(0,7mm) -| (islands.south);

\end{tikzpicture}
}
\caption{Flowchart for section \ref{secinconsistency}. \label{flowchartislandsinconsistent}}
\end{figure}

A flowchart depicting the logic of this section is provided in Figure \ref{flowchartislandsinconsistent}.

\subsection{Entanglement wedges and commuting algebras \label{entwedgeandalg}}

Islands are a special case of entanglement wedges. So we first provide a general argument that entanglement wedges correspond to underlying exact commuting algebras. This assertion is true even though entanglement wedge reconstruction is only approximate.

Entanglement wedges are byproducts of the quantum extremal surface (QES) prescription. Given a boundary region, $A$, on the conformal boundary of AdS,  whose density matrix is $\rho_{A}$ when the global state is some $|\Psi \rangle$,  its von Neumann entropy is computed through a bulk computation using \cite{Ryu:2006bv,Ryu:2006ef,Hubeny:2007xt,Engelhardt:2014gca}
\be
\label{qesprescription}
S_{A} = -\tr(\rho_{A} \ln \rho_{A}) =  \text{min}~\text{ext}\left({A(\gamma_{A}) \over 4 G} + S_{\text{bulk}} \right),
\ee
where $\gamma_{A}$ is the boundary of a bulk region, with semiclassical entropy $S_{\text{bulk}}$, whose conformal boundary is $A$.

\paragraph{Bulk reconstruction. }
Entanglement wedge reconstruction \cite{Jafferis:2015del,Faulkner:2017vdd} tells us that bulk operators in the entanglement wedge can be reconstructed from operators in $A$. This is a generalization of HKLL bulk reconstruction \cite{Hamilton:2007wj,Hamilton:2006fh,Hamilton:2005ju,Hamilton:2006az}, and like other examples of bulk reconstruction, it is only approximate. 

The precise objective \cite{Papadodimas:2015jra} of bulk reconstruction is to find CFT operators from the boundary region $A$, $\phi_{\text{CFT}}^{\rm A}(x)$ with the property that, 
\be
\label{bulkreconstruction}
\langle \Psi| \phi_{\text{CFT}}^{\rm A}(x_1) \ldots \phi^{\rm A}_{\text{CFT}}(x_n) |\Psi \rangle = G(x_1, \ldots, x_n) + {\rm O}\left({1 \over N} \right),
\ee
where on the right-hand side $G(x_1, \ldots x_n)$ is the bulk Green's function computed using the rules of QFT in curved spacetime \cite{birrell1984quantum}.  We also expect the relationship to work only when $n \ll N$.

\begin{figure}
    \centering
\includegraphics[width=0.3\linewidth]{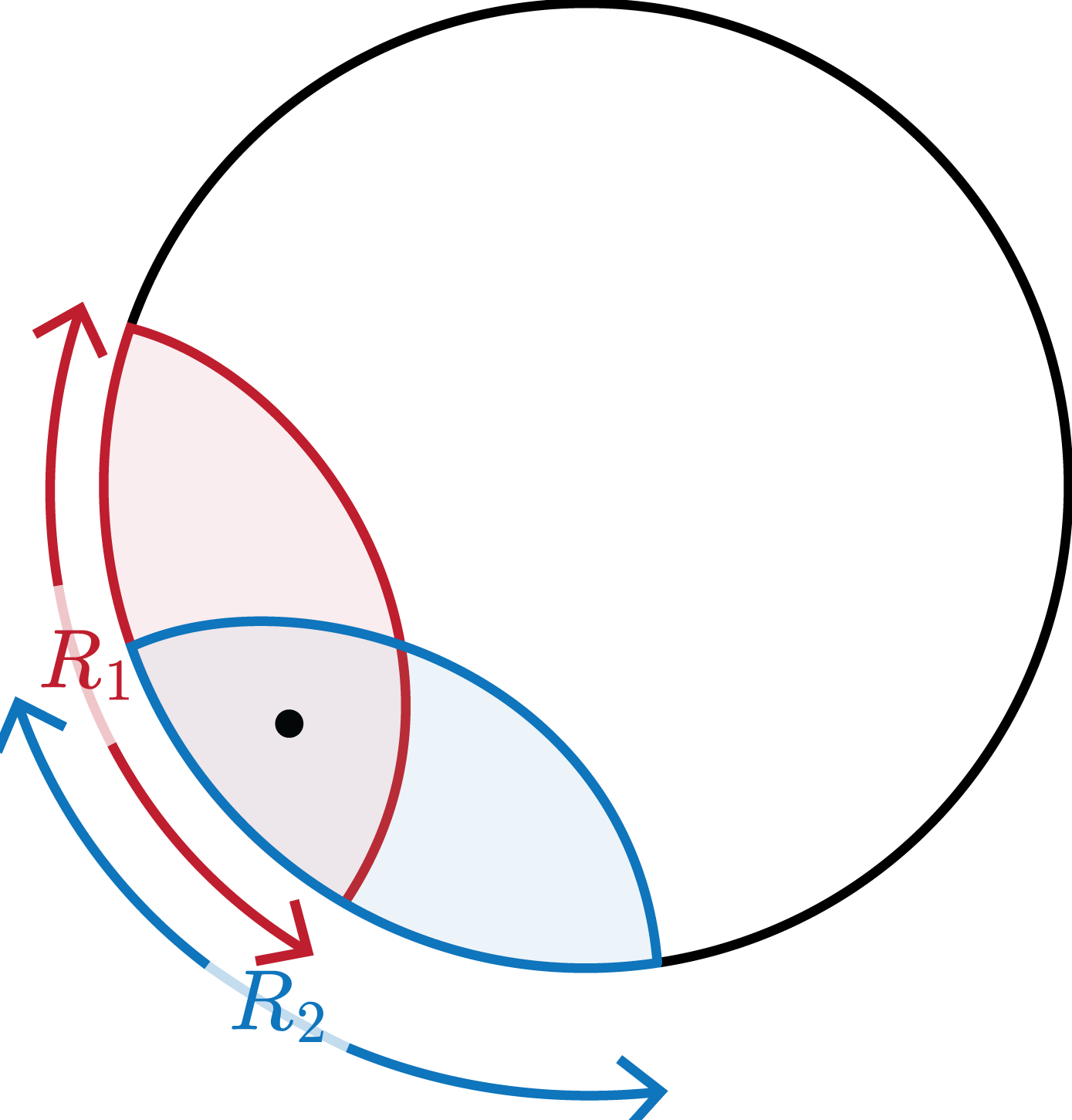}
    \caption{A point in AdS can be part of two entanglement wedges. Therefore the bulk operators in part of an entanglement wedge can be ``completed'' in multiple ways to a well-defined subalgebra of the full theory. \label{figqec}}
\end{figure}

Since we  demand that \eqref{bulkreconstruction} hold only for low-point correlators and make no demands on the behavior of $\phi_{\text{CFT}}^{A}$ in other correlators, and furthermore because we allow $\Or[{1 \over N}]$ errors in \eqref{bulkreconstruction}, it is possible to find multiple CFT operators that satisfy \eqref{bulkreconstruction}. For example, in empty AdS, the bulk field at a point can be reconstructed from multiple boundary regions as shown in Fig \ref{figqec}.  This immediately implies that entanglement-wedge reconstruction is only approximate and works only in a little Hilbert space or code subspace about $|\Psi \rangle.$

\paragraph{Underlying algebras.}
The approximate nature of bulk reconstruction might naively suggest that it is sufficient to define the algebra of observables for an entanglement wedge and its complement approximately. 

However, the formula \eqref{qesprescription} makes reference to the entanglement entropy. The entanglement entropy is a nonperturbative quantity since it is sensitive to exponentially-small corrections. This is simply because the trace in \eqref{qesprescription} is evaluated in a space whose dimension is at least $e^{S_{A}}$. Two density matrices with the property that their matrix elements disagree only at $e^{-{S_{A} \over 2}}$ can have entanglement entropies that differ by $\Or[1]$.\footnote{A related statement is that a pure state and a mixed state can agree on almost all correlators to exponential precision as reviewed in section \ref{sechawkingpage} but the entanglement entropy can easily distinguish between them.}
We do not know of any way of defining the entanglement entropy for an ``approximate algebra'' although perhaps some other quantum-information measures can be defined in that setting \cite{Ghosh:2017gtw}.

The entanglement wedge is not an arbitrary bulk region but appears when we use the gravitational path integral to compute a well-defined entropy. Since defining the underlying entropy requires an exactly factorized algebra, this leads us to the following conclusion.

\begin{conclusion}
The existence of an entanglement wedge is concomitant with the existence of a well-defined entropy, and therefore it  corresponds to an exact factorization of the set of observables into an algebra and its commutant.
\end{conclusion}

We reiterate that this conclusion is not in contradiction with the idea that bulk reconstruction is only approximate. The claim is that even though bulk reconstruction is approximate, it should be possible to ``complete'' the set of bulk operators in an entanglement wedge to a nonperturbative algebra. There might be multiple completions possible as in the case of Fig \ref{figqec}, but there should be at least one.

An analogy with effective field theory might be helpful. It is often possible to find multiple UV completions for a given effective field theory. But, sometimes, a theory might have no possible UV completion, in which case it is inconsistent even if the inconsistency is not obvious in the low-energy physics \cite{ArkaniHamed:2006dz}.

\subsection{Islands and commuting algebras}

In the presence of a nongravitational bath, the island is a compact entanglement wedge in the gravitational region  that is redundant with degrees of freedom in the radiation region, which is part of the nongravitational bath.  

Separately, it has been suggested that islands are relevant for black holes in the real world \cite{Almheiri:2019hni, Krishnan:2020fer,  Krishnan:2020oun, Ghosh:2021axl, Gautason:2020tmk}. In this setting, the radiation region is the entire asymptotic gravitational region  and it is redundant with a compact region inside the black hole \cite{Almheiri:2020cfm}.  Now, using our strict rule that islands cannot include an asymptotic piece, the union of the radiation region and this compact region in the interior is not an island. Nevertheless, we do have a compact entanglement wedge here: the complement of the island and the radiation region. So, our argument below applies to this case as well.
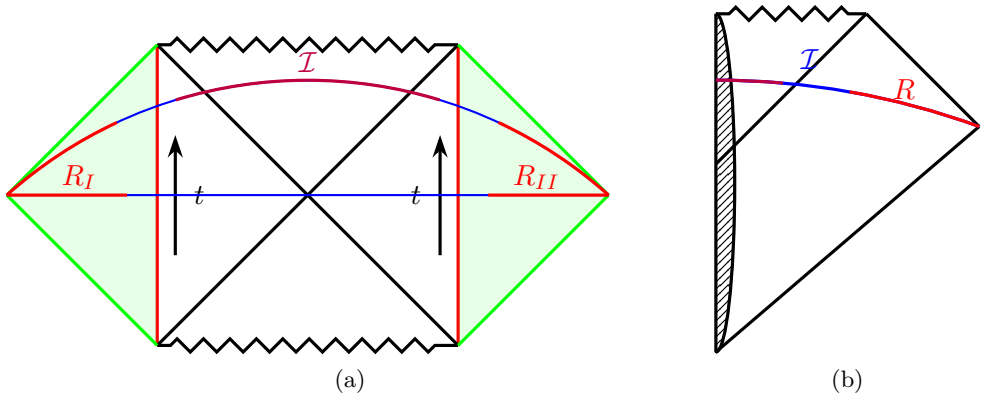
\begin{figure}[H]
\begin{center}
\subfloat[]{
  \begin{tikzpicture}[scale=0.8,decoration=snake]
       \draw[-,very thick] 
       decorate[decoration={zigzag,pre=lineto,pre length=5pt,post=lineto,post length=5pt}] {(-2.5,0) to (2.5,0)};
       \draw[-,very thick,red] (-2.5,0) to (-2.5,-5);
       \draw[-,very thick,red] (2.5,0) to (2.5,-5);
         \draw[-,very thick] 
       decorate[decoration={zigzag,pre=lineto,pre length=5pt,post=lineto,post length=5pt}] {(-2.5,-5) to (2.5,-5)};
       \draw[-,very thick] (-2.5,0) to (2.5,-5);
       \draw[-,very thick] (2.5,0) to (-2.5,-5);
       \draw[-,very thick,green] (-2.5,0) to (-5,-2.5);
       \draw[-,very thick,green] (-5,-2.5) to (-2.5,-5);
        \draw[-,very thick,green] (2.5,0) to (5,-2.5);
       \draw[-,very thick,green] (5,-2.5) to (2.5,-5);
       \draw[fill=green, draw=none, fill opacity = 0.1] (-2.5,0) to (-5,-2.5) to (-2.5,-5) to (-2.5,0);
       \draw[fill=green, draw=none, fill opacity = 0.1] (2.5,0) to (5,-2.5) to (2.5,-5) to (2.5,0);
         \draw[->,very thick,black] (-2.2,-3.5) to (-2.2,-1.5);
       \node at (-1.8,-2.5)
       {\textcolor{black}{$t$}};
        \draw[->,very thick,black] (2.2,-3.5) to (2.2,-1.5);
       \node at (1.8,-2.5)
       {\textcolor{black}{$t$}};
       \draw[-,thick,blue] (-5,-2.5) to (5,-2.5);
       \draw[-,thick,blue] (-5,-2.5) arc (90+41.81:90-41.81:7.5);
        \draw[-,very thick,red] (-5,-2.5) to (-3,-2.5);
       \draw[-,very thick,red] (5,-2.5) to (3,-2.5);
       \draw[-,very thick,red] (-5,-2.5) arc (90+41.81:115:7.5);
       \draw[-,very thick,red] (5,-2.5) arc (90-41.81:65:7.5);
 \node at (-3.8,-2.2)
       {\textcolor{red}{$R_{I}$}};
 \node at (3.8,-2.2)
       {\textcolor{red}{$R_{II}$}};
        \draw[-,very thick,purple] (0,-2.5+1.90983) arc (90:107:7.5);
          \draw[-,very thick,purple] (0,-2.5+1.90983) arc (90:73:7.5);
        \node at (0,-2.5+1.90983+0.3)
       {\textcolor{purple}{$\mathcal{I}$}};  
    \end{tikzpicture}
     \hspace{1.0cm}
\label{islandbath}}
   \subfloat[]{
\begin{tikzpicture}[scale=0.5,decoration=snake]
    \draw[-,very thick] 
       decorate[decoration={zigzag,pre=lineto,pre
       length=5pt,post=lineto,post length=5pt}] {(-2+0.5,0) to (2+0.5,0)};
       \draw[-,very thick] (-2+0.5,0) to (-2+0.5,-9);
       \draw[-,very thick] (2+0.5,0) to (-2+0.5,-4);
       \draw[-,very thick] (2+0.5,0) to (5+0.5,-3);
       \draw[-,very thick] (5+0.5,-3) to (-2+0.5,-9);
       \draw[-,very thick,black] (-2+0.5,0) arc (90:-90:0.5 and 4.5);
    \draw[thick,black,pattern=north east lines,pattern color=black] (-2+0.5,0) arc (90:-90:0.5 and 4.5)--(-2+0.5,0);
    \draw[-,very thick,blue] (5+0.5,-3) arc (70:90:20.4666308011);
    \draw[-,very thick,red] (5+0.5,-3) arc (70:80:20.4666308011);
      \draw[-,very thick,purple] (-2+0.5,-3+1.23428886496) arc (90:85:20.4666308011);
       \node at (-1+2.0,-1.2)
       {\textcolor{blue}{$\mathcal{I}$}};
       \node at (3+0.5,-2)
       {\textcolor{red}{$R$}};
\end{tikzpicture}
\label{islandnobath}
}
\caption{Two possible pictures of an island that have appeared in the literature. The subfigure on the left shows an island in a model where AdS is coupled to a nongravitational bath.  The subfigure on the right shows an island in asymptotically flat space that is redundant with a radiation region near $i^0$. Such an island has never been found in an explicit controlled computation.  \label{figpossislands}}
\end{center}
\end{figure}

The argument of section \ref{entwedgeandalg} and our definition of islands leads to the following conclusion.
\begin{conclusion}
The consistency of islands requires the existence of a compact entanglement wedge in the gravitational region.
\end{conclusion}

The complement of this wedge must include the asymptotic gravitational region. Since asymptotic bulk operators are well defined even in quantum gravity, it is natural that the algebra of the asymptotic region is completed to the algebra on the boundary of a Cauchy slice, $\alset_{\epsilon}$ defined in section \ref{secoutsidecomplete}. Therefore, the existence of the island requires the existence of an algebra $X$ with the property that
\be
\label{commutvanish}
[O, a] = 0, \forall a \in \alset_{\epsilon} \qquad \forall O \in X.
\ee
But it was shown in section \ref{secoutsidecomplete} that in a standard theory of gravity the set of asymptotic operators $\alset_{\epsilon}$ is complete.  Therefore its commutant is empty and the algebra $X$ cannot exist in a standard theory of gravity.

The argument above is effectively the same as \cite{Geng:2021hlu},  although it uses the nonperturbative results of section \ref{secoutsidecomplete}. The use of perturbation theory in \cite{Geng:2021hlu}, which is valid for a special class of states, might have suggested that our argument was limited to that class of states. But the argument is general and applies whenever the boundary theory is in a pure state. 

Returning to Figure \ref{figpossislands}, many examples of islands with nongravitational bath (as shown in Figure \ref{islandbath}) have been found through concrete computations. As we explain in section \ref{secgoodislands}, the nongravitational bath leads to a nonstandard theory of gravity in the bulk where the assumptions of section \ref{secoutsidecomplete} do not apply. On the other hand, islands corresponding to the configuration shown in Figure \ref{islandnobath} that would be relevant for the real world have not been found in any controlled computation. This is not surprising because the argument above tells us that such islands would be inconsistent.

\subsection{Relational observables \label{secrelational}}
It is instructive to consider a special class of approximately local operators, called ``relational observables'' that might appear to satisfy \eqref{commutvanish}. We now show that their existence is not in contradiction with the result reached above. The main conclusion we will reach below is as follows.
\begin{conclusion}
Relational observables lead to approximately local operators that show how local algebras re-emerge in the weak-gravitational limit. However, they cannot be used to define an algebra that satisfies \eqref{commutvanish} to the accuracy necessary to define fine-grained gravitational entropies.
\end{conclusion}

The construction of ``relational observables'' has a long history \cite{dewitt1960quantization,DeWitt:1967yk,Brown:1994py,Kuchar:1990vy,Sen:2002qa,Giddings:2005id,Marolf:2015jha, Donnelly:2016rvo,Francois:2024rdm ,Geng:2024dbl,DeVuyst:2024grw,Chataignier:2024eil,AliAhmad:2024wja, Fewster:2024pur,DeVuyst:2024uvd,Chen:2024rpx,Kudler-Flam:2024psh,Kaplan:2024xyk}. While we cannot review all possible constructions,  we will review two of them.  (1) First, we review a state-dependent ``projector-based'' construction based on the Papadodimas-Raju proposal \cite{Papadodimas:2015xma,Papadodimas:2015jra} that was also discussed in \cite{Antonini:2025sur}; we will find that this construction works well at leading order but fails at subleading order. We then turn to (2) a refinement of the projector-based construction, again based on the Papadodimas-Raju construction \cite{Papadodimas:2012aq,Papadodimas:2013wnh,Papadodimas:2013jku}. We find that this construction works perturbatively but fails nonperturbatively.

\begin{figure}[H]
\begin{tikzpicture}[node distance=14mm and 14mm]
  
  \node[box] (a) {Relational observables};
  \node[box, right=of a] (b) {Projector-based-construction};
  \node[box, right=of b] (c) {Limitations at subleading order};

  \node[box, below=of b] (d) {Tomita-Takesaki construction};
  \node[box, left=of d] (e) {Impossibility of exact algebras in bounded regions};

  \draw[arrow] (a) -- (b);
  \draw[arrow] (b) -- (c);

  \draw[arrow] (c) |- (d);
  \draw[arrow] (d) -- (e);
\end{tikzpicture}
\caption{Flowchart for subsection \ref{secrelational}\label{logicflowrel}}
\end{figure}
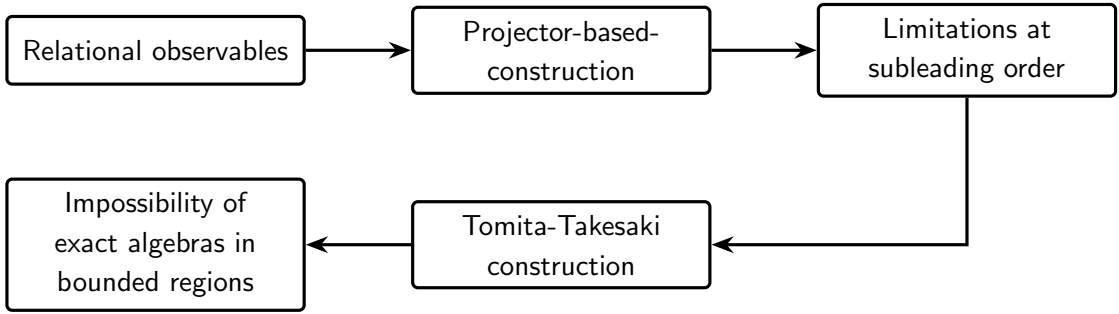

Since this subsection addresses subtle conceptual issues, we provide Figure \ref{logicflowrel} that shows the logic of this subsection.

We have seen in section \ref{secasympcomplete} that the main obstruction to finding algebras for compact regions in gravity arises because the Hamiltonian is a boundary term. So the primary objective of the relational constructions presented below, and other relational constructions, is to define bulk operators that commute with the Hamiltonian as accurately as possible.

\subsubsection{Projector-based construction of relational observables}

This construction was discussed in \cite{Antonini:2025sur}. Although \cite{Antonini:2025sur} do not emphasize this, the construction is based on the Papadodimas-Raju (PR) construction \cite{Papadodimas:2013jku}  of ``mirror operators'' in the black hole interior. 

The PR construction is usually defined by setting up linear equations that define mirror operators,  which commute with a desired set of operators, or equivalently using the Tomita-Takesaki construction \cite{takesaki2006tomita}.  However, in \cite{Papadodimas:2015xma} an alternative construction based on projectors was presented in the context of the eternal black hole. This alternative construction has found utility in recent discussions.  In \cite{Bahiru:2023zlc,Bahiru:2022oas}, it was generalized to study perturbatively localized observables in other states that might not contain black holes. In \cite{Jensen:2024dnl}, the projector-based construction was used to define the algebra of a time band on the AdS boundary in gravitational perturbation theory. It was shown that, by means of a slight change in notation, the PR-projector-based construction of relational observables leads to the same type-II algebra for the time band that was earlier described in \cite{Jensen:2023yxy}. In \cite{Chakraborty:2025izq}, it was used to define observables on the late-time slice of de Sitter space. 

We provide a lightning review of the construction here. A concise but more-detailed review of the construction can be found in Appendix B of \cite{Chakraborty:2025izq}. 

We consider a background state $|\Psi \rangle$ that is time-dependent. For instance, the state might have a planet orbiting a star, which can serve as a clock. The time-dependence implies that the state has a large spread of energies
\be
\delta E = \left( \langle \Psi | H^2 | \Psi \rangle - \langle \Psi | H | \Psi \rangle^2 \right)^{1 \over 2}.
\ee
We introduce an additional parameter $\delta T \gg {1 \over \delta E}$. $\delta T$ can be physically interpreted as the least-count of the clock provided by the background. When $|t| > \delta T$, we expect 
\be
\label{orthrelation}
\langle \Psi | e^{-i H t}  |\Psi \rangle \ll 1 , 
\ee
which is simply the assertion that the feature in the state changes significantly over a time scale ${1 \over \delta E}$ after which the state is almost orthogonal to the original state.  

In this state, we want to define relational operators in a compact region, $I$, whose complement is $R$. Since $R$ includes the asymptotic region, its algebra includes the Hamiltonian.

We define a coarse-grained set of operators from $R$ {\em excluding} the Hamiltonian,
\be
\label{algcoarsedef}
\alset_{\text{coarse}} = \text{span}\{\phi(x_1) , \phi(x_1) \phi(x_2), \ldots \phi(x_1) \phi(x_2) \ldots \phi(x_n) \}, \qquad x_i \in R\,.
\ee
We also cut off the largest polynomial that appears in this coarse-grained set so that $n$ does not scale with any power of $E$. The separation between ordinary operators and the Hamiltonian is reasonable at leading order in $G$.

Since $\alset_{\text{coarse}}$ does not include the Hamiltonian, we also have\footnote{Here, we are assuming that the operator norm of $a$ does not scale with the entropy of the background state.}
\be
\label{overlapissmallwitha}
\langle \Psi | e^{-i H t}  a |\Psi \rangle \ll 1 , \qquad \text{if } t > \delta T ~\text{and}~a \in \alset_{\text{coarse}}.
\ee

Now, consider the little Hilbert space formed by
\be
\label{hpsi}
{\cal H}_{\Psi} = \text{span}\{ e^{-i H \tau} a |\Psi \rangle \}, \qquad a \in \alset_{\text{coarse}}, \tau \in [-{\delta T \over 2}, {\delta T \over 2}].
\ee
Physically, this corresponds to states obtained by acting on the original state with a simple operator and then time translating it. Note that we have separated the action of the Hamiltonian which is not part of $\alset_{\text{coarse}}$.\footnote{Under reasonable assumptions, it may be shown \cite{Jensen:2024dnl} that this space is dense in the space obtained by first time-translating the state and then acting on it with a simple operator. This space is also dense in the space obtained by acting with operators from $R$ and $I$ rather than just operators from $R$.} Let $\hat{P}$ be the projector onto this space. 

Now, consider operator $\phi(x)$ with $x \in I$. For every such operator, consider the improved operator
\be
\label{aimproved}
\widehat{\phi}(x) = \int_{-T_{\text{cut}}}^{T_{\text{cut}}} d t e^{-i H t} \phi(x) \hat{P} e^{i H t} ,
\ee
where $T_{\text{cut}}$ is a cutoff that can be much larger than all time scales in the problem but must be smaller than $e^{{S \over 2}}$ where $S$ is the entropy of the background state.
In some treatments, a second projector is added to the left of the operator $\phi(x)$ although this projector is redundant and so we omit it. 

It is easy to check that
\be
\label{corrsimplematch}
\langle \Psi | \widehat{\phi}(x_1) \ldots \widehat{\phi}(x_n) |\Psi \rangle = \langle \Psi | \bar{\phi}(x_1) \ldots \bar{\phi}(x_n) |\Psi \rangle,
\ee
where
\be
\bar{\phi}(x) = {1 \over \delta T} \int_{-\frac{\delta T}{2}}^{\frac{\delta T}{2}} e^{i H t} \phi(x) e^{-i H t}dt,
\ee
is a slightly smeared version of the original operator. See \cite{Chakraborty:2025izq} for details of the calculation. 

Equation \eqref{corrsimplematch} tells us that, as advertised, correlators of these improved operators in simple states coincide with the original ones up to good accuracy. However
\be
\label{approxexpec}
\langle \Psi | e^{-i H t} \widehat{\phi}(x_1) \ldots \widehat{\phi}(x_n) e^{i H t} |\Psi \rangle = \langle \Psi | \widehat{\phi}(x_1) \ldots \widehat{\phi}(x_n) |\Psi \rangle + {1 \over \delta E \delta T},
\ee
which tells us that these operators commute with the Hamiltonian and time translations to good accuracy.

We may now identify the observables in $I$ as polynomials in $\widehat{\phi}(x)$ with $x \in I$. These observables commute with operators in $R$ to leading order in $G$ due to the improved commutator with the Hamiltonian, which is an observable in $R$. 

A technical remark is in order. In the original construction of \cite{Papadodimas:2015xma},  $\hat{P}$ was taken to the projector on the little Hilbert space obtained by acting on $|\Psi \rangle$ with simple operators (i.e. by setting $t = 0$ in \eqref{hpsi}). Here the projector is over a ``smeared'' little Hilbert space, which also involves an action of $e^{-i H \tau}$.   Without this smearing, the projector is too sharp and while we still get the right one-point functions in \eqref{approxexpec}, we do not get the right higher-point functions. 

The convention of using a projector that is too sharp was carried over to \cite{Bahiru:2023zlc}. This issue was first noticed and fixed in \cite{Jensen:2024dnl}.  

Subsequently, the issue was also noticed in \cite{Antonini:2025sur}, who also attempted to fix this issue. The fix proposed by \cite{Antonini:2025sur} is different from the one proposed in \cite{Jensen:2024dnl}. Rather than smearing the projector, \cite{Antonini:2025sur} propose modifying the projector with a function of the Hamiltonian. This performs a role that is similar to the smearing but it appears to us to require stronger assumptions about the distribution of energy eigenstates in $|\Psi \rangle$ and is also less physically motivated. 

\subsection{Limitations of the construction}
At the moment, it is not understood how the construction should be extended beyond leading order in $G$.  We note two important limitations.
\begin{enumerate}
\item
The construction above relies on the formula \eqref{overlapissmallwitha}. However, this formula does not hold if $a$ can contain powers of the Hamiltonian. For example, note that
\be
\langle \Psi | e^{-i H t} {1 \over H} |\Psi \rangle = \langle \Psi | \left(-i \int_0^t e^{-i H q} d q + {1 \over H} \right) |\Psi \rangle = \Or[{1 \over \delta E}].
\ee
In the limit where the background has a large spread in energy, this term is subleading but it is important if we go to higher orders in the $G$ expansion. 
\item
Also, while the construction above improves the commutator of operators with the Hamiltonian, it does not fix nonzero commutators between other operators inside $I$ and operators outside $R$ \cite{Donnelly:2016rvo}. These commutators must exist because of the OPE \eqref{opebdry} which tells us that the OPEs of two operators in $R$ can generate $H$. But this means that
\be
[\phi(x), \phi(y_1) \phi(y_2)] \neq 0,
\ee
when $x \in I$ and $y_1, y_2 \in R$ at subleading order in $G$. The formula \eqref{aimproved} does not ameliorate this issue, contrary to the assertion in \cite{Antonini:2025sur}.
\end{enumerate}

We do not object to the idea that this construction might be improved to higher orders in perturbation theory. In the next section, we discuss one such approach to a generalization. However, such a generalization has not yet been achieved and we caution against premature claims in the literature to this effect.

\subsection{Tomita-Takesaki construction}
As explained above, the PR construction was originally based on a Tomita-Takesaki-like construction  and the projector-based construction was only introduced as an elegant alternative in the case of the eternal black hole. As also discussed in \cite{Bahiru:2023zlc}, the Tomita-Takesaki based construction might provide an alternative to construct relational operators that commute with the Hamiltonian to all orders in perturbation theory. 

To start, we define a slightly enlarged coarse-grained algebra that includes not only the elements of $\alset_{\text{coarse}}$ but also the Hamiltonian. 
\be
\alset_{\text{coarse}}^{H} = \text{span}\{a_1 e^{-i H t_1},  a_1 e^{-i H t_1} a_2 e^{-i H t_2},  \ldots a_1 e^{-i H t_1} a_n e^{-i H t_n} \}, \qquad a_i \in \alset_{\text{coarse}}.
\ee
Once again we must limit the most complicated operator that appears above by limiting the value of $n$ and also the range of $t_i$.

The virtue of this enlarged algebra is that it removes the distinction between the Hamiltonian and other elements of the algebra of $R$. However, the cutoff is important to ensure that
\be
\label{separatingeq}
a |\Psi \rangle \neq 0, \qquad \forall a \in \alset_{\text{coarse}}^{H}.
\ee
The words used for \eqref{separatingeq} is the state $|\Psi \rangle$ is ``separating'' with respect to $\alset_{\text{coarse}}^H$. 
Next we define the operator
\be
\label{sdef}
S a | \Psi \rangle = a^{\dagger} |\Psi \rangle, \qquad a \in \alset_{\text{coarse}}^{H}\,,
\ee
where $a$ is any element of the set of bulk operators, including the Hamiltonian,  that is not exponentially complicated. Such an operator can always be defined because of \eqref{separatingeq}.

$S$ is an anti-linear operator and we have
\be
S^{\dagger} S = \Delta,
\ee
where $\Delta$ is called the modular operator with the property that
\be
\langle \Psi | a \Delta b |\Psi \rangle = \langle \Psi | b a | \Psi \rangle.
\ee

Finally, we define
\be
J = S \Delta^{-\frac{1}{2}}\,. 
\ee
This operator is anti-unitary and has the property that
\be
J^2 = 1\,.
\ee

For the operators in the region $I$ we now define
\be
\tphi(x) = J \phi(x) J\,.
\ee

It can be easily checked that this operator commutes with all the original operators in the following sense
\be
\tphi(x) a  |\Psi \rangle = a \tphi(x) | \Psi \rangle \Rightarrow [\tphi,a] |\Psi \rangle = 0.
\ee
This means that the commutator annihilates the original state. A short calculation shows the commutator annihilates any state in the little Hilbert space.
\be
[\tphi, a] b |\Psi \rangle = 0.
\ee
However, we see that the action of $\tphi(x)$ is different from the original operator.
\be
\label{actionoftildeonorig}
\tphi | \Psi \rangle = \Delta^{-\frac{1}{2}} \phi(x) \Delta^{1 \over 2} |\Psi \rangle.
\ee

This construction ensures that the improved operators $\tphi(x)$ are in the commutant of the set of operators $\alset_{\text{coarse}}^{H}$. Since this does not rely on a relation of the form \eqref{overlapissmallwitha}, we see that this result continues to hold at subleading order in ${1 \over N}$. 

The downside of this construction is the presence of the modular operator in \eqref{actionoftildeonorig}. It is not clear how to remove these factors, since the conjugation by $\Delta^{1 \over 2}$ is what ensures that  $\tphi(x)$ is Hermitian if $\phi(x)$ is Hermitian.
 
For example, consider a different operator $\tphi'(x) = S \phi(x) S$. This operator acts like the original operator and commutes with elements of the original algebra since we have
\be
\label{actonket}
\tphi'(x) a |\Psi \rangle = a \tphi'(x) | \Psi \rangle = a \phi(x) |\Psi \rangle. 
\ee
On the other hand, if we assume $\tphi'(x)$ is Hermitian then we have
\be
\label{actonbra}
\langle \Psi| \tphi'(x) a |\Psi \rangle = \left(\tphi'(x) |\Psi \rangle, a |\Psi \rangle \right) = \langle \Psi | \phi(x) a | \Psi \rangle.
\ee
But \eqref{actonbra} and \eqref{actonket} are inconsistent whenever
\be
\langle \Psi| [\phi(x), a] |\Psi \rangle \neq 0.
\ee
Therefore $\tphi'(x)$ defined in this manner is not Hermitian.

We do not understand the structure of the modular operator for general states and it is not clear if the map so obtained maps local operators to local operators except in special states where the modular operator has a nice form.
With operators in the region $I$ defined in this manner, one might also worry about the stress-tensor at the interface between $I$ and $R$ since 
\[
\lim_{x_{\text{in}} \rightarrow x_{\text{out}}} \langle \Psi |\tphi(x_{\text{in}}) \phi(x_{\text{out}}) | \Psi \rangle
\]
might not satisfy the Hadamard condition when we take the limit with $x_{\text{in}} \in I$ and $x_{\text{out}} \in R$.

\subsubsection{Relational observables and small commutators}
Both of the relational constructions above were based on the PR construction. However, the PR construction had two aspects that were emphasized in the original papers \cite{Papadodimas:2013jku,Papadodimas:2013wnh}:
\begin{enumerate}
\item
It is possible to define operators in a bounded region that approximately commute with operators outside the region. 
\item
It is impossible to define operators in a bounded region that exactly commute with operators outside the region to the accuracy necessary to define entanglement entropies for the outside and the inside. 
\end{enumerate}
Although \cite{Antonini:2025sur} review the first point, they do not mention the second point above. But the second point above was the key to the PR-proposal for resolving the fuzzball/firewall paradox as originally formulated by Mathur and AMPS \cite{Mathur:2012np,Almheiri:2012rt}. Mathur, and subsequently AMPS, pointed out that if the algebras of operators just inside the black hole, just outside the black hole, and in a distant region were taken to be independent, this would lead to a contradiction with the strong subadditivity of von Neumann entropy.

It was argued in \cite{Papadodimas:2012aq,Papadodimas:2013jku,Papadodimas:2013wnh} that the mirror operators relevant for the black hole interior were redundant with operators in the exterior at a nonperturbative level. Observables outside the horizon can be entangled with operators at infinity but also with observables in the interior because one should identify observables at infinity with operators in the interior. 

It is easy to see this aspect of the construction. The construction above crucially relies on \eqref{separatingeq}. But if we take $B$ to a suitable complicated operator with linear combinations of up to $S$ external field operators, turned to exponential accuracy in $S$, we can arrange that
\be
\label{failuresep}
B |\Psi \rangle = 0; \qquad B^{\dagger} |\Psi \rangle \neq 0.
\ee
For example, we can take $B$ to be a product of a large number of annihilation operators. But then the definition \eqref{sdef} is inconsistent since it equates the action of $S$ on 0 to a nonzero vector. Therefore, the construction of $\tphi$ does not make sense in the presence of insertions of operators like $B$. 

\paragraph{Other relational constructions.}
Our discussion is not restricted to the PR-construction of relational operators; the nonperturbative difficulty described above is very generic.  For instance, some constructions use a set of background fields to define a coordinate system \cite{dewitt1960quantization,Sen:2002qa,Giddings:2005id,Marolf:1994wh}. However, the background fields themselves constitute an excitation in the Hilbert space. Once we start probing the system with operators that are sensitive to the microstate of the background, the construction breaks down. 

The basic mathematical obstruction to constructing a commutant is again \eqref{failuresep}. If the background state (comprising fields or some other excitation) is drawn from an ensemble of $e^{S}$ states, we expect that a polynomial, $B$,  with insertions of $S$ asymptotic operators and coefficients tuned to accuracy $e^{-S}$ can be found that will annihilate the state as in \eqref{failuresep}.

Now, we also want the putative relational operator $\tphi(x)$ to act in a specified way on the background state. For instance, we might demand that it create a  a small excitation at $x$. Without specifying this action in detail, we can write this as
\be
\tphi(x) |\Psi \rangle = |\Psi' \rangle
\ee
However, we now see that
\be
[\tphi(x), B] |\Psi \rangle = 0 \Rightarrow B |\Psi' \rangle = 0.
\ee
But since B is finely tuned to annihilate $|\Psi \rangle$, it generically does not annihilate $|\Psi' \rangle$.  Therefore, an operator that commutes with $B$ and acts in a specified manner on the background cannot generically be found. 

This should not be surprising in light of the result proved in section \ref{secoutsidecomplete}. The principle of holography of information should lead us to anticipate that all relational constructions for compactly supported observables will fail nonperturbatively.

To summarize: relational observables are defined with respect to a background. They cannot be used to define local algebras where the commutator between operators from a compact region and its complement are smaller than $\Or[e^{-S}]$, where $S$ is the entropy of the background.  Therefore, we cannot define local algebras corresponding to the interior of a black hole to the accuracy required to obtain the Page curve, at least in the absence of additional background features whose entropy is larger than the entropy of the black hole. 

\section{Islands in nonstandard gravity and behind double horizons \label{secgoodislands}}

In this section, we describe some possibilities where islands can consistently appear in gravitating regions. In each of these cases, one or more of the assumptions of \ref{secoutsidecomplete} are violated.
\begin{itemize}
\item
Possibility 1: The theory of gravity is not a standard theory of gravity that obeys the assumptions of section \ref{secoutsidecomplete}. For instance, in the case with the nongravitational bath, the bulk theory of gravity is an open quantum system that we refer to as  massive gravity.  The Gauss law does not hold and the projector on the vacuum \eqref{projvac} is not an element of $\alset$. 
\item
Possibility 2: The algebra of operators in the island does not belong to the gravitational theory at all and acts on a completely different Hilbert space. This can be the case when islands appear behind double horizons. In this case the gravitational theory is in a state that is not part of the Hilbert space defined in section \ref{secoutsidecomplete}.
\end{itemize}

We also discuss entanglement wedges that are the {\em union} of an asymptotic region and a compact bulk region. Such entanglement wedges are not ``islands'' according to our definition of the term in this paper or our previous paper \cite{Geng:2021hlu}. Such wedges appear when we consider detectors with ``blind spots'' and the arguments of section \ref{secinconsistency} do not apply to them.

\subsection{Nonstandard theories of gravity}
When a gravitational theory in AdS is coupled to a nongravitational bath, compact entanglement wedges are consistent because the bulk graviton becomes massive.

The fastest way to see the origin of the graviton mass is via a dual boundary argument \cite{Aharony:2006hz}. In the presence of standard boundary conditions, the boundary stress-tensor is conserved. Therefore
\be
\partial_{\mu} T^{\mu \nu}(x) = 0.
\ee
This equation has a natural interpretation in terms of representations of the conformal algebra. The stress-tensor belongs to a ``short representation'' of the boundary conformal algebra \cite{Mack:1974jjo}. The partial derivative is related to the action of the momentum generator, which is an element of the conformal algebra, via
\be
\partial_{\mu} T^{\mu \nu} = -i [P_{\mu}, T^{\mu \nu}(x)] = 0.
\ee
The conformal dimension of short representations of the algebra is fixed by algebraic considerations. Therefore, if we define
\be
[D, T^{\mu \nu}(x)] = \Delta T^{\mu \nu}(x),
\ee
where $D$ is the dilatation operator, conformal representation theory fixes $\Delta = d$. 

On the other hand, in the presence of transparent boundary conditions, the stress-tensor is not conserved. This means that it must pick up an anomalous dimension i.e. $\Delta = d+\delta$ for some $\delta$.  Hence, the bulk dual corresponding to the stress-tensor must be a massive spin-2 field: a massive graviton.  Porrati verified the presence of the mass through a direct one-loop bulk calculation  \cite{Porrati:2003sa,Porrati:2001gx,Porrati:2024zvi}.

 We will adopt the following definition for the purposes of the current discussion.
\begin{definition}
{\bf Massive Gravity:} The bulk theory of gravity is said to be massive when the boundary stress-tensor is not conserved, and therefore picks up an anomalous dimension.
\end{definition}
Alternate definitions of the graviton mass might be relevant in other discussions. 

Furthermore, contrary to the suggestion made in \cite{Antonini:2025sur}, it is evident that, in the presence of transparent boundary conditions, the argument of section \ref{secasympcomplete} breaks down. 

In \cite{Geng:2021hlu}, we gave a perturbative argument for the breakdown of the bulk Gauss law, which suggested that the Hamiltonian would not remain a boundary term. This is also shown in 
 section 4 of \cite{Geng:2025gqu} and an analogous discussion for massive vector fields is in \cite{Kabat:2012av}. We refer the reader to  \cite{Chakravarty:2023cll,Geng:2023zhq,Geng:2025rov,Geng:2025bcb,Geng:2025byh} for additional discussion.  
 
From the more-refined perspective adopted in this paper, we can see that two assumptions made in section \ref{secoutsidecomplete} break down in the presence of the bath.
\begin{enumerate}
\item
Since the gravitating region is only one part of the full spacetime, the energy, as defined by the original boundary Hamiltonian, is no longer bounded below. This is simply because, in quantum field theory, the Hamiltonian is not positive when it is restricted to only part of the spacetime. Therefore the projector on the vacuum of the ADM Hamiltonian used in section \ref{secasympcomplete} is ill defined.
\item
The state of the system coupled to a bath does not lie in the Hilbert space \eqref{superselectedh}. It cannot be obtained by acting on the vacuum of the gravitational system with operators from the asymptotic gravitational region.
\end{enumerate}

This is also clear from the point of view of AdS/CFT. When the boundary theory is coupled to a bath, the original boundary Hamiltonian is no longer sufficient to evolve operators into the future.  Consequently, \eqref{evolvewithH} does not hold and there is no difficulty in producing  operators that satisfy \eqref{commutvanish}.

The original asymptotic operators are all captured in a defect that lives on the boundary of this bath. The bath and the defect together are captured by a nongravitational BCFT that obeys microcausality. Therefore, it is natural for operators in the radiation region --- which is simply part of the bath --- to commute with operators in the defect. In the gravitational description, operators in the island are redundant with operators in the radiation region, and so they also commute with operators in the defect.

\subsubsection{Karch-Randall braneworld models}
The picture presented above is general, and holds whenever the theory is coupled to a bath. However, if the matter content of the bulk theory is chosen correctly, the entire setup including the bath might have a holographic dual. In these so-called ``doubly holographic'' models \cite{Karch:2000ct,Karch:2000gx}, we have a system that can be viewed from three equivalent perspectives
\begin{enumerate}
\item
As a gravitational theory in AdS$_{d+1}$ coupled to a $d+1$-dimensional bath.
\item
As a $d+1$-dimensional BCFT.
\item
As a gravitational theory in AdS$_{d+2}$ bounded by a AdS$_{d+1}$ brane. 
\end{enumerate}

The explanation provided above applies to the duality between perspective 1 and perspective 2 above. But the graviton mass, the appearance of the island and also the fact that the island is consistent are particularly clear from perspective 3  \cite{Chen:2020uac, Chen:2020hmv, Geng:2025rov,Geng:2025byh}. 

Quantized fluctuations in AdS$_{d+2}$ can be dimensionally reduced and viewed as fields on the AdS$_{d+1}$ brane. A remarkable property of this setup \cite{Karch:2000ct} is that the lightest graviton mode is both massive and has a wavefunction that is concentrated near the brane. This provides a bulk explanation of the graviton mass.

\begin{figure}[H]
\begin{centering}
\begin{tikzpicture}[scale=1.4]
\draw[-,very thick,black!40] (-2.5,0) to (0,0);
\draw[-,very thick,black!40] (0,0) to (2.5,0);
\draw[pattern=north west lines,pattern color=black!200,draw=none] (0,0) to (-2.5,-1.875) to (-2.5,0) to (0,0);
\draw[-,dashed,color=black!50] (0,0) to (-1.875,-2.5); 
\draw[-,dashed,color=black!50] (0,0) to (0,-2.875);
\draw[-,dashed,color=black!50] (0,0) to (1.875,-2.375); 
\draw[-,dashed,color=black!50] (0,0) to (2.5,-1.6875); 
\draw[-,very thick,color=green!!50] (0,0) to (-2.5,-1.875);
\node at (0,0) {\textcolor{red}{$\bullet$}};
\node at (0,0) {\textcolor{black}{$\circ$}};
\draw[-,dashed,color=black!50] (-2,-1.5) arc (-140:-2:2.5);
\draw[-,dashed,color=black!50] (-1.5,-1.125) arc (-140:-2:1.875);
\draw[-,dashed,color=black!50] (-1,-0.75) arc (-140:-2:1.25);
\node at (-1.25,-0.9375) {\textcolor{black}{$\times$}};
\node at (-1.15,-1.3) {\textcolor{black}{$\hat{O}(x)$}};
\draw[-,very thick,snake it,color=black!50] (-1.25,-0.9375) arc (-135:-5:1.56);
\end{tikzpicture}
\caption{A demonstration of the dressing for operators inside the island on the brane. The dressing is done using the ambient bulk gravitational Wilson line \cite{Geng:2025rov}.}
\label{pic:braneVdress}
\end{centering}
\end{figure}
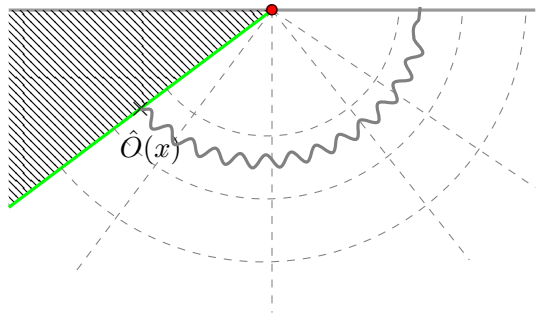

The island is simply the entanglement wedge of a part of the nongravitational boundary as shown in Figure \ref{figislandkr}. It is clear that operators in the island can be dressed to the boundary, while completely bypassing the defect as shown in the Figure \ref{pic:braneVdress} \cite{Geng:2025rov}. Therefore, there is no inconsistency between the gravitational constraints and islands in this setup.
\begin{figure}[H]
\begin{center}
\begin{tikzpicture}[scale=1.4]
\draw[-,very thick,black!100] (-2,0) to (0,0);
\draw[-,very thick,black!100] (0,0) to (1.25,0);
\draw[-,very thick,orange] (1.25,0) to (2.45,0);
\draw[pattern=north west lines,pattern color=purple!200,draw=none] (0,0) to (-2,-1.5) to (-2,0) to (0,0);
\draw[-,very thick,color=green!!50] (0,0) to (-2,-1.5);
\node at (0,0) {\textcolor{red}{$\bullet$}};
\node at (0,0) {\textcolor{black}{$\circ$}};
\draw[-,very thick,blue] (-1,-0.75) to (-2,-1.5);
\node at (-1,-0.75) {\textcolor{black}{$\bullet$}};
\node at (-0.9,-1) {\textcolor{black}{$\partial\mathcal{I}$}};
\draw[-,thick,color=red] (-1,-0.75) arc (-140:-1:1.25);
\draw[-,thick,color=red] (1.22,0) to (1.23,-1.5);
\node at (0,-1.49) {\textcolor{red}{$\gamma_{I}$}};
\node at (1.5,-0.75) {\textcolor{red}{$\gamma_{II}$}};
\end{tikzpicture}
\caption{An island in a KR-braneworld model. When the RT surface $\gamma_{I}$ dominates there is an island on the KR-brane.  \label{figislandkr}}
\end{center}
\end{figure}
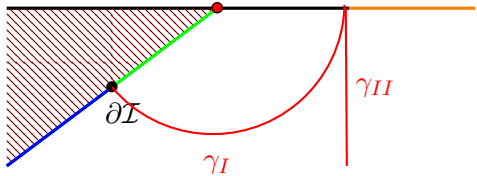

\subsection{Islands behind double horizons}
In a standard theory of gravity, it is still possible to have an island behind a double horizon \cite{Balasubramanian:2020coy,Miyata:2021ncm,Balasubramanian:2021wgd,Miyata:2026mmo}. 

\begin{figure}[ht]
\begin{centering}
\subfloat[Step 1: Coupling an AdS black hole to the bath produces an island.\label{pic:island}]
{\begin{tikzpicture}[scale=0.6,decoration=snake]
       \draw[-,very thick] 
       decorate[decoration={zigzag,pre=lineto,pre
       length=5pt,post=lineto,post length=5pt}] {(-2.5,0) to (2.5,0)};
       \draw[-,very thick,black] (-2.5,0) to (-2.5,-7);
       \draw[-,very thick,blue] (2.5,0) to (2.5,-7);
         \draw[-,very thick] 
       (-2.5,-7) to (2.5,-7);
       \draw[-,very thick,red] (2.5,0) to (-2.5,-5);
        \draw[-,very thick,green] (2.5,0) to (6,-3.5);
         \draw[-,very thick,green] (6,-3.5) to (2.5,-7);

       \draw[fill=green, draw=none, fill opacity = 0.1] (2.5,0) to (6,-3.5) to (2.5,-7) to (2.5,0);
       \draw[-,very thick,black] (-2.5,0) arc (90:0:1.5 and 7);
\draw[thick,black,pattern=north east lines,pattern color=black] (-2.5,0) arc (90:0:1.5 and 7)--(-2.5,-7)--(-2.5,0);
\draw[-,very thick,black!20!green] (6,-3.5) ..controls (2.5,-1.5).. (-2.5,-3.5+2);
\draw[-,very thick,black!40!orange] (-2.5,-3.5+2) to (0,-1.5);
\node at (-0.5,-1) {\textcolor{orange}{$\mathcal{I}$}};
    \end{tikzpicture}
}
\hspace{1cm}
\subfloat[Step 2: The bath is decoupled.\label{pic:decouple}]
{\begin{tikzpicture}[scale=0.6,decoration=snake]
       \draw[-,very thick] 
       decorate[decoration={zigzag,pre=lineto,pre
       length=5pt,post=lineto,post length=5pt}] {(-2.5,0) to (2.5,0)};
       \draw[-,very thick,black] (-2.5,0) to (-2.5,-7);
       \draw[-,very thick,blue] (2.5,0) to (2.5,-7);
         \draw[-,very thick] 
       (-2.5,-7) to (2.5,-7);
       \draw[-,very thick,red] (2.5,0) to (-2.5,-5);
        \draw[-,very thick,green] (2.8,0) to (6,-3.5);
         \draw[-,very thick,green] (2.8,0) to (2.8,-1.7);
         \draw[-,very thick,green] (2.8,-1.7) arc (0:-180:0.15);
         \draw[-,very thick,green] (6,-3.5) to (2.5,-7);

       \draw[fill=green, draw=none, fill opacity = 0.1] (2.8,0) to (6,-3.5) to (2.5,-7) to (2.5,-1.7) arc (-180:0:0.15) to (2.8,0);
       \draw[-,very thick,black] (-2.5,0) arc (90:0:1.5 and 7);
\draw[thick,black,pattern=north east lines,pattern color=black] (-2.5,0) arc (90:0:1.5 and 7)--(-2.5,-7)--(-2.5,0);
\draw[-,very thick,black!20!green] (6,-3.5) ..controls (2.5,-1.5).. (-2.5,-3.5+2);
\draw[-,very thick,black!40!orange] (-2.5,-3.5+2) to (0,-1.5);
\draw[-,very thick,black!40!orange] (-2.5,-3.5+2) to (0,-1.5);
\draw[->,very thick,purple] (2.5,-1.7) to (1.65,-0.85);
\draw[-,very thick,purple] (1.65,-0.85) to (0.8,0);
\node at (-0.5,-1) {\textcolor{orange}{$\mathcal{I}$}};
    \end{tikzpicture}}
    \hspace{1cm}
    \subfloat[Step 3: Evolve back with the decoupled Hamiltonian]
{\begin{tikzpicture}[scale=0.6,decoration=snake]
       \draw[-,very thick] 
       decorate[decoration={zigzag,pre=lineto,pre
       length=5pt,post=lineto,post length=5pt}] {(-3.5,0) to (3.5,0)};
       \draw[-,very thick,black] (-3.5,0) to (-3.5,-7);
       \draw[-,very thick] 
       decorate[decoration={zigzag,pre=lineto,pre
       length=5pt,post=lineto,post length=5pt}] {(-3.5,-7) to (3.5,-7)};
       \draw[-,very thick,blue] (3.5,0) to (3.5,-7);
       \draw[-,very thick,red] (3.5,0) to (-3.5,-7);
       \draw[-,very thick,red] (-3.5,0) to (3.5,-7);
       \draw[-,very thick,black] (-3.5,0) arc (90:-90:1.5 and 3.5);
\draw[thick,black,pattern=north east lines,pattern color=black] (-3.5,0) arc (90:-90:1.5 and 3.5)--(-3.5,-7)--(-3.5,0);
\draw[-,very thick,black!20!green] (3.5,-3.5) to (-3.5,-3.5);
\draw[-,very thick,black!30!orange] (-1,-3.5) to (-3.5,-3.5);
\node at (-1.4,-2.9) {\textcolor{orange}{$\mathcal{I}$}};
    \end{tikzpicture}}
    \caption{A three-step process. We form an island $\mathcal{I}$, decouple the bath (which will generate a shockwave), and then evolve the state back in time with the decoupled Hamiltonian to obtain an island behind the horizon (the red lines are horizons). \label{figthreestepprocess}}
    \end{centering}
\end{figure}
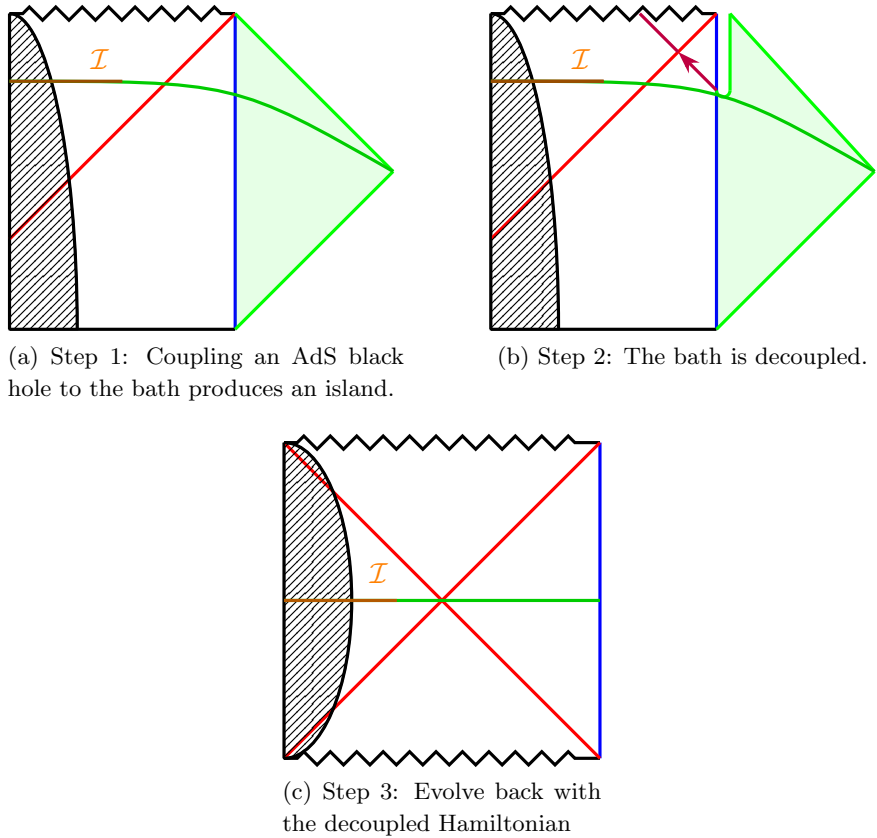

One algorithm to prepare such an island is as follows.
\begin{enumerate}
\item
We couple a standard theory of gravity to a bath, leading to the formation of an island. 
\item
We then decouple the theory. This leads to a shock-wave in the bulk. The boundary theory is still in a mixed state since it is entangled with the bath.
\item
We now evolve the state of the boundary theory back in time using the {\em decoupled} Hamiltonian.
\end{enumerate}
This algorithm leads to the geometry shown in Figure \ref{figthreestepprocess}.

Here, the operators $O$ that act on the island are still associated with the bath. Therefore,  \eqref{commutvanish} is feasible since the operator $O$ does not act on the Hilbert space defined in section \ref{secoutsidecomplete} at all.  More specifically, the assumption of the argument of section \ref{secasympcomplete} that is violated is that we are no longer within the superselection sector \eqref{superselectedh} because the theory is in a mixed state. 

As shown in  Figure \ref{figthreestepprocess}, such an island must necessarily be behind a double horizon. This is because \eqref{commutvanish} tells us that it is impossible to act with a unitary operator in the island that affects anything on the asymptotic boundary. Similarly, it is impossible to act with a unitary operator on the boundary that affects anything in the island. 

However, these islands are irrelevant for an infalling observer who can never reach them. Moreover, the information paradox is concerned with black holes formed from the collapse of a pure state but islands behind a double horizon never appear for such black holes.

\subsection{Noncompact entanglement wedges}
The phrase ``island'' is sometimes used more generally than in our paper to refer to entanglement wedges that have a disconnected piece but also include a part of the asymptotic boundary. An example of this kind is described in \cite{Antonini:2025sur} corresponding to the Page curve discussed in section \ref{pageevaporatingads}. 

\begin{figure}
    \centering
    \includegraphics[width=0.5\linewidth]{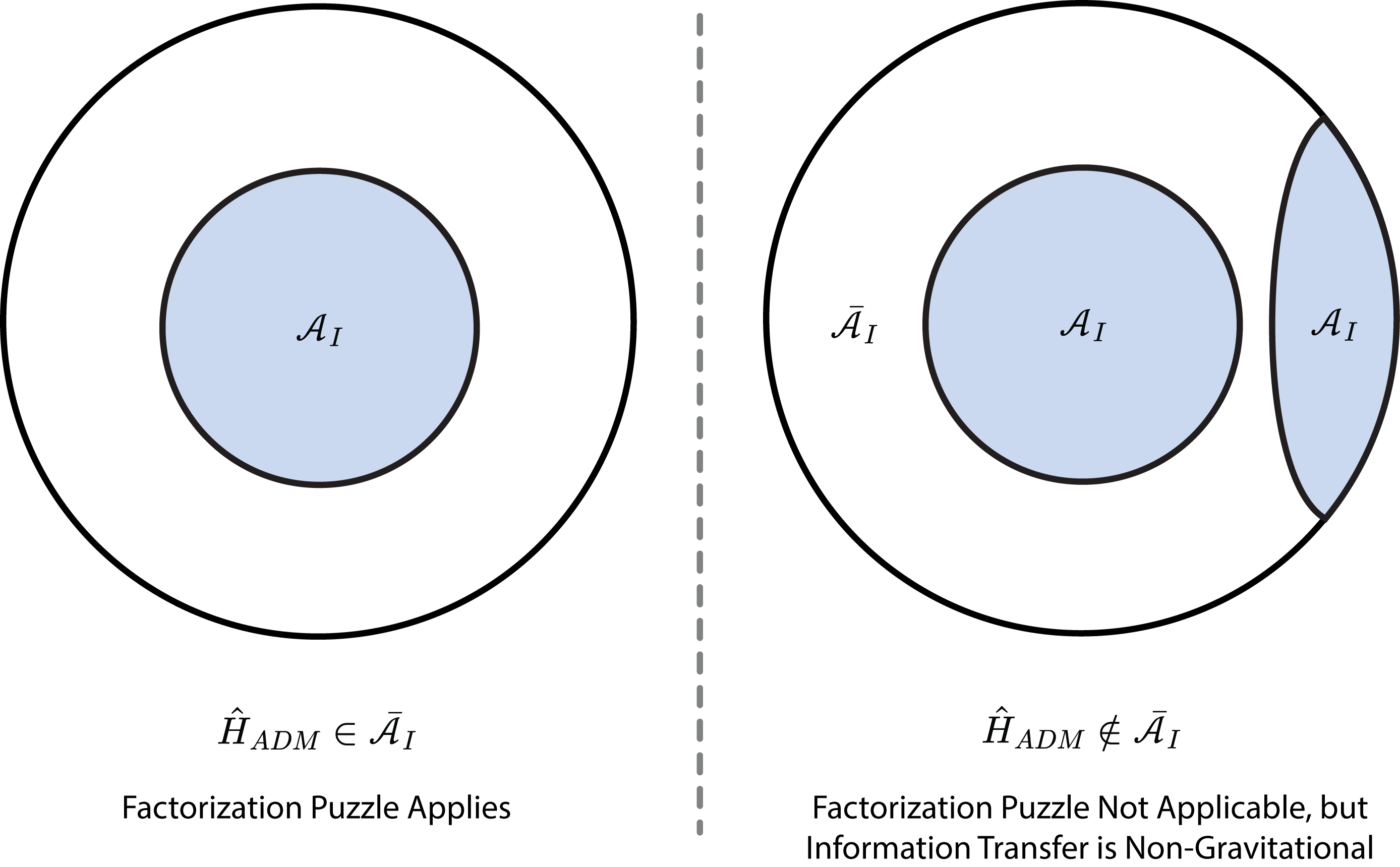}
    \caption{In a time-dependent situation, the entanglement wedge of a boundary subregion may contain a disconnected piece. Such a wedge is not compact and the arguments against the consistency of islands do not apply to this wedge.}
    \label{figislandpiece}
\end{figure}

For the purpose of our discussion, the more-complicated configuration of \cite{Antonini:2025sur} is similar to the configuration shown in Figure \ref{figislandpiece}, which shows an entanglement wedge that is the union of two disconnected parts.

It is clear that the argument of section \ref{secinconsistency} does {\em not} apply to such islands. This is because the Hamiltonian is not part of either the island or its complement. Therefore, the arguments of section \ref{secoutsidecomplete} cannot be used to conclude that the algebra of the complement of the island is complete. It is evident that such wedges do {\em not} provide an example of a compact entanglement wedge; although the asymptotic portion looks ``small'' on the Penrose diagram, it has infinite physical size. 

Just as we discussed in section \ref{pageevaporatingads}, these islands arise because the asymptotic detector has a ``blind spot''. Therefore, while such islands are relevant for the study of information transfer between two parts of the asymptotic boundary, they do not contradict the idea that information is always at infinity in a theory of gravity.

\section{Information transfer in theories with a bath \label{secinformtransfer}}

In this brief section, we turn to models in which  gravity in AdS is coupled to a nongravitational bath.  We will ask the following question: ``what is the mechanism by which information from the black hole interior enters the bath''?
Here, we show that the principle of holography of information  --- which implies that information is always available outside the black hole in a standard gravitational theory --- is helpful for answering this question.

Using a very simple argument, we will show that the information has to flow through the boundary of the gravitational region to enter the radiation region of the bath. Therefore information about the interior is also present on the timelike boundary of the gravitational region, consistent with the physics that we have explained in previous sections.

\begin{figure}[ht]
\begin{centering}
\subfloat
{\begin{tikzpicture}[scale=0.6,decoration=snake]
       \draw[-,very thick] 
       decorate[decoration={zigzag,pre=lineto,pre
       length=5pt,post=lineto,post length=5pt}] {(-2.5,0) to (2.5,0)};
       \draw[-,very thick,black] (-2.5,0) to (-2.5,-7);
       \draw[-,very thick,blue] (2.5,0) to (2.5,-7);
         \draw[-,very thick] 
       (-2.5,-7) to (2.5,-7);
       \draw[-,very thick,red] (2.5,0) to (-2.5,-5);
       \draw[-,very thick,black] (-2.5,0) arc (90:0:1.5 and 7);
\draw[thick,black,pattern=north east lines,pattern color=black] (-2.5,0) arc (90:0:1.5 and 7)--(-2.5,-7)--(-2.5,0);
\draw[-,very thick,black!20!green] (2.5,-4.5) to (-2.5,-4.5);
\node at (2.9,-4.5) {\textcolor{green}{$t_{1}$}};
    \end{tikzpicture}
}
\hspace{1cm}
\subfloat
{\begin{tikzpicture}[scale=0.6,decoration=snake]
       \draw[-,very thick] 
       decorate[decoration={zigzag,pre=lineto,pre
       length=5pt,post=lineto,post length=5pt}] {(-2.5,0) to (2.5,0)};
       \draw[-,very thick,black] (-2.5,0) to (-2.5,-7);
       \draw[-,very thick,blue] (2.5,0) to (2.5,-7);
         \draw[-,very thick] 
       (-2.5,-7) to (2.5,-7);
       \draw[-,very thick,red] (2.5,0) to (-2.5,-5);
        \draw[-,very thick,green] (2.5,0) to (6,-3.5);
         
        \draw[-,very thick,green] (6,-3.5) to (2.8+0.3,-7);
         \draw[-,very thick,green] (2.8+0.3,-7) to (2.8+0.3,-4-0.3);
           \draw[-,very thick,green] (2.8+0.3,-4-0.3) arc (0:180:0.3);
       \draw[fill=green, draw=none, fill opacity = 0.1] (2.5,0) to (6,-3.5) to (2.8+0.3,-7) to (2.8+0.3,-4-0.3) arc (0:180:0.3) to (2.5,0);
       \draw[-,very thick,black] (-2.5,0) arc (90:0:1.5 and 7);
\draw[thick,black,pattern=north east lines,pattern color=black] (-2.5,0) arc (90:0:1.5 and 7)--(-2.5,-7)--(-2.5,0);
\draw[-,very thick,black!20!green] (6,-3.5) ..controls (2.5,-1.5).. (-2.5,-3.5+2);
\draw[-,very thick,black!40!orange] (-2.5,-3.5+2) to (0,-1.5);
\draw[-,very thick,black!40!orange] (-2.5,-3.5+2) to (0,-1.5);
\draw[-,very thick,black!20!green] (2.5,-4.5) to (-2.5,-4.5);
\node at (2.8,-4.5) {\textcolor{green}{$t_{1}$}};
\node at (-0.5,-1) {\textcolor{orange}{$\mathcal{I}$}};
    \end{tikzpicture}}
    \caption{A two step process in which a black hole is first formed and, only later, coupled to a bath. The principle of holography of information is important for understanding information transfer from the black hole interior to the bath. The sudden coupling to the bath will generate a shockwave sent into the bulk. We suppress this effect to avoid cluttering the figure.  \label{figinfotobath}}
    \end{centering}
\end{figure}
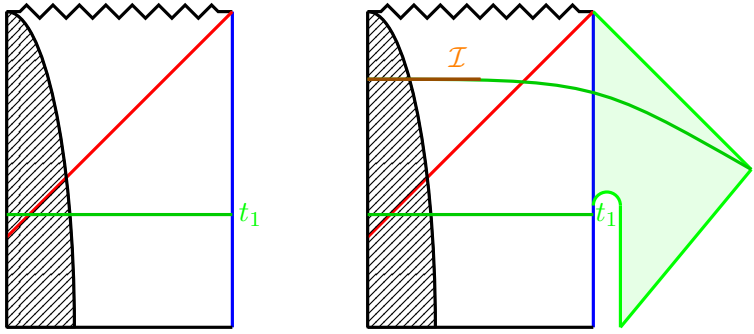

\paragraph{A two-step process.}
Consider the two-step process shown in Figure \ref{figinfotobath}. In the first step we prepare a large black hole in AdS. This black hole does not evaporate. After the black hole has settled down, we couple the boundary to a bath. We denote the time of coupling by $t =0$.

The coupling causes the black hole to evaporate. It is convenient for our purpose to consider the case where at late times the gravitational system is in equilibrium with the bath. If one considers a fixed ``radiation'' region in the bath, then the entropy of this region increases and then saturates at a constant value. Let us call the time at which this happens $t = t_{\text{page}}$. Beyond this time, degrees of freedom in the radiation region are redundant with those in an ``island'' in the gravitational region \cite{Almheiri:2019psf}.

In particular, this means that every operator in the island region, $\phi(I)$, can be reconstructed from the radiation region at time $t=t_{\text{page}}$ We can write this schematically in the form
\be
\label{bulkreconsngbath}
\phi(I) = \sum_i \int_{R} d x K_i(x) O_i(t_{\text{page}}, x),
\ee
where $O_i$ is a basis of local operators in the radiation region, $K_i$ are c-number functions whose precise form is not relevant for us, and the integral over the spatial coordinate $x$ runs over the radiation region.

\paragraph{Information from the boundary.}

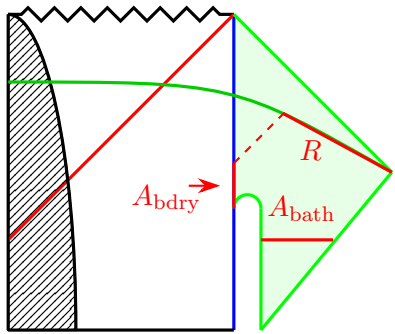
\begin{figure}[!ht]
\begin{centering}
\begin{tikzpicture}[scale=0.6,decoration=snake]
       \draw[-,very thick] 
       decorate[decoration={zigzag,pre=lineto,pre
       length=5pt,post=lineto,post length=5pt}] {(-2.5,0) to (2.5,0)};
       \draw[-,very thick,black] (-2.5,0) to (-2.5,-7);
       \draw[-,very thick,blue] (2.5,0) to (2.5,-7);
         \draw[-,very thick] 
       (-2.5,-7) to (2.5,-7);
       \draw[-,very thick,red] (2.5,0) to (-2.5,-5);
        \draw[-,very thick,green] (2.5,0) to (6,-3.5);
         
        \draw[-,very thick,green] (6,-3.5) to (2.8+0.3,-7);
         \draw[-,very thick,green] (2.8+0.3,-7) to (2.8+0.3,-4-0.3);
           \draw[-,very thick,green] (2.8+0.3,-4-0.3) arc (0:180:0.3);
       \draw[fill=green, draw=none, fill opacity = 0.1] (2.5,0) to (6,-3.5) to (2.8+0.3,-7) to (2.8+0.3,-4-0.3) arc (0:180:0.3) to (2.5,0);
       \draw[-,very thick,black] (-2.5,0) arc (90:0:1.5 and 7);
\draw[thick,black,pattern=north east lines,pattern color=black] (-2.5,0) arc (90:0:1.5 and 7)--(-2.5,-7)--(-2.5,0);
\draw[-,very thick,black!20!green] (6,-3.5) ..controls (2.5,-1.5).. (-2.5,-3.5+2);
\draw[-,very thick,red] (3.1,-5) to (4.7,-5);
\node at (4,-4.3) {\textcolor{red}{$A_{\text{bath}}$}};
\draw[-,very thick,red] (2.5,-4.3) to (2.5,-3.3);
\draw[->,thick,red] (1.5,-3.8) to (2.2,-3.8);
\node at (1,-4.1) {\textcolor{red}{$A_{\text{bdry}}$}};
\draw[-,thick,dashed,red] (2.5,-3.3) to (2.5+1.2,-3.3+1.2);
\draw[-,very thick,red] (2.5+1.1,-3.3+1.1) to (6,-3.5);
\node at (4+0.2,-3.2+0.2) {\textcolor{red}{$R$}};
    \end{tikzpicture}
    \caption{A model where the bath is coupled at $t=0$. All operators in the radiation region can be traced back to operators in its causal past on the interface, and operators in its causal past in the bath. We suppress the shockwave generated by the coupling as in Figure \ref{figinfotobath}. \label{figcausalprop}}
    \end{centering}
\end{figure}

Next, we note that since the bath is a nongravitational theory, microcausality holds as an exact statement for all operators in the bath, including the radiation region.  Therefore, referring to Figure \ref{figcausalprop}, we see that all operators in the radiation region at $t = t_{\text{page}}$ can be written in terms of operators from the bath at $t = -\epsilon$ (before the coupling) from the causal past of the radiation region and  operators at the AdS boundary that also lie in this causal past. Let us call the algebras of these two regions $A_{\text{bath}}$ and $A_{\text{bdry}}$ respectively. 

Therefore, by microcausality in the bath, every operator in the radiation region at late times  is an element of this algebra
\be
O_i(t_{\text{page}}, x) \in A_{\text{bath}} \otimes A_{\text{bdry}}.
\ee

However, since $A_{\text{bath}}$ contains operators in the bath {\em before} it was coupled to the AdS region, these operators cannot carry any information about the black hole interior.  This leads us to the conclusion that they must drop out of the right-hand side of the formula \eqref{bulkreconsngbath}, which leads to
\be
\label{islandinbdry}
\phi(I) \in A_{\text{bdry}}.
\ee

In words, this is the following conclusion.
\begin{conclusion}
The information in the radiation region must have been present at the boundary of AdS at an earlier time.
\end{conclusion}

If one takes the factorization of the bulk Hilbert space  seriously, it would be puzzling that information from the island has entered the boundary  \cite{Martinec:2022lsb}.  The island is spacelike to the boundary. So if the algebra of the island had commuted with $\al_{\text{bdry}}$, this would have been  in contradiction with \eqref{islandinbdry}. 

 This is simply a version of Hawking's information paradox, with the asymptotic algebra at ${\cal I}^{+}$ replaced with the algebra of a finite segment of the AdS boundary. Therefore, the assumption that the bulk Hilbert space factorizes is in contradiction with the successful calculations of the Page curve of the radiation region.

It is sometimes believed that information passes from the island to the radiation region via a ``wormhole''. These suggestions conflate the replica wormholes that enter the computation of the entanglement entropy in the Euclidean path integral with physical wormholes that can transmit information from the interior to the radiation region. However, we see that \eqref{islandinbdry} rules out such a mechanism. The information in the island has to {\em pass through} the boundary before reaching the radiation region and it cannot bypass the boundary through some exotic mechanism. 

The principle of holography of information provides a natural resolution to the puzzle above. The information about the black hole interior was present in the boundary at all times. So it propagates only causally in the bath from the boundary.

As already discussed above, this is not in contradiction with bulk causality. The information in the island entered the island from the exterior, which is in causal contact with the boundary.  This is clear from Figure \ref{figcausalprop}.

Note that the principle of holography of information is sufficient but not necessary to resolve the puzzle above. The puzzle only shows that operators in ${\cal A}_{\text{bdry}}$ must have information about the black hole interior, but does not show that they are restricted to an $\epsilon$-time band on the boundary.

\paragraph{Information in Hawking radiation.}
The discussion above emphasizes that fluctuations of the metric and of other degrees of freedom at the boundary of the gravitational region have information about the interior. Therefore, the radiation that enters the bath is sensitive to this information.

While the statement above is precise, it pertains to radiation that originates at the interface between AdS and the bath. It is tempting to extend these words to the horizon of an evaporating black hole. One might say that the fluctuations of the metric and other degrees of freedom at the horizon have information about the interior. So it is natural that the Hawking radiation produced near the horizon also gains access to this information.

The problem in making this proposal precise is that the horizon is part of a fluctuating bulk spacetime. The virtue of the setup with the bath is that the radiation entering the bath originates at an asymptotic boundary.

\section{Discussion \label{secdiscussion}}
The points made in this paper can be summarized as follows. 

Imagine an idealized detector that surrounds the black hole completely and makes observations in the exterior region.  Such a detector probes the algebra of the black hole's exterior. If the detector lacks sufficient precision, it will register only thermal Hawking radiation, which is insensitive to the microstate of the black hole interior.  However, if the detector is sufficiently accurate, then observations in the exterior will be sensitive to information inside the black hole at all times. This is a unique physical effect that appears in quantum gravity --- observables in the exterior encode complete information about the interior.

This effect, called the principle of holography of information, can be established under only weak assumptions about the UV-complete theory, as reviewed in section \ref{secoutsidecomplete}. Once this feature of gravity is taken into account, Hawking’s argument for information loss --- which is based on an assumed factorization of the bulk Hilbert space --- no longer applies.

Although Page's analysis for the entanglement entropy of a subsystem led to an important result for nongravitational systems, it did not account for the principle of  holography of information. Therefore, while it is true that coarse-grained observations lead to an entropy that follows the Hawking curve, fine-grained observations lead to an entropy that vanishes at all times, rather than a Page curve that first rises and then falls. 

Nevertheless, if we insist on realizing a Page curve, it is possible to invent models that display this behavior. In models with a nongravitational bath, a Page curve emerges because gravity ``switches off'' at the interface between AdS and the bath. However, this Page curve is telling us about transfer of information between parts of the bath and not between the black hole and the exterior. 

Recently \cite{Antonini:2025sur} obtained a Page curve by dividing the boundary into parts and studying transfer of information from one part to another. Physically, this is like considering a detector with a blind spot. We then create a very atypical black hole that is located near the blind spot. The Page curve so obtained is simply a measure of the information that is lost due to this blind spot.

Page curves can be obtained by dividing the boundary into two parts even for a large black hole in AdS that never evaporates.  The entanglement entropy of one part of the boundary follows a Page curve as a function of its size.  While the Page curve for an evaporating black hole is technically harder to compute, it tells us as little or as much about the information paradox as this Page curve for a static black hole.

The paper \cite{Antonini:2025sur} also reviewed the Page curve described in \cite{Laddha:2020kvp, Raju:2020smc}, where it was shown that at ${\cal I}^{+}$, it is possible to discard information about the energy of the black hole, leaving behind an algebra whose entropy follows a Page curve. While this is mathematically consistent, natural observables such as the Riemann tensor carry information about the energy. Therefore, in this example as well, a Page curve only emerges because we choose not to access some information that is naturally accessible to an observer in the exterior.

Islands, by which we mean entanglement wedges whose overlap with the gravitational region is compact, often accompany Page curves. However, islands are inconsistent in a standard theory of gravity in a pure state. They can appear consistently in theories with a nongravitational bath --- where the graviton picks up a mass --- or behind double horizons.  It is also possible to have entanglement wedges that comprise a union of a compact region and an asymptotic region. Such islands emerge when we study detectors with blind spots discussed above. 

While Page curves can always be obtained by dividing a nongravitational region into two parts --- whether the boundary or the bath --- an interesting open question is whether we can define a Page curve for a small evaporating black hole in AdS while allowing observations on the entire boundary. This setup provides a precise model of a black hole in flat space that is observed using a detector at large but finite distance from the black hole. Unlike ${\cal I}^{+}$, it is not straightforward to drop the Hamiltonian from the asymptotic algebra since it reappears in the OPE of other operators. In this paper we discussed various proposals to obtain this Page curve but found that none of them achieves this objective. We leave the question of whether we can consistently remove our blinders and still see a Page curve to future work.

\section*{Acknowledgments} We are grateful to Stefano Antonini, Simon Caron-Huot, Chang-Han Chen, Daniel Harlow,  Daniel Jafferis,  Alok Laddha, Juan Maldacena, Henry Maxfield, Jacob McNamara, Kyriakos Papadodimas, Geoff Penington, Pushkal Shrivastava and Neeraj Tata for helpful discussions.  HG, LR and MR are supported by the Gravity, Spacetime, and Particle Physics (GRASP) Initiative from Harvard University. The work of AK was supported in part by DOE grant DE-SC0022021 and by a grant from the Simons Foundation (Grant 651678, AK). Research at ICTS-TIFR is supported by the Department of Atomic Energy, Government of India, under Project Identification No. RTI4001. The work of CP was supported by the U.S. Department of Energy under grant number DESC0019470 and by the Heising-Simons Foundation under the ``Observational Signatures
of Quantum Gravity” collaboration grants 2021-2818 and 2024-5305.

\bibliographystyle{utphys}
\bibliography{references}

\providecommand{\href}[2]{#2}\begingroup\raggedright\begin{thebibliography}{100}

\bibitem{Page:1993df}
D.~N. Page, ``{Average entropy of a subsystem},''
  \href{http://dx.doi.org/10.1103/PhysRevLett.71.1291}{{\em Phys. Rev. Lett.}
  {\bfseries 71} (1993) 1291--1294},
  \href{http://arxiv.org/abs/gr-qc/9305007}{{\ttfamily arXiv:gr-qc/9305007}}.

\bibitem{Page:1993wv}
D.~N. Page, ``{Information in black hole radiation},''
  \href{http://dx.doi.org/10.1103/PhysRevLett.71.3743}{{\em Phys. Rev. Lett.}
  {\bfseries 71} (1993) 3743--3746},
  \href{http://arxiv.org/abs/hep-th/9306083}{{\ttfamily arXiv:hep-th/9306083}}.

\bibitem{Almheiri:2019psf}
A.~Almheiri, N.~Engelhardt, D.~Marolf, and H.~Maxfield, ``{The entropy of bulk
  quantum fields and the entanglement wedge of an evaporating black hole},''
  \href{http://dx.doi.org/10.1007/JHEP12(2019)063}{{\em JHEP} {\bfseries 12}
  (2019) 063}, \href{http://arxiv.org/abs/1905.08762}{{\ttfamily
  arXiv:1905.08762 [hep-th]}}.

\bibitem{Penington:2019npb}
G.~Penington, ``{Entanglement Wedge Reconstruction and the Information
  Paradox},'' \href{http://dx.doi.org/10.1007/JHEP09(2020)002}{{\em JHEP}
  {\bfseries 09} (2020) 002}, \href{http://arxiv.org/abs/1905.08255}{{\ttfamily
  arXiv:1905.08255 [hep-th]}}.

\bibitem{Almheiri:2019qdq}
A.~Almheiri, T.~Hartman, J.~Maldacena, E.~Shaghoulian, and A.~Tajdini,
  ``{Replica Wormholes and the Entropy of Hawking Radiation},''
  \href{http://dx.doi.org/10.1007/JHEP05(2020)013}{{\em JHEP} {\bfseries 05}
  (2020) 013}, \href{http://arxiv.org/abs/1911.12333}{{\ttfamily
  arXiv:1911.12333 [hep-th]}}.

\bibitem{Almheiri:2019hni}
A.~Almheiri, R.~Mahajan, J.~Maldacena, and Y.~Zhao, ``{The Page curve of
  Hawking radiation from semiclassical geometry},''
  \href{http://dx.doi.org/10.1007/JHEP03(2020)149}{{\em JHEP} {\bfseries 03}
  (2020) 149}, \href{http://arxiv.org/abs/1908.10996}{{\ttfamily
  arXiv:1908.10996 [hep-th]}}.

\bibitem{Almheiri:2020cfm}
A.~Almheiri, T.~Hartman, J.~Maldacena, E.~Shaghoulian, and A.~Tajdini, ``{The
  entropy of Hawking radiation},''
  \href{http://dx.doi.org/10.1103/RevModPhys.93.035002}{{\em Rev. Mod. Phys.}
  {\bfseries 93} no.~3, (2021) 035002},
  \href{http://arxiv.org/abs/2006.06872}{{\ttfamily arXiv:2006.06872
  [hep-th]}}.

\bibitem{Geng:2024xpj}
H.~Geng, ``{Replica wormholes and entanglement islands in the Karch-Randall
  braneworld},'' \href{http://dx.doi.org/10.1007/JHEP01(2025)063}{{\em JHEP}
  {\bfseries 01} (2025) 063}, \href{http://arxiv.org/abs/2405.14872}{{\ttfamily
  arXiv:2405.14872 [hep-th]}}.

\bibitem{Hawking:1976ra}
S.~Hawking, ``{Breakdown of Predictability in Gravitational Collapse},''
  \href{http://dx.doi.org/10.1103/PhysRevD.14.2460}{{\em Phys. Rev. D}
  {\bfseries 14} (1976) 2460--2473}.

\bibitem{Geng:2020qvw}
H.~Geng and A.~Karch, ``{Massive islands},''
  \href{http://dx.doi.org/10.1007/JHEP09(2020)121}{{\em JHEP} {\bfseries 09}
  (2020) 121}, \href{http://arxiv.org/abs/2006.02438}{{\ttfamily
  arXiv:2006.02438 [hep-th]}}.

\bibitem{Geng:2020fxl}
H.~Geng, A.~Karch, C.~Perez-Pardavila, S.~Raju, L.~Randall, M.~Riojas, and
  S.~Shashi, ``{Information Transfer with a Gravitating Bath},''
  \href{http://arxiv.org/abs/2012.04671}{{\ttfamily arXiv:2012.04671
  [hep-th]}}.

\bibitem{Geng:2021hlu}
H.~Geng, A.~Karch, C.~Perez-Pardavila, S.~Raju, L.~Randall, M.~Riojas, and
  S.~Shashi, ``{Inconsistency of islands in theories with long-range
  gravity},'' \href{http://dx.doi.org/10.1007/JHEP01(2022)182}{{\em JHEP}
  {\bfseries 01} (2022) 182}, \href{http://arxiv.org/abs/2107.03390}{{\ttfamily
  arXiv:2107.03390 [hep-th]}}.

\bibitem{Laddha:2020kvp}
A.~Laddha, S.~G. Prabhu, S.~Raju, and P.~Shrivastava, ``{The Holographic Nature
  of Null Infinity},''
  \href{http://dx.doi.org/10.21468/SciPostPhys.10.2.041}{{\em SciPost Phys.}
  {\bfseries 10} no.~2, (2021) 041},
  \href{http://arxiv.org/abs/2002.02448}{{\ttfamily arXiv:2002.02448
  [hep-th]}}.

\bibitem{Raju:2020smc}
S.~Raju, ``{Lessons from the information paradox},''
  \href{http://dx.doi.org/10.1016/j.physrep.2021.10.001}{{\em Phys. Rept.}
  {\bfseries 943} (2022) 1--80},
  \href{http://arxiv.org/abs/2012.05770}{{\ttfamily arXiv:2012.05770
  [hep-th]}}.

\bibitem{Raju:2021lwh}
S.~Raju, ``{Failure of the split property in gravity and the information
  paradox},'' \href{http://dx.doi.org/10.1088/1361-6382/ac482b}{{\em Class.
  Quant. Grav.} {\bfseries 39} no.~6, (2022) 064002},
  \href{http://arxiv.org/abs/2110.05470}{{\ttfamily arXiv:2110.05470
  [hep-th]}}.

\bibitem{Antonini:2025sur}
S.~Antonini, C.-H. Chen, H.~Maxfield, and G.~Penington, ``{An apologia for
  islands},'' \href{http://arxiv.org/abs/2506.04311}{{\ttfamily
  arXiv:2506.04311 [hep-th]}}.

\bibitem{Geng:2025byh}
H.~Geng, ``{Making the Case for Massive Islands},''
  \href{http://arxiv.org/abs/2509.22775}{{\ttfamily arXiv:2509.22775
  [hep-th]}}.

\bibitem{Papadodimas:2015xma}
K.~Papadodimas and S.~Raju, ``{Local Operators in the Eternal Black Hole},''
  \href{http://dx.doi.org/10.1103/PhysRevLett.115.211601}{{\em Phys. Rev.
  Lett.} {\bfseries 115} no.~21, (2015) 211601},
  \href{http://arxiv.org/abs/1502.06692}{{\ttfamily arXiv:1502.06692
  [hep-th]}}.

\bibitem{Papadodimas:2015jra}
K.~Papadodimas and S.~Raju, ``{Remarks on the necessity and implications of
  state-dependence in the black hole interior},''
  \href{http://dx.doi.org/10.1103/PhysRevD.93.084049}{{\em Phys. Rev. D}
  {\bfseries 93} no.~8, (2016) 084049},
  \href{http://arxiv.org/abs/1503.08825}{{\ttfamily arXiv:1503.08825
  [hep-th]}}.

\bibitem{Bahiru:2023zlc}
E.~Bahiru, A.~Belin, K.~Papadodimas, G.~Sarosi, and N.~Vardian, ``{Holography
  and localization of information in quantum gravity},''
  \href{http://dx.doi.org/10.1007/JHEP05(2024)261}{{\em JHEP} {\bfseries 05}
  (2024) 261}, \href{http://arxiv.org/abs/2301.08753}{{\ttfamily
  arXiv:2301.08753 [hep-th]}}.

\bibitem{Bahiru:2022oas}
E.~Bahiru, A.~Belin, K.~Papadodimas, G.~Sarosi, and N.~Vardian,
  ``{State-dressed local operators in the AdS/CFT correspondence},''
  \href{http://dx.doi.org/10.1103/PhysRevD.108.086035}{{\em Phys. Rev. D}
  {\bfseries 108} no.~8, (2023) 086035},
  \href{http://arxiv.org/abs/2209.06845}{{\ttfamily arXiv:2209.06845
  [hep-th]}}.

\bibitem{Jensen:2024dnl}
K.~Jensen, S.~Raju, and A.~J. Speranza, ``{Holographic observers for time-band
  algebras},'' \href{http://dx.doi.org/10.1007/JHEP06(2025)242}{{\em JHEP}
  {\bfseries 06} (2025) 242}, \href{http://arxiv.org/abs/2412.21185}{{\ttfamily
  arXiv:2412.21185 [hep-th]}}.

\bibitem{Hawking:1974sw}
S.~Hawking, ``{Particle Creation by Black Holes},''
  \href{http://dx.doi.org/10.1007/BF02345020}{{\em Commun. Math. Phys.}
  {\bfseries 43} (1975) 199--220}. [Erratum: Commun.Math.Phys. 46, 206 (1976)].

\bibitem{Marolf:2008mf}
D.~Marolf, ``{Unitarity and Holography in Gravitational Physics},''
  \href{http://dx.doi.org/10.1103/PhysRevD.79.044010}{{\em Phys. Rev. D}
  {\bfseries 79} (2009) 044010},
  \href{http://arxiv.org/abs/0808.2842}{{\ttfamily arXiv:0808.2842 [gr-qc]}}.

\bibitem{Marolf:2013iba}
D.~Marolf, ``{Holography without strings?},''
  \href{http://dx.doi.org/10.1088/0264-9381/31/1/015008}{{\em Class. Quant.
  Grav.} {\bfseries 31} (2014) 015008},
  \href{http://arxiv.org/abs/1308.1977}{{\ttfamily arXiv:1308.1977 [hep-th]}}.

\bibitem{Karch:2000ct}
A.~Karch and L.~Randall, ``{Locally localized gravity},''
  \href{http://dx.doi.org/10.1088/1126-6708/2001/05/008}{{\em JHEP} {\bfseries
  05} (2001) 008}, \href{http://arxiv.org/abs/hep-th/0011156}{{\ttfamily
  arXiv:hep-th/0011156}}.

\bibitem{Porrati:2001gx}
M.~Porrati, ``{Mass and gauge invariance 4. Holography for the Karch-Randall
  model},'' \href{http://dx.doi.org/10.1103/PhysRevD.65.044015}{{\em Phys. Rev.
  D} {\bfseries 65} (2002) 044015},
  \href{http://arxiv.org/abs/hep-th/0109017}{{\ttfamily arXiv:hep-th/0109017}}.

\bibitem{Aharony:2006hz}
O.~Aharony, A.~B. Clark, and A.~Karch, ``{The CFT/AdS correspondence, massive
  gravitons and a connectivity index conjecture},''
  \href{http://dx.doi.org/10.1103/PhysRevD.74.086006}{{\em Phys. Rev. D}
  {\bfseries 74} (2006) 086006},
  \href{http://arxiv.org/abs/hep-th/0608089}{{\ttfamily arXiv:hep-th/0608089}}.

\bibitem{Geng:2023ynk}
H.~Geng, ``{Open AdS/CFT via a double-trace deformation},''
  \href{http://dx.doi.org/10.1007/JHEP09(2024)012}{{\em JHEP} {\bfseries 09}
  (2024) 012}, \href{http://arxiv.org/abs/2311.13633}{{\ttfamily
  arXiv:2311.13633 [hep-th]}}.

\bibitem{Geng:2023zhq}
H.~Geng, ``{Graviton Mass and Entanglement Islands in Low Spacetime
  Dimensions},'' \href{http://dx.doi.org/10.21468/SciPostPhys.19.6.146}{{\em
  SciPost Phys.} {\bfseries 19} (2025) 146},
  \href{http://arxiv.org/abs/2312.13336}{{\ttfamily arXiv:2312.13336
  [hep-th]}}.

\bibitem{Geng:2025rov}
H.~Geng, ``{The Mechanism behind the Information Encoding for Islands},''
  \href{http://arxiv.org/abs/2502.08703}{{\ttfamily arXiv:2502.08703
  [hep-th]}}.

\bibitem{Geng:2025gqu}
H.~Geng, D.~Jafferis, P.~Shrivastava, and N.~Tata, ``{The Fate of Information
  Localizability and Holography in Quantum Gravity},''
  \href{http://arxiv.org/abs/2512.18912}{{\ttfamily arXiv:2512.18912
  [hep-th]}}.

\bibitem{lloyd1988black}
S.~Lloyd, {\em Black Holes, Demons and the Loss of Coherence: How complex
  systems get information, and what they do with it}.
\newblock PhD thesis, Rockefeller University, 1988.

\bibitem{Raju:2018xue}
S.~Raju and P.~Shrivastava, ``{Critique of the fuzzball program},''
  \href{http://dx.doi.org/10.1103/PhysRevD.99.066009}{{\em Phys. Rev. D}
  {\bfseries 99} no.~6, (2019) 066009},
  \href{http://arxiv.org/abs/1804.10616}{{\ttfamily arXiv:1804.10616
  [hep-th]}}.

\bibitem{Bekenstein:1973ur}
J.~D. Bekenstein, ``{Black holes and entropy},''
  \href{http://dx.doi.org/10.1103/PhysRevD.7.2333}{{\em Phys. Rev. D}
  {\bfseries 7} (1973) 2333--2346}.

\bibitem{Gibbons:1976ue}
G.~W. Gibbons and S.~W. Hawking, ``{Action Integrals and Partition Functions in
  Quantum Gravity},'' \href{http://dx.doi.org/10.1103/PhysRevD.15.2752}{{\em
  Phys. Rev. D} {\bfseries 15} (1977) 2752--2756}.

\bibitem{Mathur:2009hf}
S.~D. Mathur, ``{The Information paradox: A Pedagogical introduction},''
  \href{http://dx.doi.org/10.1088/0264-9381/26/22/224001}{{\em Class. Quant.
  Grav.} {\bfseries 26} (2009) 224001},
  \href{http://arxiv.org/abs/0909.1038}{{\ttfamily arXiv:0909.1038 [hep-th]}}.

\bibitem{Jacobson:2012gh}
T.~Jacobson, ``{Boundary unitarity and the black hole information paradox},''
  \href{http://dx.doi.org/10.1142/S0218271813420029}{{\em Int. J. Mod. Phys.}
  {\bfseries D22} (2013) 1342002},
\href{http://arxiv.org/abs/1212.6944}{{\ttfamily arXiv:1212.6944 [hep-th]}}.

\bibitem{Jacobson:2019gnm}
T.~Jacobson and P.~Nguyen, ``{Diffeomorphism invariance and the black hole
  information paradox},''
  \href{http://dx.doi.org/10.1103/PhysRevD.100.046002}{{\em Phys. Rev. D}
  {\bfseries 100} no.~4, (2019) 046002},
  \href{http://arxiv.org/abs/1904.04434}{{\ttfamily arXiv:1904.04434 [gr-qc]}}.

\bibitem{deMelloKoch:2022sul}
R.~de~Mello~Koch and G.~Kemp, ``{Holography of information in AdS/CFT},''
  \href{http://dx.doi.org/10.1007/JHEP12(2022)095}{{\em JHEP} {\bfseries 12}
  (2022) 095}, \href{http://arxiv.org/abs/2210.11066}{{\ttfamily
  arXiv:2210.11066 [hep-th]}}.

\bibitem{deMelloKoch:2024juz}
R.~de~Mello~Koch, G.~Kemp, and H.~J.~R. Van~Zyl, ``{Bilocal holography and
  locality in the bulk},''
  \href{http://dx.doi.org/10.1007/JHEP04(2024)079}{{\em JHEP} {\bfseries 04}
  (2024) 079}, \href{http://arxiv.org/abs/2403.07606}{{\ttfamily
  arXiv:2403.07606 [hep-th]}}.

\bibitem{Maldacena:1997re}
J.~M. Maldacena, ``{The Large $N$ limit of superconformal field theories and
  supergravity},'' \href{http://dx.doi.org/10.4310/ATMP.1998.v2.n2.a1}{{\em
  Adv. Theor. Math. Phys.} {\bfseries 2} (1998) 231--252},
  \href{http://arxiv.org/abs/hep-th/9711200}{{\ttfamily arXiv:hep-th/9711200}}.

\bibitem{Witten:1998qj}
E.~Witten, ``{Anti de Sitter space and holography},''
  \href{http://dx.doi.org/10.4310/ATMP.1998.v2.n2.a2}{{\em Adv. Theor. Math.
  Phys.} {\bfseries 2} (1998) 253--291},
  \href{http://arxiv.org/abs/hep-th/9802150}{{\ttfamily arXiv:hep-th/9802150}}.

\bibitem{Gubser:1998bc}
S.~S. Gubser, I.~R. Klebanov, and A.~M. Polyakov, ``{Gauge theory correlators
  from noncritical string theory},''
  \href{http://dx.doi.org/10.1016/S0370-2693(98)00377-3}{{\em Phys. Lett. B}
  {\bfseries 428} (1998) 105--114},
  \href{http://arxiv.org/abs/hep-th/9802109}{{\ttfamily arXiv:hep-th/9802109}}.

\bibitem{Haag:1992hx}
R.~Haag, {\em {Local quantum physics: Fields, particles, algebras}}.
\newblock Springer, 1992.

\bibitem{Skenderis:2000in}
K.~Skenderis, ``{Asymptotically Anti-de Sitter space-times and their stress
  energy tensor},'' \href{http://dx.doi.org/10.1142/S0217751X0100386X}{{\em
  Int. J. Mod. Phys. A} {\bfseries 16} (2001) 740--749},
  \href{http://arxiv.org/abs/hep-th/0010138}{{\ttfamily arXiv:hep-th/0010138}}.

\bibitem{Strominger:2017zoo}
A.~Strominger, ``{Lectures on the Infrared Structure of Gravity and Gauge
  Theory},'' \href{http://arxiv.org/abs/1703.05448}{{\ttfamily arXiv:1703.05448
  [hep-th]}}.

\bibitem{Arnowitt:1962hi}
R.~L. Arnowitt, S.~Deser, and C.~W. Misner, ``{The Dynamics of general
  relativity},'' \href{http://dx.doi.org/10.1007/s10714-008-0661-1}{{\em Gen.
  Rel. Grav.} {\bfseries 40} (2008) 1997--2027},
  \href{http://arxiv.org/abs/gr-qc/0405109}{{\ttfamily arXiv:gr-qc/0405109}}.

\bibitem{Regge:1974zd}
T.~Regge and C.~Teitelboim, ``{Role of Surface Integrals in the Hamiltonian
  Formulation of General Relativity},''
  \href{http://dx.doi.org/10.1016/0003-4916(74)90404-7}{{\em Annals Phys.}
  {\bfseries 88} (1974) 286}.

\bibitem{Bousso:2017xyo}
R.~Bousso, V.~Chandrasekaran, I.~F. Halpern, and A.~Wall, ``{Asymptotic Charges
  Cannot Be Measured in Finite Time},''
  \href{http://dx.doi.org/10.1103/PhysRevD.97.046014}{{\em Phys. Rev. D}
  {\bfseries 97} no.~4, (2018) 046014},
  \href{http://arxiv.org/abs/1709.08632}{{\ttfamily arXiv:1709.08632
  [hep-th]}}.

\bibitem{Ashtekar:1981sf}
A.~Ashtekar, ``{Asymptotic Quantization of the Gravitational Field},''
  \href{http://dx.doi.org/10.1103/PhysRevLett.46.573}{{\em Phys. Rev. Lett.}
  {\bfseries 46} (1981) 573--576}.

\bibitem{Ashtekar:1981hw}
A.~Ashtekar, ``{Radiative Degrees of Freedom of the Gravitational Field in
  Exact General Relativity},'' \href{http://dx.doi.org/10.1063/1.525169}{{\em
  J. Math. Phys.} {\bfseries 22} (1981) 2885--2895}.

\bibitem{Ashtekar:1987tt}
A.~Ashtekar, {\em {Asymptotic Quantization: based on 1984 Naples Lectures}}.
\newblock Bibliopolis,
1987.
\newblock

\bibitem{Ashtekar:2018lor}
A.~Ashtekar, M.~Campiglia, and A.~Laddha, ``{Null infinity, the BMS group and
  infrared issues},'' \href{http://dx.doi.org/10.1007/s10714-018-2464-3}{{\em
  Gen. Rel. Grav.} {\bfseries 50} no.~11, (2018) 140--163},
  \href{http://arxiv.org/abs/1808.07093}{{\ttfamily arXiv:1808.07093 [gr-qc]}}.

\bibitem{Geng:2024dbl}
H.~Geng, ``{Quantum rods and clock in a gravitational universe},''
  \href{http://dx.doi.org/10.1103/kpqb-jrc8}{{\em Phys. Rev. D} {\bfseries 112}
  no.~2, (2025) 026009}, \href{http://arxiv.org/abs/2412.03636}{{\ttfamily
  arXiv:2412.03636 [hep-th]}}.

\bibitem{Raju:2024gvc}
S.~Raju, ``{How does information emerge from a black hole?},'' in {\em The
  Black Hole Information Paradox: A Fifty-Year Journey}, A.~Akil and C.~Bambi,
  eds.
\newblock Springer, 2025.
\newblock \href{http://arxiv.org/abs/2404.00374}{{\ttfamily arXiv:2404.00374
  [gr-qc]}}.

\bibitem{Chowdhury:2020hse}
C.~Chowdhury, O.~Papadoulaki, and S.~Raju, ``{A physical protocol for observers
  near the boundary to obtain bulk information in quantum gravity},''
  \href{http://dx.doi.org/10.21468/SciPostPhys.10.5.106}{{\em SciPost Phys.}
  {\bfseries 10} no.~5, (2021) 106},
  \href{http://arxiv.org/abs/2008.01740}{{\ttfamily arXiv:2008.01740
  [hep-th]}}.

\bibitem{Chowdhury:2021nxw}
C.~Chowdhury, V.~Godet, O.~Papadoulaki, and S.~Raju, ``{Holography from the
  Wheeler-DeWitt equation},''
  \href{http://dx.doi.org/10.1007/JHEP03(2022)019}{{\em JHEP} {\bfseries 03}
  (2022) 019}, \href{http://arxiv.org/abs/2107.14802}{{\ttfamily
  arXiv:2107.14802 [hep-th]}}.

\bibitem{Chowdhury:2022wcv}
C.~Chowdhury and O.~Papadoulaki, ``{Recovering information in an asymptotically
  flat spacetime in quantum gravity},''
  \href{http://dx.doi.org/10.1088/1361-6382/aca192}{{\em Class. Quant. Grav.}
  {\bfseries 39} no.~24, (2022) 245012},
  \href{http://arxiv.org/abs/2203.07449}{{\ttfamily arXiv:2203.07449
  [hep-th]}}.

\bibitem{Gaddam:2024mqm}
N.~Gaddam and {Ashik H.}, ``{Holography of information in a ball of finite
  radius},'' \href{http://dx.doi.org/10.1007/JHEP06(2025)034}{{\em JHEP}
  {\bfseries 06} (2025) 034}, \href{http://arxiv.org/abs/2410.17316}{{\ttfamily
  arXiv:2410.17316 [hep-th]}}.

\bibitem{Harlow:2018tng}
D.~Harlow and H.~Ooguri, ``{Symmetries in quantum field theory and quantum
  gravity},'' \href{http://dx.doi.org/10.1007/s00220-021-04040-y}{{\em Commun.
  Math. Phys.} {\bfseries 383} no.~3, (2021) 1669--1804},
  \href{http://arxiv.org/abs/1810.05338}{{\ttfamily arXiv:1810.05338
  [hep-th]}}.

\bibitem{Harlow:2018jwu}
D.~Harlow and H.~Ooguri, ``{Constraints on Symmetries from Holography},''
  \href{http://dx.doi.org/10.1103/PhysRevLett.122.191601}{{\em Phys. Rev.
  Lett.} {\bfseries 122} no.~19, (2019) 191601},
  \href{http://arxiv.org/abs/1810.05337}{{\ttfamily arXiv:1810.05337
  [hep-th]}}.

\bibitem{Geng:2025gns}
H.~Geng, J.~Huertas, A.~Karch, L.~Randall, and D.~Thomas, ``{Wet Hair: Global
  Symmetries in Entanglement Islands},''
  \href{http://arxiv.org/abs/2512.11025}{{\ttfamily arXiv:2512.11025
  [hep-th]}}.

\bibitem{Geng:2025bcb}
H.~Geng, Y.~Jiang, and J.~Xu, ``{Algebras, Entanglement Islands, and
  Observers},'' \href{http://arxiv.org/abs/2506.12127}{{\ttfamily
  arXiv:2506.12127 [hep-th]}}.

\bibitem{Ryu:2006ef}
S.~Ryu and T.~Takayanagi, ``{Aspects of Holographic Entanglement Entropy},''
  \href{http://dx.doi.org/10.1088/1126-6708/2006/08/045}{{\em JHEP} {\bfseries
  08} (2006) 045}, \href{http://arxiv.org/abs/hep-th/0605073}{{\ttfamily
  arXiv:hep-th/0605073}}.

\bibitem{Ryu:2006bv}
S.~Ryu and T.~Takayanagi, ``{Holographic derivation of entanglement entropy
  from AdS/CFT},'' \href{http://dx.doi.org/10.1103/PhysRevLett.96.181602}{{\em
  Phys. Rev. Lett.} {\bfseries 96} (2006) 181602},
  \href{http://arxiv.org/abs/hep-th/0603001}{{\ttfamily arXiv:hep-th/0603001}}.

\bibitem{Hubeny:2007xt}
V.~E. Hubeny, M.~Rangamani, and T.~Takayanagi, ``{A Covariant holographic
  entanglement entropy proposal},''
  \href{http://dx.doi.org/10.1088/1126-6708/2007/07/062}{{\em JHEP} {\bfseries
  07} (2007) 062}, \href{http://arxiv.org/abs/0705.0016}{{\ttfamily
  arXiv:0705.0016 [hep-th]}}.

\bibitem{Hubeny:2013gta}
V.~E. Hubeny, H.~Maxfield, M.~Rangamani, and E.~Tonni, ``{Holographic
  entanglement plateaux},''
  \href{http://dx.doi.org/10.1007/JHEP08(2013)092}{{\em JHEP} {\bfseries 08}
  (2013) 092}, \href{http://arxiv.org/abs/1306.4004}{{\ttfamily arXiv:1306.4004
  [hep-th]}}.

\bibitem{Aharony:2005bm}
O.~Aharony, S.~Minwalla, and T.~Wiseman, ``{Plasma-balls in large N gauge
  theories and localized black holes},''
  \href{http://dx.doi.org/10.1088/0264-9381/23/7/001}{{\em Class.Quant.Grav.}
  {\bfseries 23} (2006) 2171--2210},
\href{http://arxiv.org/abs/hep-th/0507219}{{\ttfamily arXiv:hep-th/0507219}}.

\bibitem{Emparan:2009dj}
R.~Emparan and G.~Milanesi, ``{Exact Gravitational Dual of a Plasma Ball},''
  \href{http://dx.doi.org/10.1088/1126-6708/2009/08/012}{{\em JHEP} {\bfseries
  08} (2009) 012}, \href{http://arxiv.org/abs/0905.4590}{{\ttfamily
  arXiv:0905.4590 [hep-th]}}.

\bibitem{Papadodimas:2013jku}
K.~Papadodimas and S.~Raju, ``{State-Dependent Bulk-Boundary Maps and Black
  Hole Complementarity},''
  \href{http://dx.doi.org/10.1103/PhysRevD.89.086010}{{\em Phys. Rev. D}
  {\bfseries 89} no.~8, (2014) 086010},
  \href{http://arxiv.org/abs/1310.6335}{{\ttfamily arXiv:1310.6335 [hep-th]}}.

\bibitem{Plebanski:1976gy}
J.~F. Plebanski and M.~Demianski, ``{Rotating, charged, and uniformly
  accelerating mass in general relativity},''
  \href{http://dx.doi.org/10.1016/0003-4916(76)90240-2}{{\em Annals Phys.}
  {\bfseries 98} (1976) 98--127}.

\bibitem{Geiller:2024ryw}
M.~Geiller, A.~Laddha, and C.~Zwikel, ``{Symmetries of the gravitational
  scattering in the absence of peeling},''
  \href{http://dx.doi.org/10.1007/JHEP12(2024)081}{{\em JHEP} {\bfseries 12}
  (2024) 081}, \href{http://arxiv.org/abs/2407.07978}{{\ttfamily
  arXiv:2407.07978 [gr-qc]}}.

\bibitem{Hawking:2016sgy}
S.~W. Hawking, M.~J. Perry, and A.~Strominger, ``{Superrotation Charge and
  Supertranslation Hair on Black Holes},''
  \href{http://dx.doi.org/10.1007/JHEP05(2017)161}{{\em JHEP} {\bfseries 05}
  (2017) 161}, \href{http://arxiv.org/abs/1611.09175}{{\ttfamily
  arXiv:1611.09175 [hep-th]}}.

\bibitem{jaynes1957information}
E.~T. Jaynes, ``Information theory and statistical mechanics,'' {\em Physical
  review} {\bfseries 106} no.~4, (1957) 620.

\bibitem{jaynes1957informationII}
E.~T. Jaynes, ``Information theory and statistical mechanics. ii,'' {\em Phys.
  Rev.} {\bfseries 108} no.~2, (1957) 171.

\bibitem{deBoer:2026cng}
J.~de~Boer, A.~Rolph, and J.~Hollander, ``{A Reverse Black Hole Information
  Problem},'' \href{http://arxiv.org/abs/2601.22077}{{\ttfamily
  arXiv:2601.22077 [hep-th]}}.

\bibitem{Engelhardt:2014gca}
N.~Engelhardt and A.~C. Wall, ``{Quantum Extremal Surfaces: Holographic
  Entanglement Entropy beyond the Classical Regime},''
  \href{http://dx.doi.org/10.1007/JHEP01(2015)073}{{\em JHEP} {\bfseries 01}
  (2015) 073}, \href{http://arxiv.org/abs/1408.3203}{{\ttfamily arXiv:1408.3203
  [hep-th]}}.

\bibitem{Jafferis:2015del}
D.~L. Jafferis, A.~Lewkowycz, J.~Maldacena, and S.~J. Suh, ``{Relative entropy
  equals bulk relative entropy},''
  \href{http://dx.doi.org/10.1007/JHEP06(2016)004}{{\em JHEP} {\bfseries 06}
  (2016) 004},
\href{http://arxiv.org/abs/1512.06431}{{\ttfamily arXiv:1512.06431 [hep-th]}}.

\bibitem{Faulkner:2017vdd}
T.~Faulkner and A.~Lewkowycz, ``{Bulk locality from modular flow},''
  \href{http://dx.doi.org/10.1007/JHEP07(2017)151}{{\em JHEP} {\bfseries 07}
  (2017) 151}, \href{http://arxiv.org/abs/1704.05464}{{\ttfamily
  arXiv:1704.05464 [hep-th]}}.

\bibitem{Hamilton:2007wj}
A.~Hamilton, D.~N. Kabat, G.~Lifschytz, and D.~A. Lowe, ``{Local bulk operators
  in AdS/CFT and the fate of the BTZ singularity},'' {\em AMS/IP Stud. Adv.
  Math.} {\bfseries 44} (2008) 85--100,
  \href{http://arxiv.org/abs/0710.4334}{{\ttfamily arXiv:0710.4334 [hep-th]}}.

\bibitem{Hamilton:2006fh}
A.~Hamilton, D.~N. Kabat, G.~Lifschytz, and D.~A. Lowe, ``{Local bulk operators
  in AdS/CFT: A Holographic description of the black hole interior},''
  \href{http://dx.doi.org/10.1103/PhysRevD.75.106001}{{\em Phys. Rev. D}
  {\bfseries 75} (2007) 106001},
  \href{http://arxiv.org/abs/hep-th/0612053}{{\ttfamily arXiv:hep-th/0612053}}.
  [Erratum: Phys.Rev.D 75, 129902 (2007)].

\bibitem{Hamilton:2005ju}
A.~Hamilton, D.~N. Kabat, G.~Lifschytz, and D.~A. Lowe, ``{Local bulk operators
  in AdS/CFT: A Boundary view of horizons and locality},''
  \href{http://dx.doi.org/10.1103/PhysRevD.73.086003}{{\em Phys. Rev. D}
  {\bfseries 73} (2006) 086003},
  \href{http://arxiv.org/abs/hep-th/0506118}{{\ttfamily arXiv:hep-th/0506118}}.

\bibitem{Hamilton:2006az}
A.~Hamilton, D.~N. Kabat, G.~Lifschytz, and D.~A. Lowe, ``{Holographic
  representation of local bulk operators},''
  \href{http://dx.doi.org/10.1103/PhysRevD.74.066009}{{\em Phys. Rev. D}
  {\bfseries 74} (2006) 066009},
  \href{http://arxiv.org/abs/hep-th/0606141}{{\ttfamily arXiv:hep-th/0606141}}.

\bibitem{birrell1984quantum}
N.~Birrell and P.~Davies, {\em {Quantum fields in curved space}}.
\newblock Cambridge Univ Press, 1986.

\bibitem{Ghosh:2017gtw}
S.~Ghosh and S.~Raju, ``{Quantum information measures for restricted sets of
  observables},'' \href{http://dx.doi.org/10.1103/PhysRevD.98.046005}{{\em
  Phys. Rev. D} {\bfseries 98} no.~4, (2018) 046005},
  \href{http://arxiv.org/abs/1712.09365}{{\ttfamily arXiv:1712.09365
  [hep-th]}}.

\bibitem{ArkaniHamed:2006dz}
N.~Arkani-Hamed, L.~Motl, A.~Nicolis, and C.~Vafa, ``{The String landscape,
  black holes and gravity as the weakest force},''
  \href{http://dx.doi.org/10.1088/1126-6708/2007/06/060}{{\em JHEP} {\bfseries
  0706} (2007) 060},
\href{http://arxiv.org/abs/hep-th/0601001}{{\ttfamily arXiv:hep-th/0601001}}.

\bibitem{Krishnan:2020fer}
C.~Krishnan, ``{Critical Islands},''
  \href{http://dx.doi.org/10.1007/JHEP01(2021)179}{{\em JHEP} {\bfseries 01}
  (2021) 179}, \href{http://arxiv.org/abs/2007.06551}{{\ttfamily
  arXiv:2007.06551 [hep-th]}}.

\bibitem{Krishnan:2020oun}
C.~Krishnan, V.~Patil, and J.~Pereira, ``{Page Curve and the Information
  Paradox in Flat Space},'' \href{http://arxiv.org/abs/2005.02993}{{\ttfamily
  arXiv:2005.02993 [hep-th]}}.

\bibitem{Ghosh:2021axl}
K.~Ghosh and C.~Krishnan, ``{Dirichlet baths and the not-so-fine-grained Page
  curve},'' \href{http://dx.doi.org/10.1007/JHEP08(2021)119}{{\em JHEP}
  {\bfseries 08} (2021) 119}, \href{http://arxiv.org/abs/2103.17253}{{\ttfamily
  arXiv:2103.17253 [hep-th]}}.

\bibitem{Gautason:2020tmk}
F.~F. Gautason, L.~Schneiderbauer, W.~Sybesma, and L.~Thorlacius, ``{Page Curve
  for an Evaporating Black Hole},''
  \href{http://dx.doi.org/10.1007/JHEP05(2020)091}{{\em JHEP} {\bfseries 05}
  (2020) 091}, \href{http://arxiv.org/abs/2004.00598}{{\ttfamily
  arXiv:2004.00598 [hep-th]}}.

\bibitem{dewitt1960quantization}
B.~S. DeWitt, ``The quantization of geometry,'' in {\em Gravitation: an
  introduction to current research}, L.~Witten, ed.
\newblock John Wiley \& Sons, 1963.

\bibitem{DeWitt:1967yk}
B.~S. DeWitt, ``{Quantum Theory of Gravity. 1. The Canonical Theory},''
  \href{http://dx.doi.org/10.1103/PhysRev.160.1113}{{\em Phys. Rev.} {\bfseries
  160} (1967) 1113--1148}.

\bibitem{Brown:1994py}
J.~D. Brown and K.~V. Kuchar, ``{Dust as a standard of space and time in
  canonical quantum gravity},''
  \href{http://dx.doi.org/10.1103/PhysRevD.51.5600}{{\em Phys. Rev. D}
  {\bfseries 51} (1995) 5600--5629},
  \href{http://arxiv.org/abs/gr-qc/9409001}{{\ttfamily arXiv:gr-qc/9409001}}.

\bibitem{Kuchar:1990vy}
K.~V. Kuchar and C.~G. Torre, ``{Gaussian reference fluid and interpretation of
  quantum geometrodynamics},''
  \href{http://dx.doi.org/10.1103/PhysRevD.43.419}{{\em Phys. Rev. D}
  {\bfseries 43} (1991) 419--441}.

\bibitem{Sen:2002qa}
A.~Sen, ``{Time and tachyon},''
  \href{http://dx.doi.org/10.1142/S0217751X03015313}{{\em Int. J. Mod. Phys. A}
  {\bfseries 18} (2003) 4869--4888},
  \href{http://arxiv.org/abs/hep-th/0209122}{{\ttfamily arXiv:hep-th/0209122}}.

\bibitem{Giddings:2005id}
S.~B. Giddings, D.~Marolf, and J.~B. Hartle, ``{Observables in effective
  gravity},'' \href{http://dx.doi.org/10.1103/PhysRevD.74.064018}{{\em Phys.
  Rev. D} {\bfseries 74} (2006) 064018},
  \href{http://arxiv.org/abs/hep-th/0512200}{{\ttfamily arXiv:hep-th/0512200}}.

\bibitem{Marolf:2015jha}
D.~Marolf, ``{Comments on Microcausality, Chaos, and Gravitational
  Observables},'' \href{http://dx.doi.org/10.1088/0264-9381/32/24/245003}{{\em
  Class. Quant. Grav.} {\bfseries 32} no.~24, (2015) 245003},
  \href{http://arxiv.org/abs/1508.00939}{{\ttfamily arXiv:1508.00939 [gr-qc]}}.

\bibitem{Donnelly:2016rvo}
W.~Donnelly and S.~B. Giddings, ``{Observables, gravitational dressing, and
  obstructions to locality and subsystems},''
  \href{http://dx.doi.org/10.1103/PhysRevD.94.104038}{{\em Phys. Rev.}
  {\bfseries D94} no.~10, (2016) 104038},
\href{http://arxiv.org/abs/1607.01025}{{\ttfamily arXiv:1607.01025 [hep-th]}}.

\bibitem{Francois:2024rdm}
J.~T. Fran{\c{c}}ois and L.~Ravera, ``{Geometric Relational Framework for
  General-Relativistic Gauge Field Theories},''
  \href{http://dx.doi.org/10.1002/prop.202400149}{{\em Fortsch. Phys.}
  {\bfseries 73} no.~1-2, (2025) 2400149},
  \href{http://arxiv.org/abs/2407.04043}{{\ttfamily arXiv:2407.04043 [gr-qc]}}.

\bibitem{DeVuyst:2024grw}
J.~De~Vuyst, S.~Eccles, P.~A. Hoehn, and J.~Kirklin, ``{Linearization
  (in)stabilities and crossed products},''
  \href{http://arxiv.org/abs/2411.19931}{{\ttfamily arXiv:2411.19931
  [hep-th]}}.

\bibitem{Chataignier:2024eil}
L.~Chataignier, P.~A. Hoehn, M.~P.~E. Lock, and F.~M. Mele, ``{Relational
  Dynamics with Periodic Clocks},''
  \href{http://arxiv.org/abs/2409.06479}{{\ttfamily arXiv:2409.06479
  [quant-ph]}}.

\bibitem{AliAhmad:2024wja}
S.~Ali~Ahmad, W.~Chemissany, M.~S. Klinger, and R.~G. Leigh, ``{Quantum
  reference frames from top-down crossed products},''
  \href{http://dx.doi.org/10.1103/PhysRevD.110.065003}{{\em Phys. Rev. D}
  {\bfseries 110} no.~6, (2024) 065003},
  \href{http://arxiv.org/abs/2405.13884}{{\ttfamily arXiv:2405.13884
  [hep-th]}}.

\bibitem{Fewster:2024pur}
J.~C. Fewster, D.~W. Janssen, L.~D. Loveridge, K.~Rejzner, and J.~Waldron,
  ``{Quantum Reference Frames, Measurement Schemes and the Type of Local
  Algebras in Quantum Field Theory},''
  \href{http://dx.doi.org/10.1007/s00220-024-05180-7}{{\em Commun. Math. Phys.}
  {\bfseries 406} no.~1, (2025) 19},
  \href{http://arxiv.org/abs/2403.11973}{{\ttfamily arXiv:2403.11973
  [math-ph]}}.

\bibitem{DeVuyst:2024uvd}
J.~De~Vuyst, S.~Eccles, P.~A. Hoehn, and J.~Kirklin, ``{Crossed products and
  quantum reference frames: on the observer-dependence of gravitational
  entropy},'' \href{http://arxiv.org/abs/2412.15502}{{\ttfamily
  arXiv:2412.15502 [hep-th]}}.

\bibitem{Chen:2024rpx}
C.-H. Chen and G.~Penington, ``{A clock is just a way to tell the time:
  gravitational algebras in cosmological spacetimes},''
  \href{http://arxiv.org/abs/2406.02116}{{\ttfamily arXiv:2406.02116
  [hep-th]}}.

\bibitem{Kudler-Flam:2024psh}
J.~Kudler-Flam, S.~Leutheusser, and G.~Satishchandran, ``{Algebraic
  Observational Cosmology},'' \href{http://arxiv.org/abs/2406.01669}{{\ttfamily
  arXiv:2406.01669 [hep-th]}}.

\bibitem{Kaplan:2024xyk}
M.~Kaplan, D.~Marolf, X.~Yu, and Y.~Zhao, ``{De Sitter quantum gravity and the
  emergence of local algebras},''
  \href{http://dx.doi.org/10.1007/JHEP04(2025)171}{{\em JHEP} {\bfseries 04}
  (2025) 171}, \href{http://arxiv.org/abs/2410.00111}{{\ttfamily
  arXiv:2410.00111 [hep-th]}}.

\bibitem{Papadodimas:2012aq}
K.~Papadodimas and S.~Raju, ``{An Infalling Observer in AdS/CFT},''
  \href{http://dx.doi.org/10.1007/JHEP10(2013)212}{{\em JHEP} {\bfseries 10}
  (2013) 212}, \href{http://arxiv.org/abs/1211.6767}{{\ttfamily arXiv:1211.6767
  [hep-th]}}.

\bibitem{Papadodimas:2013wnh}
K.~Papadodimas and S.~Raju, ``{Black Hole Interior in the Holographic
  Correspondence and the Information Paradox},''
  \href{http://dx.doi.org/10.1103/PhysRevLett.112.051301}{{\em Phys. Rev.
  Lett.} {\bfseries 112} no.~5, (2014) 051301},
  \href{http://arxiv.org/abs/1310.6334}{{\ttfamily arXiv:1310.6334 [hep-th]}}.

\bibitem{takesaki2006tomita}
M.~Takesaki, {\em Tomita's theory of modular Hilbert algebras and its
  applications}, vol.~128.
\newblock Springer, 2006.

\bibitem{Jensen:2023yxy}
K.~Jensen, J.~Sorce, and A.~J. Speranza, ``{Generalized entropy for general
  subregions in quantum gravity},''
  \href{http://dx.doi.org/10.1007/JHEP12(2023)020}{{\em JHEP} {\bfseries 12}
  (2023) 020}, \href{http://arxiv.org/abs/2306.01837}{{\ttfamily
  arXiv:2306.01837 [hep-th]}}.

\bibitem{Chakraborty:2025izq}
T.~Chakraborty, {Ashik H.}, and S.~Raju, ``{Cosmological correlators in
  gravitationally-constrained de Sitter states},''
  \href{http://arxiv.org/abs/2507.15926}{{\ttfamily arXiv:2507.15926
  [hep-th]}}.

\bibitem{Mathur:2012np}
S.~D. Mathur, ``{The information paradox: conflicts and resolutions},''
  \href{http://dx.doi.org/10.1007/s12043-012-0417-z}{{\em Pramana} {\bfseries
  79} (2012) 1059--1073},
\href{http://arxiv.org/abs/1201.2079}{{\ttfamily arXiv:1201.2079 [hep-th]}}.

\bibitem{Almheiri:2012rt}
A.~Almheiri, D.~Marolf, J.~Polchinski, and J.~Sully, ``{Black Holes:
  Complementarity or Firewalls?},''
  \href{http://dx.doi.org/10.1007/JHEP02(2013)062}{{\em JHEP} {\bfseries 02}
  (2013) 062}, \href{http://arxiv.org/abs/1207.3123}{{\ttfamily arXiv:1207.3123
  [hep-th]}}.

\bibitem{Marolf:1994wh}
D.~Marolf, ``{Quantum observables and recollapsing dynamics},''
  \href{http://dx.doi.org/10.1088/0264-9381/12/5/011}{{\em Class. Quant. Grav.}
  {\bfseries 12} (1995) 1199--1220},
  \href{http://arxiv.org/abs/gr-qc/9404053}{{\ttfamily arXiv:gr-qc/9404053}}.

\bibitem{Mack:1974jjo}
G.~Mack, ``{Group Theoretical Approach to Conformal Invariant Quantum Field
  Theory},'' \href{http://dx.doi.org/10.1007/978-1-4615-8909-9_7}{{\em NATO
  Sci. Ser. B} {\bfseries 5} (1974) 123--157}.

\bibitem{Porrati:2003sa}
M.~Porrati, ``{Higgs phenomenon for the graviton in ADS space},''
  \href{http://dx.doi.org/10.1142/S0217732303011745}{{\em Mod. Phys. Lett. A}
  {\bfseries 18} (2003) 1793--1802},
  \href{http://arxiv.org/abs/hep-th/0306253}{{\ttfamily arXiv:hep-th/0306253}}.

\bibitem{Porrati:2024zvi}
M.~Porrati and A.~Zaffaroni, ``{A universal feature for the Higgs phenomenon in
  Anti de Sitter space},''
  \href{http://dx.doi.org/10.1088/1751-8121/ad9e59}{{\em J. Phys. A} {\bfseries
  58} no.~3, (2025) 035401}, \href{http://arxiv.org/abs/2408.16919}{{\ttfamily
  arXiv:2408.16919 [hep-th]}}.

\bibitem{Kabat:2012av}
D.~Kabat and G.~Lifschytz, ``{CFT representation of interacting bulk gauge
  fields in AdS},'' \href{http://dx.doi.org/10.1103/PhysRevD.87.086004}{{\em
  Phys. Rev. D} {\bfseries 87} no.~8, (2013) 086004},
  \href{http://arxiv.org/abs/1212.3788}{{\ttfamily arXiv:1212.3788 [hep-th]}}.

\bibitem{Chakravarty:2023cll}
J.~Chakravarty, D.~Jain, and A.~Sivakumar, ``{Holography of information in
  massive gravity using Dirac brackets},''
  \href{http://dx.doi.org/10.1007/JHEP06(2023)109}{{\em JHEP} {\bfseries 06}
  (2023) 109}, \href{http://arxiv.org/abs/2301.01075}{{\ttfamily
  arXiv:2301.01075 [hep-th]}}.

\bibitem{Karch:2000gx}
A.~Karch and L.~Randall, ``{Open and closed string interpretation of SUSY CFT's
  on branes with boundaries},''
  \href{http://dx.doi.org/10.1088/1126-6708/2001/06/063}{{\em JHEP} {\bfseries
  06} (2001) 063}, \href{http://arxiv.org/abs/hep-th/0105132}{{\ttfamily
  arXiv:hep-th/0105132}}.

\bibitem{Chen:2020uac}
H.~Z. Chen, R.~C. Myers, D.~Neuenfeld, I.~A. Reyes, and J.~Sandor, ``{Quantum
  Extremal Islands Made Easy, Part I: Entanglement on the Brane},''
  \href{http://dx.doi.org/10.1007/JHEP10(2020)166}{{\em JHEP} {\bfseries 10}
  (2020) 166}, \href{http://arxiv.org/abs/2006.04851}{{\ttfamily
  arXiv:2006.04851 [hep-th]}}.

\bibitem{Chen:2020hmv}
H.~Z. Chen, R.~C. Myers, D.~Neuenfeld, I.~A. Reyes, and J.~Sandor, ``{Quantum
  Extremal Islands Made Easy, Part II: Black Holes on the Brane},''
  \href{http://arxiv.org/abs/2010.00018}{{\ttfamily arXiv:2010.00018
  [hep-th]}}.

\bibitem{Balasubramanian:2020coy}
V.~Balasubramanian, A.~Kar, and T.~Ugajin, ``{Entanglement between two disjoint
  universes},'' \href{http://dx.doi.org/10.1007/JHEP02(2021)136}{{\em JHEP}
  {\bfseries 02} (2021) 136}, \href{http://arxiv.org/abs/2008.05274}{{\ttfamily
  arXiv:2008.05274 [hep-th]}}.

\bibitem{Miyata:2021ncm}
A.~Miyata and T.~Ugajin, ``{Evaporation of black holes in flat space entangled
  with an auxiliary universe},''
  \href{http://arxiv.org/abs/2104.00183}{{\ttfamily arXiv:2104.00183
  [hep-th]}}.

\bibitem{Balasubramanian:2021wgd}
V.~Balasubramanian, A.~Kar, and T.~Ugajin, ``{Entanglement between two
  gravitating universes},'' \href{http://arxiv.org/abs/2104.13383}{{\ttfamily
  arXiv:2104.13383 [hep-th]}}.

\bibitem{Miyata:2026mmo}
A.~Miyata and T.~Ugajin, ``{Black Hole Interior and Quantum Error Correction
  with Dynamical Gravity},'' \href{http://arxiv.org/abs/2602.01241}{{\ttfamily
  arXiv:2602.01241 [hep-th]}}.

\bibitem{Martinec:2022lsb}
E.~J. Martinec, ``{Trouble in Paradox},''
  \href{http://arxiv.org/abs/2203.04947}{{\ttfamily arXiv:2203.04947
  [hep-th]}}.

\end{thebibliography}\endgroup
\end{document}